\documentclass[12pt]{article}
\pdfoutput=1
\usepackage{makeidx}
\makeindex
\usepackage[a4paper]{geometry}
\usepackage{jheppub,amsmath,amssymb,amsfonts,amsxtra,mathrsfs,graphics,graphicx,amsthm,epsfig,ytableau,bm,longtable,float,color,tikz,mathtools,xfrac,footnote,rotating,lscape}
\usepackage{dsfont}
\usepackage{multicol}
\usepackage{lipsum}
\usepackage{textgreek}
\usetikzlibrary{decorations.pathmorphing}
\usetikzlibrary{decorations.markings}
\usetikzlibrary{quotes,arrows.meta}
\usetikzlibrary{arrows, decorations.markings, calc, fadings, decorations.pathreplacing, patterns, decorations.pathmorphing, positioning}
\usepackage{tikz-cd}

\usepackage{fixmath} 
\usepackage{scalerel}
\newlength\bshft
\def\fakebold#1{\ThisStyle{\ooalign{$\SavedStyle#1$\cr%
  \kern-\bshft$\SavedStyle#1$\cr%
  \kern\bshft$\SavedStyle#1$}}}

\usetikzlibrary{positioning,shapes}
\usetikzlibrary{chains}
\usetikzlibrary{arrows,fit,decorations.pathreplacing}
\tikzstyle{every picture}+=[remember picture]
\tikzstyle{na} = [baseline=-.5ex]

\usepackage{empheq}
\usepackage{multirow}
\usepackage{booktabs}
\usepackage[american]{babel}

\usepackage[latin1]{inputenc}

\makeatletter
\newcommand{\vast}{\bBigg@{1}}
\newcommand{\Vast}{\bBigg@{5}}
\makeatother

\usepackage{hyperref}

\numberwithin{equation}{section}

\newcommand{\ie}{\textit{i.e.}}

\numberwithin{equation}{section}

\newcommand{\be}{\begin{equation}} \newcommand{\ee}{\end{equation}}
\newcommand{\bea}{\begin{equation} \begin{aligned}} \newcommand{\eea}{\end{aligned} \end{equation}}

\def\U{\mathrm{U}}
\def\SO{\mathrm{SO}}

\def\SU{\mathrm{SU}}

\newcommand{\rd}{\mathrm{d}}

\DeclareMathOperator{\Tr}{Tr}

\newcommand{\cC}{\mathcal{C}}

\newcommand{\cF}{\mathcal{F}}

\newcommand{\cI}{\mathcal{I}}

\newcommand{\cN}{\mathcal{N}}

\newcommand{\cV}{\mathcal{V}}

\newcommand{\bC}{\mathbb{C}}

\newcommand{\bR}{\mathbb{R}}
\newcommand{\bZ}{\mathbb{Z}}

\newcommand{\fg}{\mathfrak{g}}

\newcommand{\fn}{\mathfrak{n}}

\usepackage[Symbol]{upgreek}
\usepackage{bm}

\usepackage{calligra}
\DeclareMathAlphabet{\mathcalligra}{T1}{calligra}{m}{n}

\theoremstyle{plain}

  \theoremstyle{definition}

\providecommand{\examplename}{Example}
\providecommand{\theoremname}{Theorem}

\makeatletter
\g@addto@macro\bfseries{\boldmath}
\makeatother

\makeatletter
\newcommand*{\rom}[1]{\expandafter\@slowromancap\romannumeral #1@}
\makeatother

\title{Geometry of $\mathcal{I}$-extremization \\
and black holes microstates}

\author[a]{Seyed Morteza Hosseini,}
\author[b,c]{and Alberto Zaffaroni}
\affiliation[a]{Kavli IPMU (WPI), UTIAS, The University of Tokyo, Kashiwa, Chiba 277-8583, Japan}
\affiliation[b]{Dipartimento di Fisica, Universit\`a di Milano-Bicocca, I-20126 Milano, Italy}
\affiliation[c]{INFN, sezione di Milano - Bicocca, I-20126 Milano, Italy}
\emailAdd{morteza.hosseini@ipmu.jp}
\emailAdd{alberto.zaffaroni@mib.infn.it}

\preprint{IPMU19-0051}

\abstract{The entropy of a class of asymptotically AdS$_4$ magnetically charged BPS black holes can be obtained by extremizing the topologically twisted index of the dual three-dimensional field theory. This principle is known as $\mathcal{I}$-extremization. A gravitational dual of $\mathcal{I}$-extremization for a class of theories obtained by twisted compactifications of M2-branes living at a Calabi-Yau four-fold has been recently proposed. In this paper we investigate the relation between the two extremization principles. We show that the two extremization procedures are equivalent for theories without baryonic symmetries, which include ABJM and the theory dual to the non-toric Sasaki-Einstein manifold $V^{5,2}$. We then consider a class of quivers dual to M2-branes at toric Calabi-Yau four-folds for which the $\mathcal{I}$-functional can be computed in the large $N$ limit, and depends on three mesonic fluxes. We propose a gravitational dual for this construction, that we call mesonic twist, and we show that the gravitational extremization problem and $\mathcal{I}$-extremization are equivalent. We comment on more general cases.}

\begin{document}

\setcounter{tocdepth}{2}
\maketitle

%
%

\date{Dated: \today}




\section{Introduction}
\label{sect:intro}

In this paper we are interested in theories living on M2-branes  sitting at the tip of a Calabi-Yau four-fold cone  and the relation of their twisted compactifications on a Riemann surface to AdS black holes physics. There are some interesting extremization problems that can be formulated for such theories. The theory on M2-branes at a Calabi-Yau cone $C(Y_7)$, where $Y_7$ is a seven-dimensional Sasaki-Einstein manifold, is a three-dimensional ${\cal N}=2$ superconformal field theory (SCFT) dual to AdS$_4\times Y_7$. In supersymmetric gauge theories the R-symmetry current is not necessarily unique and mixes with the global symmetry currents. It is known that  the exact R-symmetry of the three-dimensional theory is obtained by extremizing the free energy on $S^3$,  $F_{S^3}(\Delta_a)$, as a function of the R-charges of fields and, possibly, monopole operators  \cite{Jafferis:2010un}. The gravity dual of $F$-maximization is the volume minimization principle discovered in \cite{Martelli:2005tp,Martelli:2006yb}. The equivalence of the two extremization principles has been checked in many examples \cite{Herzog:2010hf,Jafferis:2011zi,Martelli:2011qj,Cheon:2011vi,Amariti:2011uw,Crichigno:2012sk,Amariti:2019pky}, although there is no general proof, since 
the large $N$ limit of the $S^3$ free energy is currently  available only for a restricted class of theories.  

On the other hand,  twisted compactifications of M2-brane theories are IR dual to  AdS$_2\times Y_9$ backgrounds in M-theory, where $Y_9$ is a fibration of $Y_7$ over a Riemann surface $\Sigma_\fg$, depending on a choice of magnetic fluxes $\fn_a$. These backgrounds can be interpreted as the horizon of asymptotically AdS$_4\times Y_7$ magnetically charged BPS black holes. An extremization principle for finding the entropy of magnetically charged black holes in AdS$_4\times Y_7$ has been proposed in \cite{Benini:2015eyy,Benini:2016rke}. It involves extremizing the logarithm of the topologically twisted index, \ie\;$\cI=\log Z_{\Sigma_\fg \times S^1}$, the partition function of the three-dimensional SCFT on $\Sigma_\fg \times S^1$ with a A-twist along $\Sigma_\fg$ \cite{Benini:2015noa}. This principle, dubbed $\cI$-extremization,  has been successfully applied to the microscopic counting of the entropy of black holes in  AdS$_4\times S^7$  \cite{Benini:2015eyy,Benini:2016rke} and in many other situations \cite{Hosseini:2016tor,Hosseini:2016ume, Cabo-Bizet:2017jsl,Azzurli:2017kxo,Hosseini:2017fjo,Benini:2017oxt,Bobev:2018uxk,Hosseini:2018qsx}.%
\footnote{See also \cite{Hosseini:2016cyf,Liu:2017vbl,Liu:2017vll,Jeon:2017aif, Halmagyi:2017hmw,Cabo-Bizet:2017xdr, Toldo:2017qsh,Gang:2018hjd,Hristov:2018lod,Hong:2018viz,Nian:2017hac,Liu:2018bac,Hosseini:2018uzp,Crichigno:2018adf,Hosseini:2018usu,Suh:2018szn,Suh:2018tul,Jain:2019lqb,Fluder:2019szh} for related works in a similar context or other dimensions.}  
A gravity  dual for $\cI$-extremization was recently proposed in a series of interesting papers \cite{Couzens:2018wnk,Gauntlett:2018dpc}. It is the purpose of this paper to investigate the relation between  the two extremization principles.  The analogous construction for D3-branes at Calabi-Yau three-fold toric cones has been recently proved to be equivalent to $c$-extremization for two-dimensional $(0,2)$ conformal field theories in full generality \cite{Couzens:2018wnk,Gauntlett:2018dpc,Hosseini:2019use}.
We already proved in \cite{Hosseini:2019use} that the construction in \cite{Couzens:2018wnk,Gauntlett:2018dpc} is equivalent off-shell to $\cI$-extremization for the class of magnetically charged BPS black holes in AdS$_4\times S^7$ studied in \cite{Benini:2015eyy,Benini:2016rke}, where the dual field theory is ABJM \cite{Aharony:2008ug}. In this paper, we extend this observation in various directions. 

We will  compare the construction of \cite{Couzens:2018wnk,Gauntlett:2018dpc} with the  field theory result in \cite{Hosseini:2016tor} which shows that, for the very same theories for which we can compute the $S^3$ free energy in the large $N$ limit, the $\cI$-functional can be written as 
\be
 \label{intro:general:index:theorem}
 \cI(\Delta_I,\fn_I) = -\frac{1}{2} (1-\fg) \bigg( 2 F_{S^3}(\Delta_I)  +\sum_I \left ( \frac{\fn_I}{1-\fg} -\Delta_I\right )  \frac{\partial F_{S^3}(\Delta_I)}{\partial \Delta_I} \bigg) \, ,
\ee
where $\Delta_I$ are the R-charges of a set of fields and monopoles, and $\fn_I$ the corresponding magnetic fluxes (which  can be set in relation to the magnetic charges $\fn_a$ of the black hole). If we choose an R-charge parameterization such that $F_{S^3}(\Delta_I)$ is homogeneous of degree two, which is generally possible, \eqref{intro:general:index:theorem} simplifies to
\bea
 \label{iextr0intro}
 \cI(\Delta_I,\fn_I) = -\frac{1}{2}  \sum_I\fn_I \frac{\partial F_{S^3}(\Delta_I)}{\partial \Delta_I} \, .
\eea
As we will review in details, the R-charges $\Delta_I$ and the fluxes $\fn_I$ parameterize  the mixing of R-symmetry with the abelian global  symmetries of the theory.  It will be important for us to distinguish between mesonic and baryonic symmetries. According to the holographic dictionary, the first are associated with the isometries of the internal manifold $Y_7$, while the second come from the reduction of the M-theory form on non-trivial five-cycles in $Y_7$.

For manifolds $Y_7$ with no non-trivial five cycles, we will show that the construction in \cite{Couzens:2018wnk,Gauntlett:2018dpc} is  completely equivalent  to the extremization of the $\cI$-functional given in \eqref{iextr0intro}, and that this equivalence holds off-shell. This class of manifolds is somehow limited but it contains $Y_7=S^7$, whose dual is the ABJM theory \cite{Aharony:2008ug}, and the non-toric manifold  $V^{5,2} =\SO(5) / \SO(3)$, whose dual field theory has been found in \cite{Martelli:2009ga}. For manifolds without five-cycles we can turn on  magnetic charges only for the mesonic symmetries associated with the isometry of $Y_7$. We will discuss in detail the $V^{5,2}$ example as a prototype of this class of compactifications.  In particular, we obtain a prediction for the entropy of the general two-parameter family of asymptotically AdS$_4\times V^{5,2}$ black holes with arbitrary magnetic charges under the Cartan subgroup of the internal isometry $\SO(5)$. It would be interesting to find these solutions explicitly.

We then consider the general case of toric Calabi-Yau cones $C(Y_7)$, for which we can use the master volume construction defined in \cite{Gauntlett:2018dpc}. In addition to the mesonic charges associated to the three $\U(1)$ isometries of $Y_7$, we can have a baryonic charge for every non-trivial five-cycle $S_\alpha \subset Y_7$. Here the situation is more complicated. We want to compare the former construction with the existing results for the large $N$ limit of the topologically twisted index, which only depend on a linear combination of the available magnetic charges, corresponding to the mesonic directions only \cite{Hosseini:2016tor,Hosseini:2016ume}. Correspondingly, we identify a three-parameter family of twisted compactifications, which we dub \emph{mesonic twist}, where the extremization is performed along the mesonic directions only. We will show that, for such compactifications, the construction in \cite{Couzens:2018wnk,Gauntlett:2018dpc} is equivalent off-shell to the extremization of the $\cI$-functional \eqref{iextr0intro}. This result is general and it can be proved for all toric Calabi-Yau cones $C(Y_7)$ for which the equivalence between $F$-maximization and volume minimization is valid. In particular, all the existing field theory computations to date  have a counterpart in the framework of \cite{Couzens:2018wnk,Gauntlett:2018dpc}.

Our results for toric Calabi-Yau cones are restricted to a particular class of twisted compactifications. This should be contrasted with the equivalence of  the construction in \cite{Couzens:2018wnk,Gauntlett:2018dpc} with $c$-extremization, which can be proved for an arbitrary toric $Y_7$ and an arbitrary choice of fluxes \cite{Hosseini:2019use}. The reason for such restriction lies in our incomplete understanding of the large $N$ limit of the $S^3$ free energy and the topologically twisted index for quivers with a holographic dual. It is known, for example, that the computation performed in \cite{Jafferis:2011zi} and \cite{Hosseini:2016tor,Hosseini:2016ume} only works for quivers with vector-like matter fields.%
\footnote{More precisely, the bi-fundamental fields must transform in a real representation of the gauge group and the total number of fundamentals must be equal to the total number of anti-fundamentals.}%
\footnote{For an attempt to circumvent this problem see \cite{Amariti:2011jp,Gang:2011jj,Amariti:2011uw}.}
Moreover, there are accidental flat directions at large $N$ and the R-charges parameterizing the baryonic symmetries disappear from the free energy functional $F_{S^3}(\Delta_I)$ \cite{Jafferis:2011zi}. Correspondingly, the topologically twisted index \eqref{iextr0intro} only depends on a combination of magnetic charges corresponding to the mesonic directions \cite{Hosseini:2016tor,Hosseini:2016ume}. It is not clear to us if this is just due to our ignorance about more general saddle points in the large $N$ limit or it is the signal of something deeper. The construction in \cite{Couzens:2018wnk,Gauntlett:2018dpc} seems to work for a generic twist. This does not necessarily guarantee that a corresponding  supergravity solution really exists.%
\footnote{Examples of possible obstructions are discussed in \cite{Couzens:2018wnk}.}
Nevertheless, there are certainly examples of consistent supergravity solutions with {\it only baryonic charges} \cite{Donos:2012sy,Azzurli:2017kxo}. A field theory computation for such solutions is still missing. 

In our analysis of the mesonic twist we uncover some general geometric relations on the Sasakian volumes of cycles in $Y_7$ that deserve attention in their own right and are discussed in section \ref{mesTgeo}
and demonstrated with examples in section \ref{sec:Examples}.

The paper is organized as follows. In section \ref{CFT} we discuss general features of three-dimensional toric quivers and their twisted compactifications. We review the equivalence between $F$-maximization and volume minimization, and the construction in \cite{Couzens:2018wnk,Gauntlett:2018dpc}. In section \ref{1-cycle} we show that the formalism \cite{Couzens:2018wnk,Gauntlett:2018dpc} for manifolds $Y_7$ without five-cycles (theories without baryonic symmetries) is equivalent off-shell to the $\cI$-extremization principle for black holes in AdS$_4\times Y_7$. We consider in details the example of $V^{5,2}$. In section \ref{sec:mesonictwist}, we focus on the case of toric $Y_7$. We discuss explicitly the case of the so-called universal twist \cite{Benini:2015bwz} and we define a three-parameter generalization, the mesonic twist, where again we can show that the formalism \cite{Couzens:2018wnk,Gauntlett:2018dpc} is equivalent off-shell to  $\cI$-extremization.  For the convenience of the reader, the technical aspects of the proof are deferred to appendix \ref{sec:app}. In section \ref{mesTgeo} we discuss some geometrical aspects and the field theory interpretation underlying the mesonic twist.   In section \ref{sec:Examples} we  present several examples based on quivers for  toric $Y_7$ that have been discussed in the literature.
We conclude with discussions and comments in section \ref{discussion}.

\paragraph*{Note added:} while we were writing this work, we became aware of \cite{Gauntlett:2019roi} which has some overlaps with the results presented here.

\section{Extremization principles and their geometric duals}
\label{CFT}

In this section we first review the construction of \cite{Martelli:2005tp} for computing the volume of a toric Sasaki-Einstein seven-manifold.
We then discuss its relation to the R-symmetry and the free energy of the holographic dual SCFT on a three-sphere, as shown in \cite{Herzog:2010hf,Martelli:2011qj,Jafferis:2011zi}.
In the second part of this section we give an overview of the geometric dual of the $\cI$-extremization principle \cite{Benini:2015eyy} obtained recently in \cite{Couzens:2018wnk,Gauntlett:2018dpc}.

\subsection[\texorpdfstring{$F$}{F}-maximization from geometry]{$F$-maximization from geometry}
\label{Fmax}

We are interested in gauge theories that are holographically dual to AdS$_4 \times Y_7$ backgrounds in M-theory, where $Y_7$ is a  Sasaki-Einstein manifold.
The holographic dictionary relates the volume of the Sasaki-Einsten manifold to the $S^3$ free energy of the dual CFT \cite{Herzog:2010hf}
\bea
 F_{S^3}  = N^{3/2} \sqrt{\frac{2 \pi^6}{27 \text{Vol}(Y_7)}} \, .
\eea
Many $\cN=2$ quivers describing such theories have been proposed in the literature. Most of them are obtained by dimensionally reducing a {\it parent} four-dimensional quiver gauge theory with bi-fundamentals and adjoints with an AdS$_5\times Y_5$ dual, and then adding Chern- Simons terms (whose levels sum to zero) and flavoring with fundamentals.

The value of $F_{S^3}$ and the exact R-symmetry of a three-dimensional $\cN=2$ theory can be found by extremizing the $S^3$ free energy as a function of the R-charges $\Delta_I$ of the chiral  elementary fields and monopoles \cite{Jafferis:2010un}. This procedure is called  \emph{$F$-maximization}. The R-charges $\Delta_I$ parameterize the mixing of the R-symmetry with the abelian global symmetries of the theory. We will distinguish between mesonic and baryonic symmetries. For theories dual to AdS$_4 \times Y_7$, mesonic symmetries are associated with the isometries of $Y_7$, which we take to be $\U(1)^s$, with $s\ge 1$. We will be mostly interested in the toric case where $s=4$, but we will also consider non-toric examples.   One of the mesonic symmetries is  the exact R-symmetry and the other $s-1$ are global symmetries. In addition,  we have a baryonic symmetry for each   non-trivial five-cycle $S_\alpha$ of $Y_7$. They are holographically dual to the  gauge fields that we obtain by reducing the M-theory six-form potential on the five-cycles $S_\alpha$ \cite{Klebanov:1998hh,Gubser:1998fp,Fabbri:1999hw}. In supergravity language, the corresponding vector multiplets are called Betti multiplets.%
\footnote{Notice that what we call baryonic not always reflects the field theory notion of baryonic symmetry for the parent four-dimensional quiver. For example, the ABJM theory is dual to AdS$_4\times S^7$ and there are no nontrivial five-cycles and therefore no baryonic symmetries in our sense. The baryonic symmetry of the parent  four-dimensional quiver, which is the well-known Klebanov-Witten theory \cite{Klebanov:1998hh}, corresponds to an isometry of $S^7$ and it is a mesonic symmetry in our language.}  In most of the known examples, the three-dimensional $\cN=2$ theories are quiver gauge theories that can have $\U(N)^G$ or $\U(1)\times \SU(N)^G$ gauge groups, where $G$ is the number of nodes, depending on the choice of quantization \cite{Aharony:2008ug,Benishti:2010jn}.%
\footnote{There are also quivers with orthogonal and symplectic gauge groups, see for example \cite{Aharony:2008gk}.}
When the gauge group is $\U(1)\times \SU(N)^G$ we have baryonic operators obtained by wrapping  M5-branes on the five-cycles and this is the case we will be mostly interested in.%
\footnote{The number of global symmetries is the same for both choices of gauge group, $\U(N)^G$ or $\U(1)\times \SU(N)^G$ \cite{Aharony:2008ug,Benishti:2010jn}. In the case of $\U(N)^G$, there are no baryons but we have instead monopole operators obtained by wrapping M2-branes.}

The gravitational dual of $F$-maximization is the {\it volume minimization} found in   \cite{Martelli:2005tp,Martelli:2006yb}, which works as follows.  One can relax the Einstein condition on the metric and write the volumes of a generic Sasaki manifold $Y_7(b_i)$ and of its five-cycles $S_\alpha(b_i)$ as functions of the Reeb vector $b=(b_1,b_2,\ldots ,b_s)$. Supersymmetry requires $b_1=4$. In general, the cone $C(Y_7)$ is a Calabi-Yau three-fold. As shown in \cite{Martelli:2005tp}, the extremization of the function $\text{Vol}_{\text{S}}(Y_7 (b_i))$ reproduces the Reeb vector $\bar b=(\bar b_1,\bar b_2,\ldots ,\bar b_s)$  and the volumes of the Sasaki-Einstein manifold $Y_7$.
Remarkably, for a large class of examples, the volume functional agrees \emph{off-shell} with the $S^3$ free energy \cite{Martelli:2011qj,Jafferis:2011zi}
\bea\label{free}
 F_{S^3} (\Delta_I) = N^{3/2} \sqrt{\frac{2 \pi^6}{27 \text{Vol}_{\text{S}}(Y_7(b_i))}} \, ,
\eea
with a suitable parameterization of the R-charges $\Delta_I(b_i)$ of fields and monopoles. In order to find the right parameterization, one considers all baryonic operators made with elementary fields and basic monopoles.  These correspond to M5-branes wrapped on linear combinations of the five-cycles and their dimension can be computed from the corresponding volumes.\footnote{Although there is no general prescription for arbitrary  $Y_7$, not even for toric manifolds, this can be done in general for a large class of models including those in \cite{Hanany:2008cd,Hanany:2008fj,Martelli:2008si,Benini:2009qs,Jafferis:2009th}, using perfect matchings and the symplectic quotient descriptions of $Y_7$. 
} 
In all known examples,  the R-charges $\Delta_I$ of fields and monopoles can be computed as  linear integer combinations of basic R-charges, corresponding to M5-branes wrapped over $\U(1)^s$ invariant five-cycles $S_a$ and  expressed in terms of the Sasaki volumes \cite{Gubser:1998fp,Fabbri:1999hw}
\be\label{Rcharges}
 \Delta_a ( b_i ) \equiv \frac{2 \pi}{3 b_1} \frac{\text{Vol}_{\text{S}}(S_a(b_i))}{\text{Vol}_{\text{S}}(Y_7(b_i))}  \, .
\ee
In the toric case there are quite explicit expressions for these volumes, which are associated with the vectors $v_a$, $a=1,\ldots, d$, of the toric diagram \cite{Martelli:2005tp,Hanany:2008fj} 
\bea
 \label{Sasaki:volumes}
 &\text{Vol}_{\text{S}}(Y_7(b_i)) = \frac{\pi}{3 b_1} \sum_{a=1}^d \text{Vol}_{\text{S}}(S_a(b_i)) \, , \\
 & \text{Vol}_{\text{S}}(S_a(b_i)) = \pi^3 \sum_{k = 2}^{\ell_a-1} \frac{ (v_{a}, w_{k-1},w_{k},w_{k+1}) (v_{a}, w_{k},w_{1},w_{\ell_a})}{(v_{a}, b,w_{k},w_{k+1})(v_{a}, b,w_{k-1},w_{k})(v_{a}, b,w_{1},w_{\ell_a})}  \, ,
\eea
where $w_a$,  $k = 1 , \ldots , \ell_a$, is a  counterclockwise ordered sequence of vectors  adjacent to $v_a$.  
There is also an alternative way of computing the volumes using the Hilbert series, for which we refer to \cite{Martelli:2006yb,Hanany:2008fj,Amariti:2011uw}. This method can be used also for non-toric manifolds.  We will see an example in section  \ref{sec:exampleV52}.

Notice that, in the large $N$ limit, the free energy  $F_{S^3}$ only depends on a set of linear combinations of the $\Delta_I$ equal to the number of independent parameters $b_i$ and corresponding to the mixing of the R-symmetry with the mesonic symmetries. Indeed,  as shown in \cite{Jafferis:2011zi}, there are accidental flat directions at large $N$  in $F_{S^3}(\Delta_I)$ and the R-charges parameterizing the baryonic symmetries disappear from the free energy functional. This should be contrasted with the case of $a$-maximization  and its relation to volume minimization \cite{Butti:2005vn}, where the baryonic symmetries explicitly enter in the trial $a$-charge. Extremizing  $F_{S^3}(\Delta_I)$, we can only predict the exact R-charges of the mesonic operators of the theory.  However, when the gauge group is $\U(1)\times \SU(N)^G$, consistency of the solution allows to derive constraints  expressing some of the remaining R-charges in terms of those appearing in $F_{S^3}(\Delta_I)$ \cite{Jafferis:2011zi}. This can be used to compute the exact R-charge of some baryonic operators in the theory. We will see examples of such constraints in section \ref{sec:Examples}.

\subsection[\texorpdfstring{$\cI$}{I}-extremization from geometry]{$\cI$-extremization from  geometry}\label{Iextremiz}

We are actually interested in twisted compactifications of the three-dimensional $\cN=2$ theories discussed in the previous section on a Riemann surface $\Sigma_{\fg}$ of genus $\fg$. The holographic dual is an  M-theory background AdS$_2 \times Y_9$, where topologically $Y_9$  is a fibration of  $Y_7$ over $\Sigma_\fg$, and can be interpreted as the horizon of magnetically charged BPS AdS$_4$ black holes.
As in the original computation in \cite{Benini:2015eyy,Benini:2016rke} for AdS$_4\times S^7$ black holes, the entropy of magnetically charged BPS black holes should be obtained by extremizing the functional 
\bea
 \label{iextr00}
 \cI(\Delta_I,\fn_I) = \log Z_{\Sigma_\fg \times S^1} (\Delta_I,\fn_I) \, ,
\eea
where $Z_{\Sigma_\fg \times S^1}(\Delta_I,\fn_I)$ is the topologically twisted index \cite{Benini:2015noa}, $\fn_I$ denote the magnetic charges and $\Delta_I$  are chemical potentials associated with elementary fields and monopoles. We refer to this principle as $\cI$-extremization. A  field theory computation, valid for a large class of theories, shows that, in the large $N$ limit, the $\cI$-functional can be parameterized in terms of the R-charges of the fields and reads \cite{Hosseini:2016tor}
\bea
 \label{iextr0}
 \cI(\Delta_I,\fn_I) = -\frac{1}{2}  \sum_{I} \fn_I \frac{\partial F_{S^3}(\Delta_I)}{\partial \Delta_I} \, .
\eea
This  formula \eqref{iextr0} is only valid for an R-charge parameterization that makes $F_{S^3}(\Delta_I)$ homogeneous.%
\footnote{For an arbitrary parameterization one can use \eqref{intro:general:index:theorem}. The expression \eqref{intro:general:index:theorem} can be used also for parameterizations of fluxes and R-charges, where a set of $\Delta_I$ satisfies some linear constraint, provided that the corresponding $\fn_I/(1-\fg)$ satisfy the same constraint. See \cite{Hosseini:2016tor,Hosseini:2016ume,Hosseini:2016cyf,Hosseini:2019use} for details.   }
The same accidental symmetry that affects the large $N$ limit of the $S^3$ free energy appears in the computation of the topologically twisted index at large $N$, and, therefore,
the $\cI$-functional only depends on R-charges and fluxes along the mesonic directions \cite{Hosseini:2016tor}.

The gravitational dual of $\cI$-extremization has been found in \cite{Couzens:2018wnk,Gauntlett:2018dpc}. The authors of \cite{Couzens:2018wnk,Gauntlett:2018dpc} considered  a class of off-shell backgrounds by imposing supersymmetry but relaxing the equations of motion.  More precisely, they considered M-theory backgrounds of the form
\bea
 & \rd s_{11}^2 = L^2 e^{-2 B / 3} \left( \rd s_{\text{AdS}_2}^2 + \rd s_{9}^2 \right) \, , \\
 & G = L^3 \text{Vol}_{\text{AdS}_2} \wedge F \, ,
\eea
where $F$ is a closed two-form on $Y_9$ and
\be
 \rd s_9^2 = \eta^2 + e^{B} \rd s^2 \, .
\ee
Here, $\eta \equiv (\rd z + P)/b_1$, where $\rho = \rd P$ is the transverse Ricci form,
and $\rd s^2$ is a K\"ahler metric. The Reeb vector field associated with the R-symmetry reads 
\bea \xi = b_1 \partial_z = \sum_{i=1}^s b_i \partial_{\varphi_i}\, , \qquad s \geq 1\, , \eea
where $\partial_{\varphi_i}$ are real holomorphic vector fields generating the $\U(1)^s$ action on $Y_7$. When $Y_7$ is toric, $s=4$. 
Finally, the closed two-form $F$ is given by 
\be
 F = - b_1 J + \rd \left( e^{-B} \eta \right) \, ,
\ee
where $J$ is the transverse K\"ahler form.  Supersymmetry requires $b_1=1$. 

For the background of interest, the transverse K\"ahler cohomology class decomposes as
\be
 \label{J:decompose:omega}
 J = A \text{Vol}_{\Sigma_\fg} + \omega \, ,
\ee
where $\omega$ is a transverse K\"ahler form on $Y_7$ and $A > 0$ is a constant parameterizing the K\"ahler class of $\Sigma_\fg$.
We normalize $\int_{\Sigma_{\fg}} \text{Vol}_{\Sigma_\fg} = 1$. The fibration of $Y_7$ over $\Sigma_\fg$ is specified by $s$ integer magnetic fluxes $n_i$. We can introduce them through $s$  $\U(1)$ gauge fields $A_i$ on $\Sigma_\fg$
\be\label{mesonicfluxes}
 \frac{1}{2 \pi}\int_{\Sigma_{\fg}} F_i = n_i \in \bZ \, , \qquad i = 1 , \ldots , s \, ,
\ee
where $F_i = \rd A_i$.
Supersymmetry requires
\be
 \label{n1:twisting}
 n_1 = 2 - 2 \fg \, ,
\ee
that we refer to as the \emph{twisting condition}.

As shown in \cite{Couzens:2018wnk,Gauntlett:2018dpc}, the on-shell background and the exact R-symmetry vector can be found by 
extremizing the supersymmetric action 
\be
 \label{S-susy:J3}
 S_{\text{SUSY}} ( \xi ; [J]) = \frac16 \int_{Y_9} \eta \wedge \rho \wedge J^3 \, ,
\ee
which is a function of the Reeb vector and the cohomology class of $J$, with the constraint 
\be\label{constraint1}
 \int_{Y_9} \eta \wedge \rho^2 \wedge J^2 = 0 \, .
\ee
The flux quantization conditions also require\footnote{Comparing with \cite{Couzens:2018wnk,Gauntlett:2018dpc}, we set for simplicity $L^6=(2\pi l_P)^6$ and $M_a=-N \fn_a$.}
\be
 \label{AdS2:flux:quantization}
 \int_{Y_7} \eta \wedge \rho \wedge J^2= 2 N \, ,
\ee
where $N$ is the total number of M2-branes and
\be
 \label{AdS2:flux:quantizationS}
 \int_{S_\alpha\times \Sigma_\fg} \eta \wedge \rho \wedge J^2= - 2 \fn_\alpha N \, ,
\ee
where $\fn_\alpha$  is an integer, for all five-cycles $S_\alpha \subset Y_7$, and $\alpha= 1, \ldots , \text{dim}\,H_5(Y_7,\bZ)$. The integer numbers $n_i$ and $\fn_\alpha$   are interpreted as the magnetic fluxes  of the twisted compactification on $\Sigma_\fg$.  The $\fn_\alpha$ are associated with the baryonic symmetries and the $n_i$ associated with the mesonic ones.
Finally, the R-charge of an operator obtained by wrapping a M5-branes on a five-cycle $S_\alpha \subset Y_7$ is given by \cite{Couzens:2017nnr,Couzens:2018wnk}
\bea
 \label{RSa}
 R[S_\alpha]=  2 \pi  \int_{S_\alpha} \eta \wedge \omega^2 \, ,
\eea
where $\omega$ is the restriction of $J$ to $Y_7$. Since this is the R-charge of a baryonic operator, $R[S_\alpha]$ is proportional to $N$. 

As noticed in \cite{Couzens:2018wnk,Gauntlett:2018dpc} in the specific example of the magnetically charged AdS$_4$ black holes of \cite{Caldarelli:1998hg}, the on-shell value of $S_{\text{SUSY}}$ reproduces the entropy. This is also true 
for all magnetically charged black holes in AdS$_4\times S^7$, as indirectly checked in  \cite{Hosseini:2019use} by comparing with $\cI$-extremization. Since $S_{\text{SUSY}}$ is related to the holographic free energy of the horizon solution, and the latter to the entropy by general arguments \cite{Azzurli:2017kxo,Halmagyi:2017hmw,Cabo-Bizet:2017xdr}, we may expect this to be true in general. 
If this is the case, the construction in \cite{Couzens:2018wnk,Gauntlett:2018dpc} gives an efficient method to write  the entropy of a class of black holes from few geometrical data, even without knowing the explicit metric on $Y_7$. 

\section{Theories with no baryonic symmetries}\label{1-cycle}

In this section we consider the case of  manifolds $Y_7$ with no non-trivial five-cycles and therefore dual field theories with no baryonic symmetries.  
Examples include  $Y_7=S^7$, whose dual is the ABJM theory \cite{Aharony:2008ug}, and the \emph{non-toric} $V^{5,2} =\SO(5) / \SO(3)$, whose dual field theory has been found in \cite{Martelli:2009ga}.%
\footnote{See \cite{Jafferis:2009th} for an alternative model with fundamental chiral multiplets.}
For the ABJM theory, we already checked in \cite{Hosseini:2019use} that  $S (b_i , \fn_a)$ is equal off-shell to the $\cI$-functional \eqref{iextr0}.
In this section we will check that this is also true for the case of $V^{5,2}$ and, in general, for all manifolds $Y_7$ with  a $\U(1)^s$ action and no five-cycles. 
 
It is convenient to first rewrite the conditions of supersymmetry as follows \cite{Couzens:2018wnk,Gauntlett:2018dpc}.
The supersymmetric action can be written as \cite{Gauntlett:2018dpc}%
\bea
 \label{non-toric:action}
 S_{\text{SUSY}}= \frac{A}{2}  \int_{Y_7} \eta \wedge \rho \wedge \omega^2  - \frac{\pi}{3}   b_1 \nabla \int_{Y_7} \eta  \wedge \omega^3 \, ,
\eea
where we defined the operator 
\bea\label{nabla}  \nabla \equiv \sum_{i=1}^{s} n_i \partial_{b_i}\, .\eea
We can also write the constraints \eqref{constraint1} and  \eqref{AdS2:flux:quantization}  as
\bea
 \label{non-toric:constraints}
 & N = \frac12 \int_{Y_7} \eta \wedge \rho \wedge \omega^2 \, , \\
 & A \int_{Y_7} \eta \wedge \rho^2 \wedge \omega = - \pi n_1 \int_{Y_7} \eta \wedge \rho \wedge \omega^2
 + \pi b_1 \nabla \int_{Y_7} \eta \wedge \rho \wedge \omega^2 \, .
\eea
The derivation of \eqref{non-toric:action} and \eqref{non-toric:constraints} is given in \cite{Gauntlett:2018dpc} for toric $Y_5$ but extends with little modifications to seven dimensions and to the non-toric case. 

We are interested in manifolds $Y_7$ with no five-cycles.  
The supersymmetry conditions \eqref{non-toric:action} and \eqref{non-toric:constraints} only depend on the cohomology classes of $\omega$ and $\rho$ on the foliation transverse to the Reeb vector action.
Focusing, for simplicity, on the quasi regular case we can take the quotient with respect to the Reeb action, and consider the base $V = Y_7/\U(1)$.%
\footnote{For example, as a complex  manifold, $V$ can be identified with $\mathbb{P}^3$ for $Y_7=S^7$ and with the Grassmannian $\text{Gr}_{2}(\bR^5)$ of two-planes in $\mathbb{R}^5$, which admits a complex structure, for $Y_7=V^{5,2}$.} The manifold $V$ has only one two-cycle  and, therefore,
$\omega$ and $\rho$ are proportional in cohomology 
\be
 \label{omega:lambda:rho}
[ \omega ]= \frac{\lambda(b)}{2 b_1}\, [\rho] \, ,
\ee
where $\lambda(b)$ is a function to be found.
Using \eqref{omega:lambda:rho} and the first equation in \eqref{non-toric:constraints}, we then obtain
\be
 \label{lambda:one:2-cycle}
 \lambda (b) = \pm 2 b_1 \sqrt{\frac{2 N}{\int_{Y_7} \eta \wedge \rho^3}} \, .
\ee
Using the last equation in \eqref{non-toric:constraints} we find
\bea
 \frac{A \lambda}{2 b_1} \int_{Y_7} \eta \wedge \rho^3 
 = - \frac{\pi n_1 \lambda^2}{4 b_1^2} \int_{Y_7} \eta \wedge \rho^3 + 2 \pi b_1 {\nabla N} \, ,
\eea
and therefore, since $\nabla N=0$,
\be\label{rA}
 A = - \frac{\pi n_1 \lambda}{2 b_1} \, .
\ee
Now we are in a position to evaluate the functional \eqref{non-toric:action}. We may write, using the plus sign in \eqref{lambda:one:2-cycle},
\bea
 S_{\text{SUSY}} & = \frac{A \lambda^2}{8 b_1^2} \int_{Y_7} \eta \wedge \rho^3 - \frac{\pi b_1}{24} \nabla \left( \frac{\lambda^3}{b_1^3} \int_{Y_7} \eta \wedge \rho^3 \right)
 = - \frac{\pi (2 N)^{3/2}}{3 \sqrt{b_1}} \nabla \frac{b_1^{3/2}}{\sqrt{\int_{Y_7} \eta \wedge \rho^3}} \, .
\eea
We now notice that the expression 
\be
 \label{Sasaki:vol:complex:geometry}
 \text{Vol}_{{\text S}} (Y_7) = \frac{1}{48 b_1^3} \int_{Y_7} \eta \wedge \rho^3 \, ,
 \ee
is formally identical to the volume of a Sasakian manifold  $Y_7(b_i)$ with Reeb vector $b$. Indeed, this expression can be evaluated using a fixed point theorem  which only depends 
on the local complex geometry of $C(Y_7)$ near the fixed points  \cite{Couzens:2018wnk}. Hence it formally coincides with the expression for the Sasakian volume of $Y_7(b_i)$ computed in 
\cite{Martelli:2005tp,Martelli:2006yb} and given in \eqref{Sasaki:volumes} in the toric case. Form now on, we will  understand the explicit dependence on $b_i$ in \eqref{Sasaki:volumes} and just  
use the subscript S to indicate that the volumes are computed using the Sasaki metric on $Y_7(b_i)$. 
 
We conclude that  the entropy functional $S(b_i , n_i)$ is given by
\be\label{S1}
 S (b_i , n_i) \equiv 8 \pi^2 S_{\text{SUSY}} = - \frac{4}{\sqrt{b_1}} \nabla \sqrt{\frac{2 \pi^6}{27 \text{Vol}_{{\text S}}(Y_7)}} N^{3/2} \, .
\ee
Notice that, since the equations involve derivatives with respect to $b_1$, we can set $b_1=1$, as required by supersymmetry,  only at the end of the computation.

We are considering manifolds with no five-cycles. However, in analogy with the toric case that we will discuss in the next section, we can consider 
the set $S_a \subset Y_7$ of $\U(1)^s$ invariant submanifolds of dimension five and {\it formally} extend the quantization condition to these cycles.
We then also impose \cite{Gauntlett:2018dpc}
 \be
  \label{non-toric:constraints2}
  \fn_a N = - A \int_{S_a} \eta \wedge \rho \wedge \omega + \pi b_1 \nabla \int_{S_a} \eta \wedge \omega^2 \, ,
 \ee
where  $\fn_a$ are integers. Notice that, since there are no baryonic symmetries,  the $\fn_a$  must be linear combinations of the mesonic fluxes $n_i$.
 Using \eqref{lambda:one:2-cycle} and \eqref{rA} we find
\bea
 \label{flux:constraint:non-toric}
 \fn_a N
 = 2 \pi N \left( n_1 \frac{\int_{S_a} \eta \wedge \rho^2}{\int_{Y_7} \eta \wedge \rho^3} + b_1 \nabla \frac{\int_{S_a} \eta \wedge \rho^2}{\int_{Y_7} \eta \wedge \rho^3} \right) \, .
\eea
Defining normalized R-charges (see \eqref{RSa})
\be\label{V52vol}
 \Delta_a \equiv \frac{2 \pi}{N} \int_{S_a} \eta \wedge \omega^2
 = 4 \pi \frac{\int_{S_a} \eta \wedge \rho^2}{\int_{Y_7} \eta \wedge \rho^3} \, ,
\ee
we can therefore rewrite \eqref{flux:constraint:non-toric} as
\be\label{S2}
 \fn_a = \frac{1}{2} \left( n_1 \Delta_a + b_1 \nabla \Delta_a \right)
 = \frac{1}{2} \nabla (b_1 \Delta_a) \, .
\ee
In many quivers, including those for ABJM and $V^{5,2}$, the elementary fields can be associated with linear combinations of invariant five-cycles $S_a$ and their R-charges can be computed using \eqref{V52vol}. 
 
\subsection[Example: the manifold \texorpdfstring{$V^{5,2}$}{V[5,2]}]{Example: the manifold $V^{5,2}$}
\label{sec:exampleV52}

The non-toric Sasaki manifold $V^{5,2} = \SO(5) / \SO(3)$ possesses an $\SO(5) \times \U(1)_R \supset \U(1)^3$ isometry,
where $\U(1)_R$  is identified with the R-symmetry on the SCFT side.
The  base $V$ is the Grassmannian $\text{Gr}_2(\bR^5)$, which admits a complex structure, and has only one two-cycle.
The dual gauge theory  is given by the quiver \cite{Martelli:2009ga}
\bea
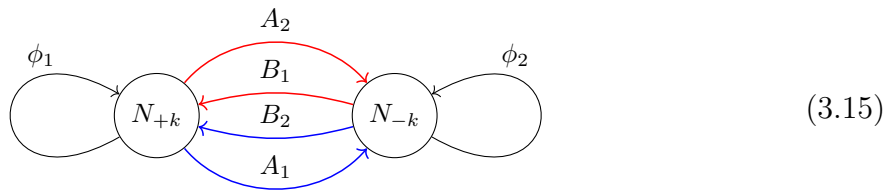
\begin{tikzpicture}[baseline, font=\footnotesize, scale=0.8]
\begin{scope}[auto,%
  every node/.style={draw, minimum size=0.5cm}, node distance=2cm];
\node[circle] (USp2k) at (-0.1, 0) {$N_{+k}$};
\node[circle, right=of USp2k] (BN)  {$N_{-k}$};
\end{scope}
\draw[draw=blue,solid,line width=0.2mm,<-]  (USp2k) to[bend right=15] node[midway,above] {$B_2 $}node[midway,above] {}  (BN) ;
\draw[draw=blue,solid,line width=0.2mm,->]  (USp2k) to[bend right=50] node[midway,above] {$A_1$}node[midway,above] {}  (BN) ; 
\draw[draw=red,solid,line width=0.2mm,<-]  (USp2k) to[bend left=15] node[midway,above] {$B_1$} node[midway,above] {} (BN) ;  
\draw[draw=red,solid,line width=0.2mm,->]  (USp2k) to[bend left=50] node[midway,above] {$A_2$} node[midway,above] {} (BN) ;    
\draw[black,-> ] (USp2k) edge [out={-150},in={150},loop,looseness=10] (USp2k) node at (-2,1) {$\phi_1$} ;
\draw[black,-> ] (BN) edge [out={-30},in={30},loop,looseness=10] (BN) node at (5.8,1) {$\phi_2$};
\end{tikzpicture}
\eea
with  superpotential 
\begin{equation}
 W = \Tr\left[ \phi_1^3 + \phi_2^3 +\phi_1(A_1 B_2 + A_2 B_1) + \phi_2 (B_2 A_1+ B_1 A_2) \right] \, .
\end{equation}
We will focus on the quiver with  $k=1$ which is dual to AdS$_4\times V^{5,2}$. The Calabi-Yau cone $C(V^{5,2})$ is described by the equation
\be
 \sum_{\ell = 0}^{4} z_\ell^2 = 0 \, ,
\ee
which has a manifest $\SO(5)$ invariance.  This arises as the solution to the $F$-term equation for the adjoint fields 
\be
 \label{F-term:phi1}
 3 \phi_1^2 + A_1 B_2 + A_2 B_1 = 0 \, , \qquad \phi_1= - \phi_2 \, ,
\ee
using the  variables
\be
 \begin{aligned}
  & z_0 \equiv \sqrt{3}\phi_1 \, , \qquad z_1 \equiv \frac12 ( A_1 + B_2 ) \, , \qquad z_2 \equiv - \frac{i}{2} ( A_1 - B_2 ) \, , \\
  & z_3 \equiv \frac12 ( A_2 + B_1 ) \, , \qquad z_4 \equiv - \frac{i}{2} ( A_2 - B_1 ) \, .
 \end{aligned}
\ee
The R-charges of the adjoint fields are constrained by the superpotential to be $\Delta_{\phi_i}=2/3$ while those of the bi-fundamental fields must satisfy
\be
 \Delta_{A_1}+\Delta_{B_2} = \Delta_{A_2}+\Delta_{B_1} =\frac43 \, .
\ee 
There are five obvious $\U(1)^3$ invariant divisors in $C(V^{5,2})$ that are obtained by setting $\phi_1=0$, $A_1=0$, $A_2=0$, $B_1=0$ and $B_2=0$, respectively; and can be associated in a one-to-one way to the fields. The restriction to $V^{5,2}$ gives five invariant five-cycles $S_a$.

The $S^3$ free energy was derived in \cite{Cheon:2011vi,Martelli:2011qj,Jafferis:2011zi,Hosseini:2016ume} from localization and reads 
\be \label{freeV52}
 F_{S^3} (\Delta_a) = \frac{4 \pi N^{3/2}}{3} \sqrt{ \Delta_{A_1} \Delta_{A_2} \Delta_{B_1} \Delta_{B_2}} \, .
\ee

In order to compute the volume \eqref{Sasaki:vol:complex:geometry} we can use the Hilbert series method \cite{Martelli:2006yb,Hanany:2008fj,Amariti:2011uw,Cremonesi:2016nbo}.\footnote{We refer to these papers  for more details on the  method. } The Hilbert series is just the generating function of holomorphic functions on the cone $C(V^{5,2})$, graded under $\U(1)^3$. We assign fugacities to the fields
\bea
 \label{action}
 \phi_1\to  e^{- \frac{\epsilon b_1}{4} \Delta_{\phi_1}} \equiv t  \, ,  & \qquad A_1\to e^{-\frac{\epsilon b_1}{4}\Delta_{A_1}}\equiv  t/y \, ,\qquad  A_2\to e^{-\frac{\epsilon b_1}{4}\Delta_{A_2}}\equiv t/x \, ,  \\ &
 \qquad B_1\to e^{-\frac{\epsilon b_1}{4} \Delta_{B_1}}\equiv t x \, ,\qquad B_2\to e^{-\frac{\epsilon b_1}{4}\Delta_{B_2}}\equiv t y \, ,
\eea
where $t$ corresponds to the R-symmetry, $x$ and $y$ are fugacities for the Cartan of $\SO(5)$, and $\epsilon$ is just a rescaling parameter that we will send to zero at the end of the computation.
Since $C(V^{5,2})$ is a complete intersection, its Hilbert series  is simply given by
\be
 H(t,x,y) = \frac{1-t^2}{(1- t)(1-t x)(1-t/x)(1-t y ) (1-t/y)} \, .
\ee
The volume \eqref{Sasaki:vol:complex:geometry} of $V^{5,2}$ can be extracted as the coefficient of the leading pole in $H(t,x,y)$ when $\epsilon\to 0$ \cite{Martelli:2006yb}, \ie\;
\be
 H(t,x,y) \sim\frac{48} {\pi^4}  \frac{\text{Vol}_{\text{S}}  (V^{5,2})}{\epsilon^4} \, , \quad \text{ as } \epsilon \to 0 \, .
\ee
Hence, we obtain
\be\label{VV52}
 \text{Vol}_{\text{S}}  (V^{5,2}) = \frac{\pi^4}{24} \left(\frac{4}{b_1}\right )^4\frac{1}{ \Delta_{A_1} \Delta_{A_2} \Delta_{B_1} \Delta_{B_2}} \, .
\ee
Comparing \eqref{freeV52} with \eqref{VV52}, we see that \eqref{free} is satisfied, when we set $b_1=4$, as appropriate for three-dimensional computations. We thus see that volume minimization is equivalent to $F$-maximization off-shell.%
\footnote{The on-shell equivalence was proved in \cite{Cheon:2011vi,Martelli:2011qj,Jafferis:2011zi}.}

The  action of the Reeb vector on the fields is the linearization of the $\U(1)_R \times \SO(5)$ action, namely $(b_1, b_2, b_3) \sim \log (t, x , y) $, and, therefore, from \eqref{action}
we read
\bea\label{D1}
 \Delta_{\phi_1}= \frac{2}{3} \, , \quad \Delta_{A_1}= \frac{2}{3} \frac{b_1-b_3}{b_1} \, , \quad \Delta_{A_2}= \frac{2}{3} \frac{b_1-b_2}{b_1} \, , \quad \Delta_{B_1}= \frac{2}{3} \frac{b_1+b_2}{b_1} \, , \quad \Delta_{B_2}= \frac{2}{3} \frac{b_1+b_3}{b_1} \, ,
\eea
where we normalized the R-charges by requiring that $ \Delta_{\phi_1}= 2 / 3$.
The same expression can be derived by evaluating the volumes \eqref{V52vol} as a limit of  the Hilbert series for divisors, using the method discussed in \cite{Hanany:2008fj}. The volume \eqref{VV52} then reads
\be\label{VV522}
 \text{Vol}_{\text{S}}  (V^{5,2}) = \frac{54 \pi^4}{ (b_1-b_2)(b_1+b_2)(b_1-b_3)(b_1+b_3)}  \, .
\ee 
From  \eqref{S2} we also find
\bea\label{D2}
 \fn_1= \frac{n_1}{3} \, , \quad \fn_2= \frac{n_1-n_3}{3} \, , \quad \fn_3= \frac{n_1-n_2}{3} \, , \quad \fn_4= \frac{n_1+n_2}{3} \, , \quad \fn_5= \frac{n_1+n_3}{3} \, ,
\eea
where we associate $a=1,2,3,4,5$ to $(\phi_1,A_1,A_2,B_1,B_2)$ in this order.

At this point it is straightforward  to check that
\be
 S (b_i , n_i) \Big |_{b_1=1} = - 4 \nabla \sqrt{\frac{2 \pi^6}{27 \text{Vol}_{{\text S}}(V^{5,2})}} N^{3/2} \bigg |_{b_1=1}
 = - \frac12  \sum_{a=1}^5 \fn_a \frac{\partial F_{S^3} (\Delta_a)}{\partial \Delta_a} \, ,
\ee
where, this time,  we set $b_1=1$ at the end of the computation. To obtain the correct normalization it is important to remember that, for black hole solutions, supersymmetry  requires $b_1 = 1$, while, for volume minimization, we impose $b_1=4$. 
We learn that the construction in  \cite{Couzens:2018wnk,Gauntlett:2018dpc} is equivalent off-shell to $\cI$-extremization
\be S(b_i,n_i) = \cI (\Delta_a,\fn_a) \, .\ee

The same analysis can be applied to other Sasaki-Einsten manifolds $Y_7$ without five-cycles. In particular, it applies  to some of the models recently discussed in \cite{Amariti:2019pky}. We can also provide a very general argument that applies to many different theories and it is based on very few assumptions. 

\subsection{A general argument}\label{subsec:general}
On general grounds, the entropy functional \eqref{S1} coincides with the $\cI$-functional \eqref{iextr0} for all quivers with no baryonic symmetries
where the off-shell equivalence of $F$-maximization and volume minimization \eqref{free} is valid.
Indeed, for any quiver gauge theory with superpotential $W=\sum_a W_a$, the R-charges of the fields must satisfy
\be\label{W1} \sum_{I\in W_a} \Delta_I = 2 \, ,\ee
for each superpotential monomial $W_a$, where the sum is restricted to the fields that appear in $W_a$.
Similarly, the fluxes are constrained to satisfy
\be
 \label{W2} \sum_{I\in W_a} \fn_I = 2 - 2 \fg \, ,
\ee
for each superpotential term. This is just the twisting condition \eqref{n1:twisting}. In theories with no baryonic symmetries, there is the same number of  independent $\Delta_I$ and $\fn_I$ as the number of independent components of the Reeb vector.
There must be then a linear relation  among the $\Delta_I$ and the $b_i$,
\be
 \label{par}
 \Delta_I =\sum_{i=1}^s \alpha_{Ii} \frac{b_i}{b_1} \, ,
\ee
that parameterizes the $\U(1)^s$ action on the R-charges. By consistency, a similar linear relation exists between the $\fn_I$ and the $n_i$
\be
 \label{par2}  \fn_I =\sum_{i=1}^s \alpha_{Ii} \frac{n_i}{2} \, ,
\ee
where the normalization is determined by comparing \eqref{W1} and \eqref{W2}, and recalling that $n_1=2 - 2 \fg$.
These expressions generalize \eqref{D1} and \eqref{D2}.  We now assume that, for our theory, $F$-maximization is equivalent to volume minimization, \ie\; 
\bea\label{free2}
 F_{S^3} (\Delta_I) = N^{3/2} \sqrt{\frac{2 \pi^6}{27 \text{Vol}_{\text{S}}(Y_7)}} \, ,
\eea
when substituting \eqref{par} and imposing $b_1=4$. We also assume that there exists a parameterization  of  R-charges such that $F_{S^3} (\Delta_I)$ is a homogeneous function of degree two. This is generally the case. 
We can then write
\be
 \label{F1}  F_{S^3} (\Delta_I) = 4 \bigg( \sum_{I\in W_a} \Delta_I \bigg)^2 f(\Delta_I) \, ,
\ee
for some choice of a term $W_a$ in the superpotential and a function $f(\Delta_I)$ homogeneous of degree zero.
Since $\text{Vol}_{\text{S}}(Y_7)$ is homogeneous of degree minus four in $b_i$, we must also have
\be
 \label{F2}
 N^{3/2} \sqrt{\frac{2 \pi^6}{27 \text{Vol}_{\text{S}}(Y_7)}} = b_1^2 f \bigg(\sum_{i=1}^s \alpha_{Ii} \frac{b_i}{b_1} \bigg) = b_1^2 f \bigg(\sum_{i=1}^s \alpha_{Ii} b_i \bigg) \, ,
\ee
for the same function $f$. Setting $b_1=4$ and using \eqref{par} and \eqref{W1} we find indeed \eqref{free2}.

It takes then a short computation to show that \eqref{F1} and \eqref{F2} imply
\be
 S (b_i , n_i) \Big |_{b_1=1} = -4 \nabla \sqrt{\frac{2 \pi^6}{27 \text{Vol}_{{\text S}}(Y_7)}} N^{3/2} \bigg |_{b_1=1}= -\frac12  \sum_{a=1}^{s} \fn_a \frac{\partial F_{S^3} (\Delta_a)}{\partial \Delta_a} = \cI (\Delta_a,\fn_a) \, ,
\ee
using \eqref{par2} and setting $b_1=1$ after taking derivatives. This confirms the off-shell equivalence of the entropy functional with the $\cI$-functional.

In quivers where the elementary fields can be associated with linear combinations of invariant five-cycles $S_a$, their R-charge can be computed using \eqref{V52vol}.
By consistency, the result must reproduce \eqref{par}. Formula \eqref{S2} is then clearly consistent with \eqref{par2}.

\section{Extremization for toric manifolds}\label{sec:mesonictwist}

In this section we consider the $F$-maximization and $\cI$-extremization principles in the case of toric manifolds. We will first review the construction of \cite{Gauntlett:2018dpc} for the toric case. We will then  see that for the  class of  compactifications with a particular  mesonic twist, the construction in  \cite{Gauntlett:2018dpc} is equivalent to $\cI$-extremization. We will also comment about the general case.

Recall that the Sasaki-Einstein manifold $Y_7$ is toric when it has  isometry $\U(1)^4$. In this case, the cone over $Y_7$, $C(Y_7)$ is a toric Calabi-Yau four-fold, which can be characterized in terms of combinatorial data associated with a toric diagram.  Indeed, $C(Y_7)$ can be described by a fan generated by $d$ four-dimensional vectors $v_a=(1,{\vec v}_a)$. The toric diagram is the three-dimensional polytope in $\mathbb{R}^3$  with vertices the integer points ${\vec v}_a$. 
The toric four-fold $C(Y_7)$ has exactly $d$ $\U(1)^4$ invariant divisors $D_a$,  one for each vertex ${\vec v}_a$. The restriction to the base $Y_7$ gives rise to $\U(1)^4$-invariant five-cycles $S_a$, $a=1, \ldots d$. Since  ${\rm dim} H_5(Y_7,\bZ)={\rm dim} H_2(Y_7,\bZ)=d-4$,  only $d-4$ of these cycles are inequivalent in cohomology.  

\subsection{The conditions of supersymmetry for toric manifolds}\label{sec:toric}

The off-shell family of supersymmetric backgrounds of \cite{Couzens:2018wnk,Gauntlett:2018dpc} can be parameterized by the Reeb vector $b=(b_1,b_2,b_3,b_4)$ with $b_1=1$ and by the cohomology class of the transverse K\"ahler form. In the toric case, this can be specified as follows \cite{Couzens:2018wnk,Gauntlett:2018dpc}.  It is useful to restrict to the quasi regular case where the quotient with respect to the Reeb action, $V = Y_7/\U(1)$, is a six-dimensional compact toric orbifold. The restriction of $\omega$ and $\rho$ to $V$ can be written in cohomology as
\be
 \label{omega:rho:toric}
[\omega] = - 2 \pi \sum_{a=1}^d \lambda_a c_a \, , \qquad [\rho] = 2 \pi \sum_{a=1}^d c_a \, ,
\ee
where $c_a$ are the Poincar\'e duals of the restriction of the $d$ toric divisors to $V$.
The parameters $\lambda_a$ parameterize the cohomology class of the transverse K\"ahler form $J$. Since  ${\rm dim} H_2(V,\bZ)=d-3$,
only $d-3$ parameters $\lambda_a$ are independent. The Sasaki case is recovered for $\lambda_a= -1/(2b_1)$.

The master volume is then defined as the volume of the dual polytope associated with the K\"ahler parameter $\lambda_a$ \cite{Gauntlett:2018dpc}:
\bea
 \label{master:volume:1D}
 \cV = \frac{1}{6} \int_{Y_7} \eta \wedge \omega^3 = \frac{(2\pi)^4}{|b|} \text{Vol} \left( \left \{ y\in H(b) \, |\, (y-y_0,v_a) \ge \lambda_a \, , a=1,\ldots, 4 \right \} \right) ,
\eea
where $H(b)$ is the hyperplane $(y,b)=1/2$ and $y_0=(1,0,0,0)/(2b_1)$. The master volume \eqref{master:volume:1D} is invariant under
\be\label{gauge} 
 \lambda_a \to \lambda_a + \sum_{i=1}^4 l_i ( b_1 v_a^i - b_i) \, , \quad \forall a = 1, \ldots, d \, ,
\ee
and this leaves indeed  $d-3$ independent $\lambda_a$ since $l_1$ does not contribute ($v_a^1=1 \, , \forall a = 1, \ldots, d$). 

The twisted compactification is specified by four mesonic fluxes $n_i$  and $d-4$ baryonic ones $\fn_\alpha$. They can be conveniently parameterized by $d$ integer magnetic fluxes $\fn_a$, one for each toric divisor. 
The supersymmetry and flux quantization conditions for the off-shell background can be then written as \cite{Gauntlett:2018dpc}%
\footnote{In order to compare with \cite{Gauntlett:2018dpc} one must  set $L^6=(2\pi l_P)^6$, $\Delta_a= \frac{R_a}{N}$ and  $\fn_a = -\frac{M_a}{N}$.} 
\bea
 \label{constraints}
 & N= - \sum_{a=1}^d \frac{\partial \cV}{\partial\lambda_a} \, , \\
 & \fn_a N = -\frac{A}{2\pi} \sum_{b=1}^d \frac{\partial^2 \cV}{\partial\lambda_a\partial \lambda_b} - b_1 \sum_{i=1}^4 n_i  \frac{\partial^2 \cV}{\partial\lambda_a\partial b_i} \, ,\\
 & \frac{A}{2 \pi} \sum_{a,b=1}^d \frac{\partial^2 \cV}{\partial\lambda_a\partial \lambda_b} = n_1 \sum_{a=1}^d \frac{\partial \cV}{\partial\lambda_a} - b_1 \sum_{i=1}^4 n_i \sum_{a=1}^d  \frac{\partial^2 \cV }{\partial\lambda_a\partial b_i} \, .
\eea
Notice that $n_i$ and $\fn_a$ are not independent.
Consistency of the equations above requires $n_i = \sum_{a=1}^d v_a^i \fn_a$. In particular, the twisting condition \eqref{n1:twisting} becomes
\be
 \label{twisting}
 \sum_{a=1}^d \fn_a = 2 - 2 \fg \, .
\ee
The total number of independent fluxes is $d-1$, corresponding to $d-4$ baryonic symmetries and three mesonic ones.  
With $n_i = \sum_{a=1}^d v_a^i \fn_a$, only $d-2$ equations in \eqref{constraints} are independent and allow to determine the $d-3$ independent $\lambda_a$ and the area $A$ as a function of $\fn_a$ and $b_i$. Notice that, since the equations involve derivatives with respect to $b_1$, we can set $b_1=1$ only at the end of the computation.
 
One can also simplify $ S_{\text{SUSY}}$,  and define  the entropy functional \cite{Gauntlett:2018dpc} 
\bea
 \label{entropy:functional}
 S (b_i,\fn_a) \equiv 8 \pi^2 S_{\text{SUSY}} = - 8 \pi^2 \bigg ( A \sum_{a=1}^d \frac{\partial \cV}{\partial \lambda_a} +2 \pi b_1 \sum_{i=1}^4 n_i  \frac{\partial \cV}{\partial b_i} \bigg ) \bigg |_{\lambda_a(b,\fn), \, A(b,\fn)} \, .
\eea
For future reference, we also define the on-shell value of the master volume \eqref{master:volume:1D}:
\bea
 \label{master:volume:on-shell}
 \cV_{\text{on-shell}}(b_i,\fn_a) \equiv \cV \Big |_{\lambda_a(b,\fn), \, A(b,\fn)} \, .
\eea

We can explicitly solve the conditions of supersymmetry for a certain class of twisted compactifications.
In order to simplify the exposition, we summarize the results and refer to appendix \ref{sec:app}  for the proof.

\subsection{The universal twist}
\label{subsec:UT}

The simplest case of twisting is when we twist the theory \emph{only} along the exact R-symmetry of the theory, \ie\;
\be
 n_i = \frac{b_i}{b_1} n_1 \, , \quad \forall i = 1,\ldots , 4 \, .
\ee
Recall that $n_1 = 2 - 2 \fg$.
This is the so-called universal twist \cite{Benini:2015bwz}. According to the general argument in \cite{Azzurli:2017kxo}, we expect that the entropy of the black holes
is given by
\be
 S_{\text{BH}} = ( \fg - 1 ) F_{S^3} \, .
\ee
This formula indeed follows from the conditions of supersymmetry \eqref{constraints}. The argument is analogous to the similar one
discussed in \cite{Gauntlett:2018dpc} in the context of $c$-extremization.

As shown in appendix \ref{sec:appUT}, there is a solution to the equations \eqref{constraints} where all the $\lambda_a$ are equal\footnote{We normalize in such a way that $\lambda=1$ corresponds to the Sasaki case.}
\be
 \label{lambda:universal:twist}
 \lambda_a = - \frac{1}{2b_1} \lambda \, , \quad \forall a = 1,\ldots , d \, .
\ee
Consistency of the equations  \eqref{constraints}  then implies that
\be\label{fluxesUT} \fn_a =\frac{n_1}{2} \left( \frac{2\pi}{3 b_1} \frac{\text{Vol}_{\text{S}}(S_a)}{\text{Vol}_{\text{S}}(Y_7)} \right) \equiv \frac{n_1}{2} \Delta_a(b_i) \, , \ee
where we introduced the set of basic R-charges \eqref{Rcharges}. Notice that $\sum_{a=1}^d \Delta_a(b_i)=2$.
Note also that $\sum_{a=1}^d \fn_a = n_1$ hence \eqref{twisting} is correctly satisfied. 

The entropy functional \eqref{entropy:functional} becomes 
\bea
 \label{SBH1}
 S (b_i,\fn_a) 
  = - \frac{8 n_1}{b_1^{3/2}} N^{3/2} \sqrt{\frac{2 \pi^6}{27 \text{Vol}_{\text{S}}(Y_7)}} \, ,
\eea
where $\text{Vol}_{\text{S}}(Y_7)$ is the Sasaki volume \eqref{Sasaki:volumes}.
The volume functional $\text{Vol}_{\text{S}}(Y_7)$ is extremized at $b_i/b_1 = \bar b_i/\bar b_1$, with $\bar b = (4,\bar b_2,\bar b_3,\bar b_4)$ being the Reeb vector of the Calabi-Yau four-fold \cite{Martelli:2005tp}.
In our case supersymmetry requires $b_1 = 1$ and we obtain $b_i = \bar b_{i}/4$, and thus, using the homogeneity of the volume function,
\be\label{rescaling}
 \text{Vol}_{\text{S}} (Y_7 ) \big|_{b = \bar b/4} = 4^4 \text{Vol}(Y_7) \, ,
\ee
where $\text{Vol}(Y_7)$ is the volume of the Sasaki-Einstein manifold.
The entropy functional \eqref{SBH1} at the extremum is then given by
\be
 S (b_i,\fn_a) \big|_{b = \bar b/4} = - \frac{n_1}{2} N^{3/2} \sqrt{\frac{2 \pi^6}{27 \text{Vol}(Y_7)}} \equiv ( \fg - 1 ) F_{S^3} \, ,
\ee
where we set $b_1=1$ and used \eqref{free} and \eqref{twisting}.
This is in agreement with the results in \cite{Hosseini:2016tor,Azzurli:2017kxo}. Note also that
\be
 \cV_{\text{on-shell}} = \frac{1}{64 \pi^3} F_{S^3} \, .
\ee

Finally notice that, since volume minimization is the dual of $F$-maximization,  at the extremum, the R-charges $\Delta(b_i)$ becomes the exact R-charges of the fields of the three-dimensional CFT.\footnote{More precisely, the R-charges of the fields are integer linear combinations of the $\Delta(b_i)$.} Therefore,  the relation \eqref{fluxesUT}
\be \fn_a = (1-\fg) \Delta_a  \big|_{b = \bar b/4}  \, , \ee
tells us that all the fluxes are proportional to the corresponding exact R-charges, hence confirming that we are dealing with the universal twist.
Since the fluxes $\fn_a$ must be integers, the universal twist is defined only for manifolds $Y_7$ where the exact R-charges $\Delta_a$
are rational and for certain values of $\fg$. Examples of Sasaki-Einstein manifolds with known duals admitting the universal twist are discussed in \cite{Azzurli:2017kxo}.

\subsection{Mesonic twist}
\label{subsec:mesonic:twist}

In this section we perform a particular topological twist that we dub {\it mesonic twist}. It depends on  three magnetic fluxes that we can take to be the $n_i$ (recall that $n_1=2 - 2 \fg$). 
It is characterized by the $d-4$ conditions
\be
 \label{Decoupling:lambda}
 \sum_{a = 1}^{d} B_a^{(i)} \lambda_a = 0 \, , \qquad \forall i=1,\ldots , d-4 \, ,
\ee
on the K\"ahler parameters $\lambda_a$. Here $B_a^{(i)}$ denotes the baryonic symmetries that can be defined  geometrically by the vector identity 
\be
 \label{baryonic}
 \sum_{a=1}^{d} B_a^{(i)} v_a = 0 \, , \qquad \forall i = 1, \ldots, d - 4 \, .
\ee
The condition \eqref{Decoupling:lambda} requires a particular choice for the fluxes $\fn_a$ that are specific functions of $b_i$, $n_i$ and $\fg$, $\fn_a=\fn_a(b_i,n_i,\fg)$. As for the universal twist, the quantization conditions
for the fluxes give constraints on the twisted compactification. In particular, the value of $b_i$ at the extremum  and $\fg$ must be  such that $\fn_a(b_i,n_i,\fg)$ is an integer.
The constraint here is milder than for the universal twist, since  the value of $b_i$ at the extremum can be tuned by varying $n_i$, but still it restricts the class of solutions.   

To study the mesonic twist,  it is convenient to use the freedom \eqref{gauge} to fix some of the $\lambda_a$ and work in the gauge
\be
 \label{gauge:lambda:main}
 \lambda_1 = \lambda_2 = \lambda_3 = 0 \, .
\ee
Furthermore, we will prove  in appendix \ref{sec:app} that there exists a solution to the set of equations \eqref{constraints}, compatible with \eqref{Decoupling:lambda}, such that
\be
 \label{lambda:solution:toric}
 \lambda_a = - \frac12 \frac{(v_1,v_2,v_3,v_a)}{(v_1,v_2,v_3,b)} \lambda \, , \quad \forall a = 1, \ldots , d \, .
\ee
Notice that \eqref{Decoupling:lambda} is satisfied as a consequence of \eqref{baryonic}. 

As in \cite{Hosseini:2019use} we define the normalized R-charges
\be
 \label{normRcharges}
 \Delta_a (b_i , \fn_a) \equiv \frac{2 \pi}{N}  \int_{S_\alpha} \eta \wedge \omega^2   \bigg|_{\lambda_a(n,\fn) , \, A(b,\fn)}  = - \frac{2}{N} \frac{\partial \cV}{\partial \lambda_a} \bigg|_{\lambda_a(n,\fn) , \, A(b,\fn)} \, , 
\ee
inspired by \eqref{RSa}.  As a consequence of the first equation in \eqref{constraints} they satisfy
\be
\sum_{a = 1}^d \Delta_a(b_i,\fn_a) = 2\, .
\ee
Quite remarkably, as shown in appendix \ref{sec:appMT}, the conditions of supersymmetry \eqref{constraints} imply
that the $\Delta_a$ are actually independent of the fluxes $\fn_a$ and are given by the Sasakian parameterization \eqref{Rcharges}
\be
 \label{Delta:J:ratio}
 \Delta_a (b_i) = \frac{2 \pi}{3 b_1} \frac{\text{Vol}_{\text{S}} (S_a)}{\text{Vol}_{\text{S}}(Y_7)} \, , \qquad \forall a = 1, \ldots, d \, ,
\ee
where the volumes are given in \eqref{Sasaki:volumes}. They satisfy the useful identity  
\be
 \label{Reeb0}
2\frac{b_k}{b_1} = \sum_{a=1}^d v_a^k \Delta_a(b_i) \, , \qquad \forall k = 1, \ldots, 4 \, .
\ee

The on-shell value of the master volume can be written as
\bea\label{masterMT}
 \cV_{\text{on-shell}} (b_i) 
 = \frac{N^{3/2}}{4 \pi^3 b_1^{3/2}}  \sqrt{\frac{2 \pi^6}{27 \text{Vol}_{\text{S}} (Y_7)}} \equiv \frac{1}{64 \pi^3} F_{S^3} (\Delta_a) \, ,
\eea
where in the last step we set $b_1=1$ and used the equivalence between volume minimization and $F$-maximization, see \eqref{free}.
More precisely, since $\text{Vol}_{\text{S}} (Y_7)$ is a homogeneous function of the $b_i$ of degree $-4$, and we can choose $F_{S^3} (\Delta_a)$ to be homogeneous of degree $2$,  we can write
\bea
 \label{FF2}
 & N^{3/2} \sqrt{\frac{2 \pi^6}{27 \text{Vol}_{\text{S}}(Y_7)}} = b_1^2 f \bigg( \frac{b_i}{b_1} \bigg) \, , \\
 & F_{S^3} (\Delta_a) = 4 \bigg( \sum_{a=1}^d \Delta_a \bigg)^2 f \bigg( \sum_{a=1}^d \frac{v_a^i \Delta_a}{2} \bigg) \, . 
\eea
for some function $f$. Setting $b_1=4$, as appropriate for $F$-maximization, we find \eqref{free}. Setting $b_1=1$, we find \eqref{masterMT}.

The entropy functional \eqref{entropy:functional} is given by 
\be\label{S10}
 S (b_i,\fn_a) = - \frac{4}{\sqrt{b_1}} \nabla \sqrt{\frac{2 \pi^6}{27 \text{Vol}_{{\text S}}(Y_7)}} N^{3/2} \, ,
\ee
where we defined the operator
\bea\label{nabla0}  \nabla \equiv \sum_{i=1}^{4} n_i \partial_{b_i}\, .\eea
Notice that, since $\nabla$ explicitly takes a derivative with respect to $b_1$, we can set $b_1=1$ only at the end of the computation.

On the other hand, the topologically twisted index of $\cN=2$ theories in the large $N$ limit reads \cite{Hosseini:2016tor}
\be
 \label{index:theorem:1D}
 \cI (\Delta_a , \fn_a) = - \frac12 \sum_{a=1}^{d} \fn_a \frac{\partial F_{S^3} (\Delta_a)}{\partial \Delta_a} \, .
\ee
It takes a short computation to show that \eqref{FF2} implies
\be
 S (b_i , n_i) \Big |_{b_1=1} = - 4 \nabla \sqrt{\frac{2 \pi^6}{27 \text{Vol}_{{\text S}}(Y_7)}} N^{3/2} \bigg |_{b_1=1}
 = - \frac12  \sum_{a=1}^{d} \fn_a \frac{\partial F_{S^3} (\Delta_a)}{\partial \Delta_a} = \cI (\Delta_a,\fn_a)  \, ,
\ee
where we used $n_i = \sum_{a=1}^d v_a^i \fn_a$.  This confirms the off-shell equivalence of the entropy functional with the $\cI$-functional.

As anticipated, the solution is only consistent if we impose the following constraints on the fluxes
\bea
 \label{S20}
 \fn_a = \frac{1}{2} \nabla ( b_1 \Delta_a) \, , \qquad \forall a = 1, \ldots, d \, ,
\eea
leaving only the $n_i$ as independent fluxes. Notice that, as required by consistency, 
\be
  \sum_{a=1}^d v_a \fn_a
 = \frac{1}{2} \sum_{a=1}^d \nabla ( b_1 v_a \Delta_a)
 = \nabla b =n \, ,
\ee
where we used \eqref{Reeb0}. 

Note that \eqref{S10} and \eqref{S20} for the mesonic twist are formally identical to the expressions that we found for theories with no baryonic symmetries, \eqref{S1} and \eqref{S2}.

\subsection{Interpreting the mesonic twist}
\label{mesTgeo}

The condition \eqref{Decoupling:lambda} and the constraints \eqref{S20} have a nice geometrical interpretation. The condition \eqref{Decoupling:lambda} is equivalent to requiring that
 the R-charges $\Delta(b_i)$ are parameterized in terms of the Sasaki volumes as in the three-dimensional $F$-maximization problem. Since \eqref{Delta:J:ratio} only depends on the Reeb vector components $b_i$, there are  $d-4$ constraints among the $\Delta(b_i)$. In all the examples that we have checked, these constraints can be written as cubic polynomials   
in the $\Delta_a$. Moreover, they can be compactly written  in terms  of derivatives of a single  auxiliary quartic polynomial.%
\footnote{We thank Francesco Sala and Yuji Tachikawa for useful discussions on this point and collaboration on a related project.}
In all known examples, there exist indeed  a quartic polynomial $a_{3\rd}(\Delta)$ that identically coincides with $F_{S^3} (\Delta_a)^2$ on the locus parameterized by \eqref{Delta:J:ratio} \cite{Amariti:2011uw,Amariti:2012tj}. We have checked that the $d-4$ constraints among R-charges can be written as
\be
 \label{Rcharge constraints}
 \sum_{a=1}^d B_a^{(i)} \frac{\partial a_{3\text{d}} (\Delta)}{\partial \Delta_a} =0 \, , \qquad \forall i = 1, \ldots, d - 4 \, .
\ee
Furthermore, we have checked that the constraints \eqref{S20} among fluxes can be written as
\be
 \label{flux constraints}
 \sum_{a,b=1}^d B_a^{(i)} \fn_b \frac{\partial^2 a_{3\rd} (\Delta)}{\partial \Delta_a \partial \Delta_b}  =0 \, , \qquad \forall i = 1, \ldots, d - 4\, ,
\ee
which is quadratic in $\Delta_a$. The last statement is equivalent to say that the flux constraints \eqref{S20} can be obtained by applying the operator  $\sum_{a=1}^d \fn_a \partial_{\Delta_a}$ to the R-charge constraints. 


There is also some interesting field theory interpretation for  some of R-charge constraints \eqref{Rcharge constraints}.  In the available computation for the large $N$ limit of the $S^3$ free energy \cite{Jafferis:2011zi} and the topologically twisted index of ${\cal N}=2$ quivers \cite{Hosseini:2016tor}, they arise  when one imposes that the theory has gauge group $\U(1)\times \SU(N)^G$. The free energy and the index are the same for $\U(N)$ and $\SU(N)$ groups and, as already discussed, depend only on a linear combination of the $\Delta_a$ parameterizing the mesonic directions. However, for $\SU(N)$ gauge groups one has to impose that the distribution of eigenvalues of the matrix model is traceless. This gives some constraints among the $\Delta_a$ that allows to fix some of R-charges of baryonic operators, as already mentioned at the end of section \ref{Fmax}.
In all known examples,  the $\SU(N)$ constraints -- when nontrivial -- coincide with a subset of the R-charge constraints \eqref{Rcharge constraints}.\footnote{Since not all baryonic symmetries are realized as ungaugings of $\U(N)$ gauge groups, not all  constraints \eqref{Rcharge constraints} arise in this way. It would be very interesting to find the field theory interpretation of the remaining constraints.}  We will see  explicit examples in section  \ref{sec:Examples}. 
 
Finally, we notice that there is a similar story for flows from four-dimensional $\cN=1$ CFTs to $(0,2)$ two-dimensional CFTs induced by twisted compactifications on $\Sigma_\fg$.
In this case, volume minimization \cite{Martelli:2005tp,Martelli:2006yb} is the dual of $a$-maximization \cite{Intriligator:2003jj} and the construction in \cite{Couzens:2018wnk,Gauntlett:2018dpc} is the dual of $c$-extremization \cite{Benini:2013cda,Benini:2015bwz}. A general solution to the equations in \cite{Couzens:2018wnk,Gauntlett:2018dpc} for arbitrary fluxes $\fn_a$ has been provided in \cite{Hosseini:2019use} together with a general formula for $\Delta_a(\fn_a,b_i)$. This solution automatically satisfies
\be
 \label{a2}
 \sum_{a=1}^d B_a^{(i)} \frac{\partial c_{2\rd}(\Delta)}{\partial \Delta_a} = \sum_{a,b=1}^d B_a^{(i)} \fn_b \frac{\partial^2 a_{4\rd} (\Delta)}{\partial \Delta_a \partial \Delta_b}  = 0 \, , \qquad \forall i = 1, \ldots, d - 3 \, ,
\ee
where $c_{2\rd}(\Delta)$ is the two-dimensional trial right-moving central charge, and $a_{4\rd}(\Delta)$ is the four-dimensional trial $a$ central charge.
Physically, this condition guarantees that the two-dimensional central charge is extremized with respect 
to the baryonic directions. Moreover, using the explicit solution in \cite{Hosseini:2019use}, it is not difficult to show that
\be
 \label{a10}
 \sum_{a=1}^d B_a^{(i)} \lambda_a = - \frac{1}{(3 \pi)^3 N} \sum_{a=1}^d B_a^{(i)} \frac{\partial a_{4\rd}(\Delta)}{\partial \Delta_a} \, , \qquad \forall i = 1, \ldots, d - 3 \, .
\ee
Also in four dimensions, we can restrict to the {\it mesonic twist}, defined 
again by the condition $\sum_a B_a^{(i)}  \lambda_a=0$. This condition, using \eqref{a10}, then becomes 
\be
 \label{a1}
 \sum_{a=1}^d B_a^{(i)} \frac{\partial a_{4\rd}(\Delta)}{\partial \Delta_a} = 0 \, , \qquad \forall i = 1, \ldots, d - 3 \, ,
\ee
and \eqref{a2} becomes a condition that expresses the baryonic fluxes in terms of the mesonic ones, $\fn_a=\fn_a(b_i,n_i)$, as in the three-dimensional case. The conditions \eqref{a1} and \eqref{a2} are the analogue of \eqref{Rcharge constraints} and \eqref{flux constraints}. Notice that \eqref{a1} can be interpreted as the extremization of  the four-dimensional $a$ central charge with respect to  the baryonic directions. As such, it enters as a decoupling condition for the baryonic symmetries in the proof of the equivalence between volume minimization and $a$-maximization  \cite{Butti:2005vn}. We see that, in the four-dimensional context, the mesonic twist corresponds to enforce the decoupling of baryonic symmetries before flowing from four to two dimensions. This explains the name and it was our original motivation for studying it. 

We expect that, similarly to the  case of $c$-extremization discussed in \cite{Hosseini:2019use},  one can find a general solution to the equations \eqref{constraints} such that it reduces to the mesonic twist when further 
imposing the decoupling condition \eqref{Decoupling:lambda}. Unfortunately, solving \eqref{constraints} in general is hard because the equations are quadratic in $\lambda_a$.
It would be very interesting to see if the quartic function $a_{3\rd}(\Delta)$ plays some role in the general solution, as its counterpart $a_{4\rd}(\Delta)$ does for $c$-extremization \cite{Hosseini:2019use}.

\section{Toric examples}
\label{sec:Examples}

In this section we consider some examples of the general construction presented in section \ref{subsec:mesonic:twist}. We thus perform a mesonic twist of the three-dimensional $\cN=2$ theories. We will use the results in  \cite{Hanany:2008fj,Benini:2009qs,Martelli:2011qj,Jafferis:2011zi} to which we refer for more details on the parameterization of R-charges and the large $N$ limit of $S^3$ free energy.

\subsection[The \texorpdfstring{$\cC \times \bC$}{conifold x C} geometry]{The $\cC \times \bC$ geometry}

Our first example is a flavored ABJM theory (with Chern-Simons levels set to zero), whose quiver is depicted below \cite{Benini:2009qs,Cremonesi:2010ae}
\bea
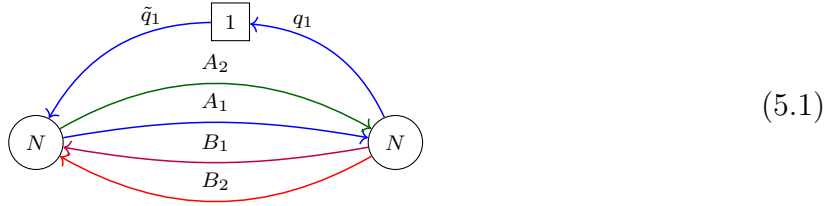
\label{conifold x C:quiver}
\begin{tikzpicture}[baseline, font=\scriptsize, scale=0.8]
\begin{scope}[auto,%
  every node/.style={draw, minimum size=0.5cm}, node distance=4cm];
\node[circle] (UN1) at (0, 0) {$N$};
\node[circle, right=of UN1] (UN2)  {$N$};
\node[rectangle] at (3.2,2.) (UNa1)  {$1$};
\end{scope}
\draw[draw=red,solid,line width=0.2mm,<-]  (UN1) to[bend right=30] node[midway,above] {$B_2 $}node[midway,above] {}  (UN2) ;
\draw[draw=blue,solid,line width=0.2mm,->]  (UN1) to[bend right=-10] node[midway,above] {$A_1$}node[midway,above] {}  (UN2) ; 
\draw[draw=purple,solid,line width=0.2mm,<-]  (UN1) to[bend left=-10] node[midway,above] {$B_1$} node[midway,above] {} (UN2) ;  
\draw[draw=black!60!green,solid,line width=0.2mm,->]  (UN1) to[bend left=30] node[midway,above] {$A_2$} node[midway,above] {} (UN2) ;   
\draw[draw=blue,solid,line width=0.2mm,->]  (UN2)  to[bend right=30] node[pos=0.9,right] {}   (UNa1);
\draw[draw=blue,solid,line width=0.2mm,->]  (UNa1) to[bend right=30] node[pos=0.1,left] {} (UN1) ; 
\node at (4.4,2.) {$q_1$};
\node at (1.9,2.1) {$\tilde{q}_1$};
\end{tikzpicture}
\eea
where a circular node denotes a $\U(N)$ gauge group and the square node denotes a $\U(1)$ flavor symmetry.
There are bi-fundamental chiral multiplets between the two gauge groups labeled by $(A_i , B_i)$, $i=1,2$; fundamental and anti-fundamental chiral multiplets labeled by $(q_1, \tilde q_1)$.
The theory has the superpotential
\be
 \label{W:conifoldxC}
 W = \Tr \left( A_1 B_1 A_2 B_2 - A_1 B_2 A_2 B_1 + q_1 A_1 \tilde q_{1} \right) \, .
\ee
The moduli space of the quiver gauge theory \eqref{conifold x C:quiver}  can be characterized in terms of the fields $A_i$, $B_i$ and two monopole operators $T_1$ and $T_2$ satisfying  $T_1 T_2= A_1$ \cite{Benini:2009qs}.
The mesonic spectrum is generated  by the gauge invariant operators $x_1=T_1 B_1$, $x_2=A_2 B_2$, $x_3=T_1 B_2$, $x_4=A_2 B_1$, $x_5=T_2$ \cite{Benini:2009qs}, subject to the relation $x_1 x_2=x_3x_4$.
Hence, the dual geometry is  $\cC \times \bC$ ($\cC$ the conifold). 
Denote the R-charges of the chiral bi-fundamental fields $(A_1,A_2,B_1,B_2)$ and the monopoles $(T_1 , T_2)$ by $(\Delta_{A_1},\Delta_{A_2},\Delta_{B_1},\Delta_{B_2})$ and $(\Delta_{T_1}, \Delta_{T_2})$, respectively.
We also define the bare monopole R-charges $\Delta_{m_1}$ and $\Delta_{m_2}$ associated with the topological symmetries \cite{Jafferis:2011zi}, satisfying
\be
 \Delta_{T_1} - \Delta_{T_2} = 2 \Delta_{m} \, , \qquad \Delta_m = \Delta_{m_1} + \Delta_{m_2} \, .
\ee
That the superpotential \eqref{W:conifoldxC} has R-charge two, imposes the following constraint on the R-charges
\be
 \Delta_{A_1} + \Delta_{A_2} + \Delta_{B_1} + \Delta_{B_2} = 2 \, .
\ee
We also introduce the magnetic fluxes $(\fn_{A_1},\fn_{A_2},\fn_{B_1},\fn_{B_2})$ and $(\fn_{m_1}, \fn_{m_2})$.
 Supersymmetry imposes the constraint
\be
 \fn_{A_1} + \fn_{A_2} + \fn_{B_1} + \fn_{B_2} = 2 - 2 \fg\, .
\ee

The dual geometry $\cC \times \bC$  is specified by the vectors
\be
 \vec{v}_1 = ( 0, 0, 0 ) \, , \quad \vec{v}_2 = ( 1, 0, 0 ) \, ,  \quad \vec{v}_3 = ( 0, 1, 0 ) \, , \quad \vec{v}_4 = ( 0, 0, 1 ) \, ,  \quad \vec{v}_{5} = ( 1, 1, 0 ) \, .
\ee
Its toric diagram is shown below
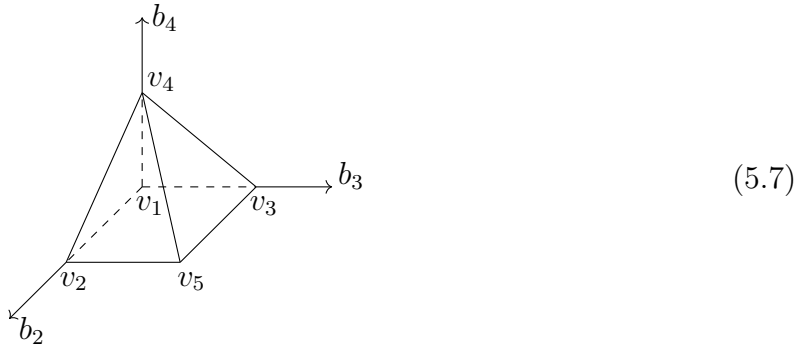
\bea
 \label{toric:conifoldxC}
 \begin{tikzpicture}
  [scale=0.5 ]
  
  \draw[-,dashed] (0,0) -- (3,0) node[below] {};
  \draw[->,solid] (3,0) -- (5,0) node[below] {};
  
  \draw[-,dashed] (0,0) -- (0,2.5) node[below] {};
  \draw[->,solid] (0,2.5) -- (0,4.5) node[below] {};
  
  \draw[-,dashed] (0,0) -- (-2,-2) node[below] {};
  \draw[->,solid] (-2,-2) -- (-3.5,-3.5) node[below] {};
  
  \draw (-2.9,-3.8) node {$b_2$};
  \draw (5.5,0.3) node {$b_3$};
  \draw (0.6,4.5) node {$b_4$};
  
  \draw[-,solid] (3.,0) -- (1,-2) node[below] {};
  \draw[-,solid] (1,-2) -- (-2,-2) node[below] {};
  \draw[-,solid] (-2,-2) -- (0,2.5) node[below] {};
  \draw[-,solid] (0,2.5) -- (1,-2) node[below] {};
  \draw[-,solid] (0,2.5) -- (3,0) node[below] {};
  
  \draw (0.2,-0.5) node {$v_1$};
  \draw (-1.8,-2.5) node {$v_2$};
  \draw (3.2,-.5) node {$v_3$};
  \draw (0.5,2.8) node {$v_4$};
  \draw (1.3,-2.5) node {$v_5$};
  
 \end{tikzpicture}
\eea
We associate R-charges $\Delta_a$ and fluxes $\fn_a$, $a=1,\ldots,5$, to the five vertices $v_a$ of the toric diagram \eqref{toric:conifoldxC}.
They are related to the R-charges and fluxes of the chiral fields and monopoles by \cite{Benini:2009qs,Jafferis:2011zi}\footnote{The $\Delta_a$ can be associated to a parameterization in terms of GLSM fields. See in particular \cite[(6.21)]{Benini:2009qs}.} 
\bea
 \Delta_{A_1} = \Delta_1 + \Delta_4 \, , ~~\quad & \Delta_{A_2} = \Delta_5 \, , \quad && \Delta_{B_1} = \Delta_2 \, , &&& \Delta_{B_2} = \Delta_3 \, , &&&&\Delta_m = \frac{1}{2} ( \Delta_1 - \Delta_4 ) \, , \\
 \fn_{A_1} = \fn_1 + \fn_4 \, , ~~ \quad & \fn_{A_2} = \fn_5 \, , \quad && \fn_{B_1} = \fn_2 \, , &&& \fn_{B_2} = \fn_3 \, , &&&& \fn_m = \frac{1}{2} ( \fn_1 - \fn_4 ) \, ,
\eea
where $\fn_m = \fn_{m_1} + \fn_{m_2}$.
The $S^3$ free energy of this theory was computed in \cite[(6.9)]{Jafferis:2011zi} and it reads\footnote{We correct a typo there.}
\be
 F_{S^3} (\Delta_a) = \frac{4 \pi \sqrt{2} N^{3/2}}{3} \sqrt{\frac{(\Delta_1+\Delta_2) (\Delta_1+\Delta_3) \Delta_4 (\Delta_2+\Delta_5) (\Delta_3+\Delta_5)}{\Delta_1+\Delta_2+\Delta_3+\Delta_5}} \, ,
\ee
In the $\SU(N) \times \SU(N) \times \U(1)$ theory one has to impose the additional constraint
\be
 \label{conifold:constrain:R-charge}
 \Delta_1 \Delta_5 - \Delta_2 \Delta_3 = 0 \, ,
\ee
as discussed in \cite{Jafferis:2011zi}. This arises due the tracelessness condition on the eigenvalues distribution.
Remarkably, this constraint is equivalent to
\be
 \sum_{a=1}^{5} B_a \frac{\partial a_{3\rd} (\Delta_a)}{\partial \Delta_a} = 0 \, ,
\ee
where $a_{3\rd} (\Delta_a)$ is given by
\bea
 a_{3\rd} (\Delta_a) & \equiv \frac{1}{24} \sum_{a , b , c , e = 1}^{5} \left| \det ( v_a , v_b , v_c , v_e ) \right| \Delta_a \Delta_b \Delta_c \Delta_e \\
 & = \left( \Delta_1 \Delta_2 \Delta_3 + \Delta_1 \Delta_3 \Delta_5 + \Delta_2 \Delta_3 \Delta_5 + \Delta_1 \Delta_2 \Delta_5 \right) \Delta_4 \, ,
\eea
and the baryonic symmetry, characterized by \eqref{baryonic}, reads
\be
 B_1 = 1 \, , \qquad B_2 = -1 \, , \qquad B_3 = -1 \, , \qquad B_4 = 0 \, , \qquad B_5 = 1 \, .
\ee
Since $F_{S^3} (\Delta_a)$ is homogeneous of degree two, the topologically twisted index of this theory is simply given by \eqref{iextr0}, \ie\;
\bea
 \cI (\Delta_a , \fn_a ) & = - \frac{\pi \sqrt{2} N^{3/2}}{3} \sqrt{\frac{(\Delta_1+\Delta_2) (\Delta_1+\Delta_3) \Delta_4 (\Delta_2+\Delta_5) (\Delta_3+\Delta_5)}{\Delta_1+\Delta_2+\Delta_3+\Delta_5}} \bigg[ \frac{\fn_4}{\Delta_4} \\
 & + \left(\frac{1}{\Delta_1+\Delta_3}-\frac{1}{\Delta_1+\Delta_2+\Delta_3+\Delta_5}+\frac{1}{\Delta_1+\Delta_2}\right) \fn_1 \\
 & + \left(\frac{1}{\Delta_2+\Delta_5}-\frac{1}{\Delta_1+\Delta_2+\Delta_3+\Delta_5}+\frac{1}{\Delta_1+\Delta_2}\right) \fn_2 \\
 & + \left(\frac{1}{\Delta_3+\Delta_5}-\frac{1}{\Delta_1+\Delta_2+\Delta_3+\Delta_5}+\frac{1}{\Delta_1+\Delta_3}\right) \fn_3 \\
 & + \left(\frac{1}{\Delta_3+\Delta_5}-\frac{1}{\Delta_1+\Delta_2+\Delta_3+\Delta_5}+\frac{1}{\Delta_2+\Delta_5}\right) \fn_5
 \bigg] \, .
\eea
The dual polytope associated with the K\"ahler parameters $\lambda_a$ is given by
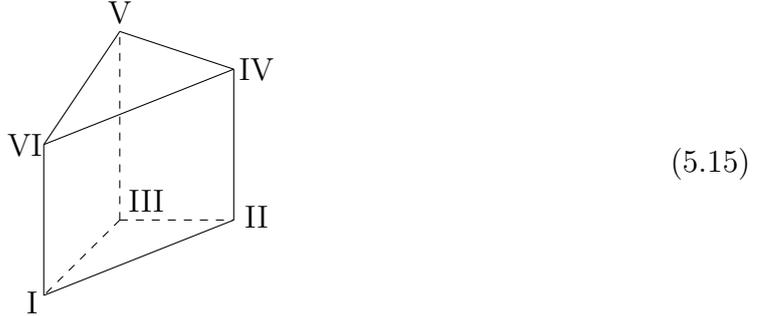
\bea
 \label{dual:polytope:conifoldxC}
 \begin{tikzpicture}
  [scale=0.5 ]
  
  \draw[-,dashed] (0,0) -- (3,0) node[below] {};
  \draw[-,dashed] (0,0) -- (0,5) node[below] {};
  \draw[-,dashed] (0,0) -- (-2,-2) node[below] {};
  
  \draw[-,solid] (3,0) -- (-2,-2) node[below] {};
  \draw[-,solid] (-2,-2) -- (-2,2) node[below] {};
  \draw[-,solid] (3,0) -- (3,4) node[below] {};
  \draw[-,solid] (3,4) -- (-2,2) node[below] {};
  \draw[-,solid] (3,4) -- (0,5) node[below] {};
  \draw[-,solid] (0,5) -- (-2,2) node[below] {};
  
  \draw (0.7,0.5) node {\rom{3}};
  \draw (3.6,0.1) node {\rom{2}};
  \draw (-2.3,-2.2) node {\rom{1}};
  \draw (0.,5.5) node {\rom{5}};
  \draw (3.6,4.) node {\rom{4}};
  \draw (-2.5,2) node {\rom{6}};
  
 \end{tikzpicture}
\eea
whose vertices can be found by solving for every distinct triple $v_a$, $v_b$, $v_c$ the equations
\be
 (y-y_0,v_a)= \lambda_a \, , \qquad (y-y_0,v_b)= \lambda_b \, , \qquad (y-y_0,v_c)= \lambda_c \, , \qquad (y-y_0,b)= 0 \, .
\ee
They are related to the facets of the toric diagram \eqref{toric:conifoldxC} by
\bea
 \rom{1} & = (321) \, , ~~~~ && \rom{2} = (431) \, , &&& \rom{3} = (412) \, , \\
 \rom{4} & = (345) \, , && \rom{5} = (254) \, , &&& \rom{6} = (352) \, .
\eea
Note, that the facet $(3521)$ in \eqref{toric:conifoldxC} corresponds to a pyramid in $\bR^4$ --- this is a singularity on the facet --- and we resolved it by blowing up the surface, \ie\;we write it as $(321) + (352)$.
The master volume \eqref{master:volume:1D} is then easily computed by splitting the dual polytope into tetrahedra and adding the corresponding volumes. It reads 
\bea
 \label{V:conifoldxC}
 \cV & = -\frac{8 \pi^4 (b_1 - b_2 - b_3 - b_4)^2 \lambda_1^3}{3 b_2 b_3 b_4}
 - \frac{8 \pi^4 \left(b_2^2 - b_3 b_2 - b_3 ( b_1 - b_3 - b_4)\right) \lambda_2^3}{3 b_3 (b_1-b_2-b_4) b_4} \\
 & - \frac{8 \pi^4 \left(b_2^2-b_1 b_2 - ( b_3 - b_4) b_2+b_3^2\right) \lambda_3^3}{3 b_2 (b_1-b_3-b_4) b_4}
 - \frac{8 \pi^4 b_4^2 ( b_1 - b_4 ) \lambda_4^3}{3 b_2 b_3 (b_1 - b_2 - b_4) ( b_1 - b_3 - b_4)} \\
 & + \frac{8 \pi^4 ( b_1 - b_2 - b_3 - b_4)^2 \lambda_5^3}{3 (b_1-b_2-b_4) (b_1-b_3-b_4) b_4}
 - \frac{8 \pi^4 b_4 \lambda_4^2 \lambda_5}{( b_1 - b_2 - b_4 ) ( b_1 - b_3 - b_4)} \\
 & - 8 \pi^4 ( b_1 - b_2 - b_3 - b_4) \left( \frac{\lambda_2}{b_3 b_4} + \frac{\lambda_3}{b_2 b_4} + \frac{\lambda_4}{b_2 b_3}\right) \lambda_1^2 \\
 & + 8 \pi^4 \left( \frac{\lambda_3}{b_4} - \frac{(b_2-b_3) \lambda_4}{b_3 ( b_1 - b_2 - b_4)} - \frac{( b_1 - b_3 - b_4) \lambda_5}{(b_1-b_2-b_4) b_4}\right) \lambda_2^2 \\
 & -  8 \pi^4 \left( \frac{b_2 \lambda_2^2}{b_3 b_4} + \left( \frac{2 \lambda_3}{b_4} + \frac{2 \lambda_4}{b_3}\right) \lambda_2
 + \frac{2 \lambda_3 \lambda_4}{b_2} + \frac{b_4 \lambda_4^2}{b_2 b_3} + \frac{b_3 \lambda_3^2}{b_2 b_4}\right) \lambda_1 \\
 & + \frac{8 \pi^4 ( b_1 - b_2 - b_3 - b_4) \lambda_4 \lambda_5^2}{(b_1 - b_3 - b_4) ( b_1 - b_2 - b_4)}
 - \frac{8 \pi^4}{b_1 - b_3 - b_4} \left( \frac{(b_1-b_2-b_4) \lambda_5}{b_4} - \frac{(b_2-b_3) \lambda_4}{b_2}\right) \lambda_3^2 \\
 & + 8 \pi^4 \left( \frac{\lambda_3^2}{b_4} - \frac{2 \lambda_5 \lambda_3}{b_4} - \frac{b_4 \lambda_4^2}{b_3 ( b_1 - b_2 - b_4)}
 + \frac{(b_1-b_2-b_3-b_4) \lambda_5^2}{(b_1-b_2-b_4) b_4} - \frac{2 \lambda_4 \lambda_5}{b_1 - b_2 - b_4}\right) \lambda_2 \\
 & - \frac{8 \pi^4}{b_1 - b_3 - b_4} \left(\frac{b_4 \lambda_4^2}{b_2} + 2 \lambda_5 \lambda_4 - \frac{(b_1-b_2-b_3-b_4) \lambda_5^2}{b_4}\right) \lambda_3 \, .
\eea
From now on, we work in the gauge \eqref{gauge:lambda:main}, \ie\;$\lambda_1 = \lambda_2 = \lambda_3 = 0$. Using \eqref{constraints} and \eqref{lambda:solution:toric}, we may fix the remaining $\lambda_a$ as
\be
 \lambda_4 = - \frac{\sqrt{N}}{2 \sqrt{2} \pi^2} \sqrt{\frac{b_2 b_3 (b_1 - b_2 - b_4) (b_1 - b_3 - b_4)}{b_1 (b_1-b_4) b_4}} \, , \qquad \lambda_5 = 0 \, .
\ee
Notice that $\sum_a B_a \lambda_a = 0$, as it is required by the mesonic twist.
The last equation in \eqref{constraints} is easily solved for $A$. Plugging the solutions for $A$ and $\lambda_a$ back into \eqref{V:conifoldxC} we obtain
\be
 \cV_{\text{on-shell}} = \frac{N^{3/2}}{6 \sqrt{2} \pi^2 b_1} \sqrt{\frac{b_2 b_3 b_4 (b_1 - b_2 - b_4) (b_1 - b_3 - b_4)}{b_1 (b_1 - b_4)}} \, .
\ee
For the entropy functional \eqref{entropy:functional} we find that
\bea
 S (b_i & , \fn_a) = -\frac{2 \sqrt{2} \pi N^{3/2}}{3} \sqrt{\frac{b_2 b_3 b_4 ( b_1 - b_2 - b_4 ) ( b_1 - b_3 - b_4 )}{b_1 ( b_1 - b_4 )}} \bigg( \frac{\fn_4}{b_4} \\
 & + \frac{\left( ( b_1 - b_4 )^2 + 2 b_2 b_3 \right) \fn_1}{( b_1 - b_4 ) ( b_1 - b_2 - b_4 ) ( b_1 - b_3 - b_4 )}
 + \frac{ \left( b_1^2 - ( b_3+ 2 b_4 ) b_1- 2 b_2 b_3 + b_4 (b_3 + b_4 ) \right) \fn_2}{b_2 (b_1 - b_4 ) ( b_1 - b_3 - b_4 )} \\
 & + \frac{\left( b_1^2 - ( b_2 + 2 b_4 ) b_1 + b_4^2 + b_2 ( b_4 - 2 b_3 ) \right) \fn_3}{b_3 ( b_1 - b_4 ) ( b_1 - b_2 - b_4 )}
 + \frac{\left( 2 b_2 b_3 + b_1 ( b_2 + b_3 ) - ( b_2 + b_3 ) b_4 \right) \fn_5}{b_2 b_3 ( b_1 - b_4 )}
 \bigg) \, .
\eea
Using the expression \eqref{normRcharges} for the R-charges, we also obtain 
\bea
 \label{Delta:b:conifoldxC}
 \Delta_1 & = \frac{2 (b_1-b_2-b_4) (b_1-b_3-b_4)}{b_1 (b_1-b_4)} \, , \qquad
 \Delta_2 = \frac{2 b_2}{b_1} \left( 1- \frac{b_3}{b_1-b_4} \right) \, , \\
 \Delta_3 &= \frac{2 b_3}{b_1} \left( 1- \frac{b_2}{b_1-b_4} \right) \, , \qquad
 \Delta_4 = \frac{2 b_4}{b_1} \, , \qquad \Delta_5 = \frac{2 b_2 b_3}{b_1 (b_1 - b_4)} \, .
\eea
Let us emphasize again that in the case of the mesonic twist the R-charges are \emph{independent} of the magnetic charges and coincide with  the Sasakian parameterization \eqref{Rcharges}. Notice that the R-charges \eqref{Delta:b:conifoldxC} automatically satisfy \eqref{conifold:constrain:R-charge}.
Finally, the second equation in \eqref{constraints} imposes the following constraint on the magnetic fluxes
\be
 \label{constraint:fluxes:conifoldxC}
 0  = \Delta_5 \fn_1 - \Delta_3 \fn_2 - \Delta_2 \fn_3 + \Delta_1 \fn_5 = \sum_{a=1}^{5} \fn_a \frac{\partial}{\partial \Delta_a} (\Delta_1 \Delta_5 - \Delta_2 \Delta_3) \, .
\ee
Notice that the constraint on fluxes can be obtained by applying the operator $\sum_{a=1}^d \fn_a \partial_{\Delta_a}$ to the R-charge constraint \eqref{conifold:constrain:R-charge}. This constraint is also equivalent to \eqref{flux constraints}. This will be true for all other examples in this paper.
It is easy to see that, using \eqref{Delta:b:conifoldxC} and \eqref{constraint:fluxes:conifoldxC}, and setting $b_1=1$,
\bea
 S (b_i , \fn_a) & = \cI (\Delta_a , \fn_a)\bigg|_{\Delta_a(b_i)}
 = - \frac{1}{2} \sum_{a=1}^{5} \fn_a \frac{\partial F_{S^3}(\Delta_a)}{\partial \Delta_a} \bigg|_{\Delta_a(b_i)} \\
 & = - \frac{2 \sqrt{2} \pi N^{3/2}}{3} \sum_{a=1}^{5} \fn_a \frac{\partial \sqrt{a_{3\rd}(\Delta_a)}}{\partial \Delta_a} \bigg|_{\Delta_a(b_i)} \, .
\eea
We thus proved the \emph{off-shell} equivalence of the $\cI$-extremization principle and its geometric dual.
It is also interesting to observe that
\be
 \cV_{\text{on-shell}} (b_i) = \frac{1}{64 \pi^3} F_{S^3}(\Delta_a) \bigg|_{\Delta_a(b_i)} = \frac{N^{3/2}}{24 \sqrt{2} \pi^2} \sqrt{a_{3\text{d}} (\Delta_a)}\bigg|_{\Delta_a(b_i)} \, .
\ee

\subsection[The cone over \texorpdfstring{$Q^{1,1,1}$}{Q[111]}]{The cone over $Q^{1,1,1}$}

Our second example is a gauge theory described by the quiver \cite{Benini:2009qs,Cremonesi:2010ae}
\bea
 
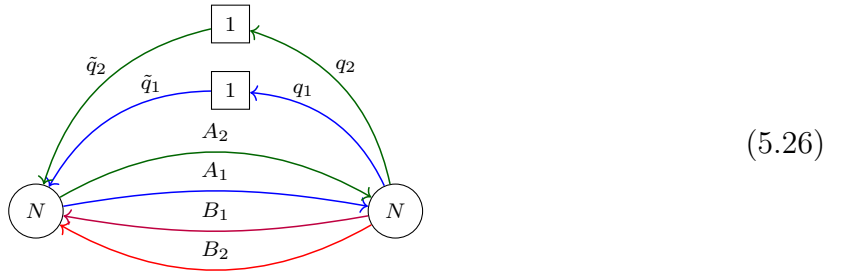
\label{Q111:quiver}
\begin{tikzpicture}[baseline, font=\scriptsize, scale=0.8]
\begin{scope}[auto,%
  every node/.style={draw, minimum size=0.5cm}, node distance=4cm];
\node[circle] (UN1) at (0, 0) {$N$};
\node[circle, right=of UN1] (UN2)  {$N$};
\node[rectangle] at (3.2,2.) (UNa1)  {$1$};
\node[rectangle] at (3.2,3.1) (UNa2)  {$1$};
\end{scope}
\draw[draw=red,solid,line width=0.2mm,<-]  (UN1) to[bend right=30] node[midway,above] {$B_2 $}node[midway,above] {}  (UN2) ;
\draw[draw=blue,solid,line width=0.2mm,->]  (UN1) to[bend right=-10] node[midway,above] {$A_1$}node[midway,above] {}  (UN2) ; 
\draw[draw=purple,solid,line width=0.2mm,<-]  (UN1) to[bend left=-10] node[midway,above] {$B_1$} node[midway,above] {} (UN2) ;  
\draw[draw=black!60!green,solid,line width=0.2mm,->]  (UN1) to[bend left=30] node[midway,above] {$A_2$} node[midway,above] {} (UN2) ;   
\draw[draw=blue,solid,line width=0.2mm,->]  (UN2)  to[bend right=30] node[pos=0.9,right] {}   (UNa1);
\draw[draw=blue,solid,line width=0.2mm,->]  (UNa1) to[bend right=30] node[pos=0.1,left] {} (UN1) ; 
\draw[draw=black!60!green,solid,line width=0.2mm,->]  (UN2)  to[bend right=30]  node[pos=0.9,right] {} (UNa2);
\draw[draw=black!60!green,solid,line width=0.2mm,->]  (UNa2) to[bend right=30] node[pos=0.1,left] {}  (UN1); 
\node at (4.4,2.) {$q_1$};
\node at (1.9,2.1) {$\tilde{q}_1$};
\node at (5.1,2.4) {$q_2$};
\node at (1.,2.4) {$\tilde{q}_2$};
\end{tikzpicture}
\eea
where the notation is understood as before.
The theory has the superpotential
\bea
 \label{W:Q111}
 W = \Tr \left( A_1 B_1 A_2 B_2 - A_1 B_2 A_2 B_1 + q_{1} A_1 \tilde q_{1} + q_{2} A_2 \tilde q_{2} \right) \, .
\eea
The manifold in this case is $Y_7 = Q^{1,1,1}$ that is defined by the coset
\be
 \frac{\SU(2) \times \SU(2) \times \SU(2)}{\U(1) \times \U(1)} \, .
\ee
The cone $C(Q^{1,1,1})$ determines a polytope with six vertices
\bea
 \vec{v}_1 & = ( 1, 0, 0 ) \, , ~~~ && \vec{v}_2 = ( 0, 1, 0 ) \, , &&& \vec{v}_3 = ( 0, 0, 1 ) \, , \\
 \vec{v}_4 & = ( 1, 0, 1) \, , && \vec{v}_5 = ( 1, 1, 0) \, , &&& \vec{v}_6 = ( 0, 1, 1 ) \, .
\eea
The toric diagram is depicted below
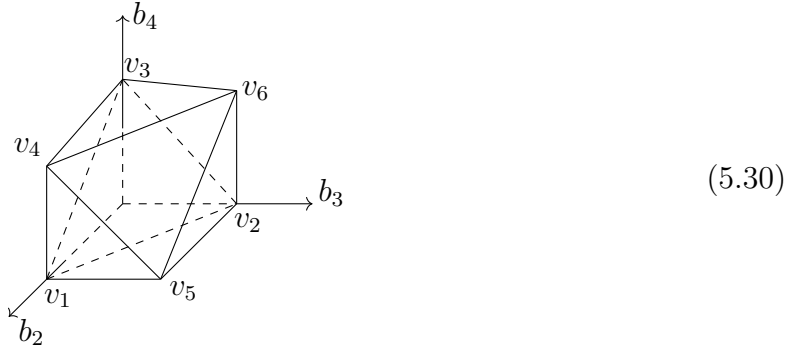
\bea
 \label{toric:Q111}
 \begin{tikzpicture}
  [scale=0.5 ]
  
  \draw[-,dashed] (0,0) -- (3,0) node[below] {};
  \draw[->,solid] (3,0) -- (5,0) node[below] {};
  
  \draw[-,dashed] (0,0) -- (0,2) node[below] {};
  \draw[->,solid] (0,2) -- (0,5) node[below] {};
  
  \draw[-,dashed] (0,0) -- (-1.6,-1.6) node[below] {};
  \draw[->,solid] (-1.6,-1.6) -- (-3.,-3.) node[below] {};
  
  \draw (-2.4,-3.4) node {$b_2$};
  \draw (5.5,0.3) node {$b_3$};
  \draw (0.6,5) node {$b_4$};
    
  \draw[-,solid] (-2,1.) -- (1.,-2) node[below] {};
  \draw[-,solid] (-2,1.) -- (3,3) node[below] {};
  \draw[-,solid] (1.,-2) -- (3,3) node[below] {};
  \draw[-,solid] (-2,-2) -- (1.,-2) node[below] {};
  \draw[-,solid] (-2,-2) -- (-2,1.) node[below] {};
  \draw[-,dashed] (-2,-2) -- (3,0) node[below] {};
  \draw[-,solid] (3,3) -- (3,0) node[below] {};
  \draw[-,solid] (3,0) -- (1.,-2) node[below] {};
  \draw[-,solid] (0,3.3) -- (3,3) node[below] {};
  \draw[-,solid] (0,3.3) -- (-2,1.) node[below] {};
  \draw[-,dashed] (-2,-2) -- (0,3.3) node[below] {};
  \draw[-,dashed] (3,0) -- (0,3.3) node[below] {};
  
  \draw (-1.7,-2.5) node {$v_1$};
  \draw (3.3,-0.5) node {$v_2$};
  \draw (0.4,3.6) node {$v_3$};
  \draw (-2.5,1.5) node {$v_4$};
  \draw (1.6,-2.3) node {$v_5$};
  \draw (3.5,3) node {$v_6$};
  
 \end{tikzpicture}
\eea
Since $\text{dim} H_2(Q^{1,1,1},\bZ) = 2$, there are two baryonic symmetries $\U(1)_{B_1} \times \U(1)_{B_2}$ that are characterized by \eqref{baryonic}. They are given by
\bea
 B_1^{(1)} = 1 \, , ~~\quad & B_2^{(1)} = -1 \, , ~~&& B_3^{(1)} = 0 \, , &&& B_4^{(1)} = -1 \, , &&&& B_5^{(1)} = 0 \, , &&&&& B_6^{(1)} = 1 \, , \\
 B_1^{(2)} = 0 \, , ~~\quad & B_2^{(2)} = -1 \, , ~~&& B_3^{(2)} = 1 \, , &&& B_4^{(2)} = -1 \, , &&&& B_5^{(2)} = 1 \, , &&&&& B_6^{(2)} = 0 \, .
\eea
Thus, the symmetries of the model is $\SU(2)^3 \times \U(1)_R \times \U(1)_{B_1} \times \U(1)_{B_2}$, where $\SU(2)^3 \times \U(1)_R$ are the isometries of $Q^{1,1,1}$ (mesonic symmetries). The moduli space of the quiver gauge theory \eqref{Q111:quiver}  can be characterized in terms of the fields $A_i$, $B_i$ and two monopole operators $T_1$ and $T_2$ satisfying  $T_1 T_2= A_1 A_2$ \cite{Benini:2009qs}.  The mesonic spectrum is generated by the gauge invariants  $B_i A_j$ and $B_i T_j$, transforming as $( \bm{2}, \bm{2}, \bm{2} )$ of $\SU(2)^3$. Using \cite[(6.28)]{Benini:2009qs}, and assigning R-charges  $\Delta_{T_i}$ to the monopoles, we find the relation between the chiral fields R-charges $\Delta_I$ and the basic R-charges $\Delta_a$ associated with toric geometry
\bea &\Delta_{B_1}= \Delta_1 \, ,\qquad \,\,\,\,\,\, \,\,\,\,\,\,\, \,\, \Delta_{B_2}= \Delta_6 \, , \qquad &\Delta_{T_1} = \Delta_3 + \Delta_4 \, , \\
& \Delta_{A_1}= \Delta_2+\Delta_3  \, ,\qquad \Delta_{A_2}= \Delta_4+\Delta_5  \, , \qquad &\Delta_{T_2} = \Delta_2 + \Delta_5 \, .
\eea
The dual polytope associated with the K\"ahler parameters $\lambda_a$ is given by
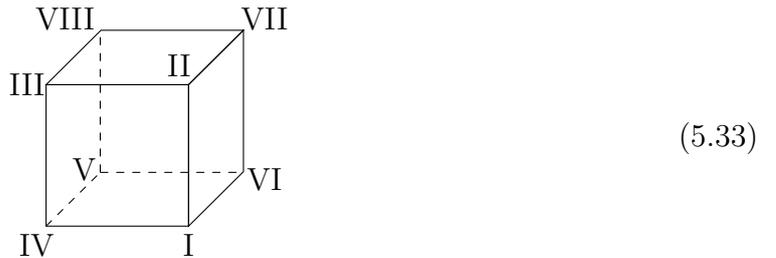
\bea
 \label{dual:polytope:Q111}
 \begin{tikzpicture}
 [scale=1.25 ]
  \pgfmathsetmacro{\cubex}{1.5}
  \pgfmathsetmacro{\cubey}{1.5}
  \pgfmathsetmacro{\cubez}{1.5}
  \draw[black] (0,0,0) -- ++(-\cubex,0,0) -- ++(0,-\cubey,0) -- ++(\cubex,0,0) -- cycle;
  \draw[black] (0,0,0) -- ++(0,0,-\cubez) -- ++(0,-\cubey,0) -- ++(0,0,\cubez) -- cycle;
  \draw[black] (0,0,0) -- ++(-\cubex,0,0) -- ++(0,0,-\cubez) -- ++(\cubex,0,0) -- cycle;
    
  \draw[-,dashed] (-1.5,-1.5) -- (-0.93,-0.93) node[below] {};
  \draw[-,dashed] (-0.93,-0.93) -- (0.55,-0.93) node[below] {};
  \draw[-,dashed] (-0.93,-0.93) -- (-0.93,0.55) node[below] {};
  
  \draw (0,-1.7) node {\rom{1}};
  \draw (-0.1,0.2) node {\rom{2}};
  \draw (-1.7,0.) node {\rom{3}};
  \draw (-1.6,-1.7) node {\rom{4}};
  \draw (-1.1,-.9) node {\rom{5}};
  \draw (0.8,-1) node {\rom{6}};
  \draw (0.8,0.7) node {\rom{7}};
  \draw (-1.3,0.7) node {\rom{8}};

 \end{tikzpicture}
\eea
whose vertices correspond to the facets of the toric diagram \eqref{toric:Q111} as follows
\bea
 \rom{1} & = (645) \, , ~~~~ && \rom{2} = (634) \, , &&& \rom{3} = (431) \, , &&&& \rom{4} = (415) \, , \\
 \rom{5} & = (251) \, , && \rom{6} = (265) \, , &&& \rom{7} = (623) \, , &&&& \rom{8} = (132) \, .
\eea
The master volume can be easily computed from \eqref{dual:polytope:Q111}. Since the resulting expression is too long we report it in the appendix  --- see \eqref{Q111:master:volume}.
As before we work in the gauge \eqref{gauge:lambda:main}. Using \eqref{constraints} and \eqref{lambda:solution:toric} we may fix $\lambda_a$ and $A$.
Their explicit expressions are quite long and we shall not present them here.
Employing the parameterization \eqref{normRcharges} for the R-charges, we obtain
{\footnotesize \bea
 \label{Delta:b:Q111}
 \Delta_1 & = \frac{2 b_2 (b_1-b_3) (b_1-b_4) (b_3+b_4) (2 b_1 - b_2 - b_3 - b_4)}{b_1 \left(2 (b_3 b_4+b_2 (b_3+b_4)) b_1^2-\left((b_3+b_4) b_2^2+\left(b_3^2+6 b_4 b_3+b_4^2\right) b_2+b_3 b_4 (b_3+b_4)\right) b_1+2 b_2 b_3 b_4 (b_2+b_3+b_4)\right)} \, , \\
 \Delta_2 & = \frac{2 (b_1-b_2) b_3 (b_1-b_4) (2 b_1-b_2-b_3-b_4) (b_2+b_4)}{b_1 \left(2 (b_3 b_4+b_2 (b_3+b_4)) b_1^2-\left((b_3+b_4) b_2^2+\left(b_3^2+6 b_4 b_3+b_4^2\right) b_2+b_3 b_4 (b_3+b_4)\right) b_1+2 b_2 b_3 b_4 (b_2+b_3+b_4)\right)} \, , \\
 \Delta_3 & = \frac{2 (b_1-b_2) (b_1-b_3) (b_2+b_3) b_4 (2 b_1- b_2-b_3-b_4)}{b_1 \left(2 (b_3 b_4+b_2 (b_3+b_4)) b_1^2-\left((b_3+b_4) b_2^2+\left(b_3^2+6 b_4 b_3+b_4^2\right) b_2+b_3 b_4 (b_3+b_4)\right) b_1+2 b_2 b_3 b_4 (b_2+b_3+b_4)\right)} \, , \\
 \Delta_4 & = - \frac{2 b_2 (b_1-b_3) b_4 (2 b_1-b_2-b_4) (b_1-b_2-b_3-b_4)}{b_1 \left(2 (b_3 b_4+b_2 (b_3+b_4)) b_1^2-\left((b_3+b_4) b_2^2+\left(b_3^2+6 b_4 b_3+b_4^2\right) b_2+b_3 b_4 (b_3+b_4)\right) b_1+2 b_2 b_3 b_4 (b_2+b_3+b_4)\right)} \, , \\
 \Delta_5 & = - \frac{2 b_2 b_3 (2 b_1-b_2-b_3) (b_1-b_4) (b_1-b_2-b_3-b_4)}{b_1 \left(2 (b_3 b_4+b_2 (b_3+b_4)) b_1^2-\left((b_3+b_4) b_2^2+\left(b_3^2+6 b_4 b_3+b_4^2\right) b_2+b_3 b_4 (b_3+b_4)\right) b_1+2 b_2 b_3 b_4 (b_2+b_3+b_4)\right)} \, \\
 \Delta_6 & = - \frac{2 (b_1-b_2) b_3 (2 b_1-b_3-b_4) (b_1-b_2-b_3-b_4) b_4}{b_1 \left(2 (b_3 b_4+b_2 (b_3+b_4)) b_1^2-\left((b_3+b_4) b_2^2+\left(b_3^2+6 b_4 b_3+b_4^2\right) b_2+b_3 b_4 (b_3+b_4)\right) b_1+2 b_2 b_3 b_4 (b_2+b_3+b_4)\right)} \, .
\eea}
Notice that the R-charges \eqref{Delta:b:Q111} are independent of the fluxes and they satisfy
\be
 \sum_{a=1}^{6} B_a^{(i)} \frac{\partial a_{3\rd} (\Delta_a)}{\partial \Delta_a} = 0 \quad \text{ for } i = 1 , 2 \, ,
\ee
with \cite{Amariti:2011uw}
\bea
 a_{3\rd} (\Delta_a) & \equiv \frac{1}{24} \sum_{a , b , c , e = 1}^{6} \left| \det ( v_a , v_b , v_c , v_e ) \right| \Delta_a \Delta_b \Delta_c \Delta_e \\
 & + \frac{1}{2} ( \Delta_2 \Delta_4 + \Delta_3 \Delta_5 + \Delta_1 \Delta_6 )^2 - ( \Delta_2 \Delta_4 )^2 - ( \Delta_3 \Delta_5 )^2 - ( \Delta_1 \Delta_6)^2 \, .
\eea
Explicitly, we only write one of the two constraints that we will use later on, \ie\,
\be
 \label{Q111:Delta:constraint}
 \Delta_5 + \frac{(\Delta_1 \Delta_6 - \Delta_2 \Delta_4) (\Delta_1+\Delta_2+2 \Delta_3+\Delta_4+\Delta_6)}{2 (\Delta_1 \Delta_6-\Delta_2 \Delta_4) + \Delta_3 (\Delta_1-\Delta_2-\Delta_4+\Delta_6)} = 0 \, .
\ee
The associated constraint on the magnetic fluxes follows from the second equation in \eqref{constraints} and it can be written as
\bea
 \label{Q111:flux:constraint}
 0 = & \sum_{a=1}^{6} \fn_a \frac{\partial}{\partial \Delta_a} \left[ \text{LHS of } \eqref{Q111:Delta:constraint} \right] \\
 & = - \left( \Delta_6^2 + 2 \Delta_1 \Delta_6 + \Delta_2 \Delta_6 + 2 \Delta_3 \Delta_6 + \Delta_4 \Delta_6 + 2 \Delta_5 \Delta_6 - \Delta_2 \Delta_4 + \Delta_3 \Delta_5 \right) \fn_1 \\
 & + \left(\Delta_4^2+\Delta_1 \Delta_4+2 \Delta_2 \Delta_4+2 \Delta_3 \Delta_4+2 \Delta_5 \Delta_4+\Delta_6 \Delta_4+\Delta_3 \Delta_5-\Delta_1 \Delta_6\right) \fn_2 \\
 & + \left(2 \Delta_2 \Delta_4+\Delta_5 \Delta_4-\Delta_1 \Delta_5+\Delta_2 \Delta_5-2 \Delta_1 \Delta_6-\Delta_5 \Delta_6\right) \fn_3 \\
 & + \left(\Delta_2^2+\Delta_1 \Delta_2+2 \Delta_3 \Delta_2+2 \Delta_4 \Delta_2+2 \Delta_5 \Delta_2+\Delta_6 \Delta_2+\Delta_3 \Delta_5-\Delta_1 \Delta_6\right) \fn_4 \\
 & - \left( \Delta_1 \Delta_3 - \Delta_2 \Delta_3 - \Delta_4 \Delta_3 + \Delta_6 \Delta_3 - 2 \Delta_2 \Delta_4 + 2 \Delta_1 \Delta_6 \right) \fn_5 \\
 & - \left( \Delta_1^2 + \Delta_2 \Delta_1 + 2 \Delta_3 \Delta_1 + \Delta_4 \Delta_1 + 2 \Delta_5 \Delta_1 + 2 \Delta_6 \Delta_1 - \Delta_2 \Delta_4 + \Delta_3 \Delta_5 \right) \fn_6 \, .
\eea
To compare with the results in  \cite{Jafferis:2011zi,Hosseini:2016ume}, using the symmetries of the quiver \eqref{Q111:quiver}, we restrict
\bea
 \label{Q111:limit}
 & \Delta_1 = \Delta_6 \equiv 1 - \Delta \, , ~~ && \Delta_2 = \Delta_5 \equiv \frac{1}{2} ( \Delta -\Delta_m ) \, , &&& \Delta_3 = \Delta_4 \equiv \frac{1}{2} ( \Delta + \Delta_m ) \, , \\
 & \fn_1 = \fn_6 \equiv ( 1 - \fg ) ( 1 - \fn ) \, , ~~&& \fn_2 = \fn_5 \equiv \frac{1 - \fg}{2} ( \fn - \fn_m ) \, , &&& \fn_3 = \fn_4 \equiv \frac{1 - \fg}{2} ( \fn + \fn_m ) \, ,
\eea
where we defined
\bea
 & \Delta_{T_1} - \Delta_{T_2} = 2 \Delta_{m} \, , \qquad && \Delta_m = \Delta_{m_1} + \Delta_{m_2} \, , \\
 & \fn_{T_1} - \fn_{T_2} = 2 \fn_{m} \, , \qquad && \fn_m = \fn_{m_1} + \fn_{m_2} \, .
\eea
Hence, the constraint \eqref{Q111:Delta:constraint} becomes
\be
 \label{Q111:Delta:sim:const}
 \Delta = \frac{2}{3 - \Delta_m^2} \, .
\ee
The decoupling condition \eqref{Q111:flux:constraint} is also simplified to
\be
 \label{Q111:flux:sim:const}
 2 \Delta ( 3 - \Delta_m \fn_m ) + \left( 3 - \Delta_m^2 \right) \fn - 6 = 0 \, .
\ee
The $S^3$ free energy of this theory was computed in \cite[(6.15)]{Jafferis:2011zi} and it is given by 
\be
 F_{S^3} (\Delta_m) = \frac{4 \pi N^{3/2}}{3} \frac{|1 - \Delta_m^2|}{\sqrt{3 - \Delta_m^2}} \, ,
\ee
together with constraint \eqref{Q111:Delta:sim:const}, that exists in the $\SU(N) \times \SU(N) \times \U(1)$ theory, see \cite[(6.16)]{Jafferis:2011zi}.
The topologically twisted index was also computed in \cite[(5.47)]{Hosseini:2016ume} and it reads
\be
 \label{Q111:index}
 \cI (\Delta_m , \fn_m) = \frac{2 \pi ( 1 - \fg ) N^{3/2}}{3 \left( 3 - \Delta_m^2 \right)^{3/2}}
 \left( \Delta_m ( 5 - \Delta_m^2 ) \fn_m - \Delta_m^4 + 3 \Delta_m^2 - 6 \right)
 \, .
\ee
It is easy to evaluate the entropy functional \eqref{entropy:functional}; using \eqref{Delta:b:Q111}, \eqref{Q111:limit}, \eqref{Q111:flux:sim:const} and setting $b_1=1$, we find that
\be
 S (b_i , \fn_a) = \cI (\Delta_m , \fn_m)\Big|_{\Delta_m(b_i)} \, .
\ee
It is also interesting to observe that
\be
 \cV_{\text{on-shell}} (b_i) = \frac{1}{64 \pi^3} F_{S^3}(\Delta_m) \bigg|_{\Delta_m(b_i)} = \frac{N^{3/2}}{24 \sqrt{2} \pi^2} \sqrt{a_{3\text{d}} (\Delta_m)} \bigg|_{\Delta_m(b_i)} \, .
\ee
Finally, extremizing the index \eqref{Q111:index} with respect to $\Delta_m$ we obtain
\be
 \fn_m = -\frac{\Delta_m^3 \left( \Delta_m^2 - 9 \right)}{15 + \Delta_m^2} \, .
\ee
Using \eqref{Q111:Delta:sim:const} and \eqref{Q111:flux:sim:const}, we also find that
\be
 \fn = \frac{2}{9 -3 \Delta_m^2} + \frac{200}{3 \left( 15 + \Delta_m^2 \right)} - 4 \, .
\ee
As can be seen from the above equations, one can find many twisted compactifications where the quantization conditions on the fluxes $(\fn , \fn_m)$ are fulfilled.
In particular, it is enough to have a rational value for $\Delta_m$ at the extremum since then $(\fn , \fn_m)$ are also rational.
We then see from \eqref{Q111:limit} that we can make all the $\fn_a$ integer by taking the genus $\fg$ large enough.

\subsection[Flavoring \texorpdfstring{$\cN = 8$}{N=8} SYM]{Flavoring $\cN = 8$ SYM}

We consider the flavored $\cN=8$ super Yang-Mills whose quiver description is given by \cite{Benini:2009qs}
\bea
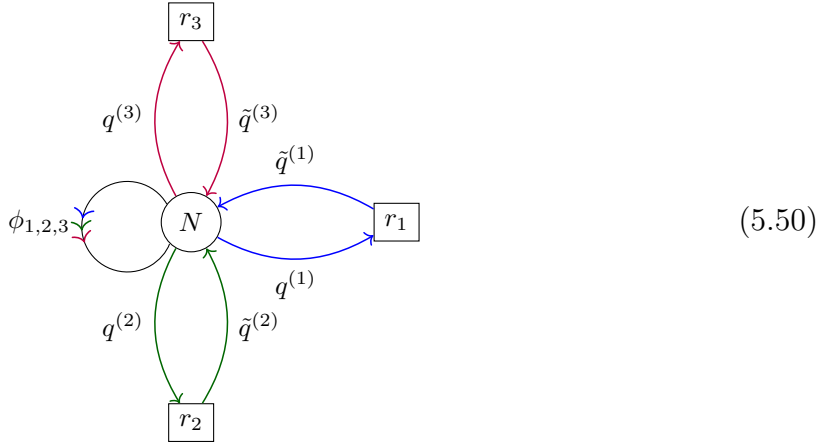
\begin{tikzpicture}[baseline, font=\footnotesize, scale=0.8]
\begin{scope}[auto,%
  every node/.style={draw, minimum size=0.5cm}, node distance=2cm];
\node[circle]  (UN)  at (0.3,1.7) {$N$};
\node[rectangle, right=of UN] (Ur1) {$r_1$};
\node[rectangle, below=of UN] (Ur2) {$r_2$};
\node[rectangle, above=of UN] (Ur3) {$r_3$};
\end{scope}
\draw[decoration={markings, mark=at position 0.45 with {\arrow[blue,scale=1.5]{>}}, mark=at position 0.5 with {\arrow[black!60!green,scale=1.5]{>}}, mark=at position 0.55 with {\arrow[purple,scale=1.5]{>}}}, postaction={decorate}, shorten >=0.7pt] (-0.1,2) arc (30:340:0.75cm);
\draw[draw=blue,solid,line width=0.2mm,->]  (UN) to[bend right=30] node[midway,below] {$q^{(1)}$}node[midway,above] {}  (Ur1) ; 
\draw[draw=blue,solid,line width=0.2mm,<-]  (UN) to[bend left=30] node[midway,above] {$\tilde{q}^{(1)}$} node[midway,above] {} (Ur1) ;    
\draw[draw=black!60!green,solid,line width=0.2mm,->]  (UN) to[bend right=30] node[midway,left] {$q^{(2)}$}node[midway,above] {}  (Ur2) ; 
\draw[draw=black!60!green,solid,line width=0.2mm,<-]  (UN) to[bend left=30] node[midway,right] {$\tilde{q}^{(2)}$} node[midway,above] {} (Ur2) ;  
\draw[draw=purple,solid,line width=0.2mm,->]  (UN) to[bend left=30] node[midway,left] {$q^{(3)}$}node[midway,above] {}  (Ur3) ; 
\draw[draw=purple,solid,line width=0.2mm,<-]  (UN) to[bend right=30] node[midway,right] {$\tilde{q}^{(3)}$} node[midway,above] {} (Ur3) ;    
\node at (-2.2,1.7) {$\phi_{1,2,3}$};
\end{tikzpicture}
\eea
where the loop around the circular node denotes the adjoint chiral multiplets $\phi_i$, $i=1,2,3$, and the rest of the notation is understood as before.
The theory has the superpotential
\begin{equation}
 \label{W:flavored:SYM}
 W = \Tr \bigg( \phi_1\left[\phi_2,\phi_3\right]
 + \sum_{j=1}^{r_1} q_j^{(1)} \phi_1 \tilde q_{j}^{(1)}
 + \sum_{j=1}^{r_2} q_j^{(2)} \phi_2 \tilde q_{j}^{(2)}
 + \sum_{j=1}^{r_3} q_j^{(3)} \phi_3 \tilde q_{j}^{(3)} \bigg) \, .
\end{equation}
The quantum corrected moduli space of vacua is a toric Calabi-Yau cone, parameterized by the complex coordinates $\phi_i$ and the monopole operators $T_1$, $T_2$ fulfilling the constraint
\be\label{quantumcorrected}
 T_1 T_2 = \phi_1^{r_1} \phi_2^{r_2} \phi_3^{r_3} \, .
\ee
We assign R-charges $(\Delta_{\phi_i},\Delta_{q_{j}^{(i)}},\Delta_{\tilde q_{j}^{(i)}} )$ to $( \phi_i, q_{j}^{(i)}, \tilde q_{j}^{(i)} )$, and $(\Delta_{T_1} , \Delta_{T_2})$ to the monopoles $(T_1 , T_2)$, respectively.
We also define the bare monopole R-charge  
\be
 \Delta_{T_1} - \Delta_{T_2} = 2 \Delta_{m} \, .  
\ee
The superpotential \eqref{W:flavored:SYM} and \eqref{quantumcorrected} impose the constraints
\be
 \sum_{i = 1}^{3} \Delta_{\phi_i} = 2 \, , \qquad \Delta_{T_1} + \Delta_{T_2} = \sum_{i = 1}^{3} r_i \Delta_{\phi_i} \, , \qquad \Delta_{q_{j}^{(i)}} + \Delta_{\tilde q_{j}^{(i)}} + \Delta_{\phi_i} = 2 \, .
\ee
Similar constraints exist on the fluxes.
The toric cone is determined by the vectors
\bea
 \vec{v}_1 & = (0, 0, 0) \, , ~~~ && \vec{v}_2 = (0, 1, 0) \, , &&& \vec{v}_3 = (1, 0, 0) \, , \\
 \vec{v}_4 & = (0, 0, r_1) \, , && \vec{v}_5 = (0, 1, r_2) \, , &&& \vec{v}_6 = (1, 0, r_3) \, .
\eea
The toric diagram is shown below.
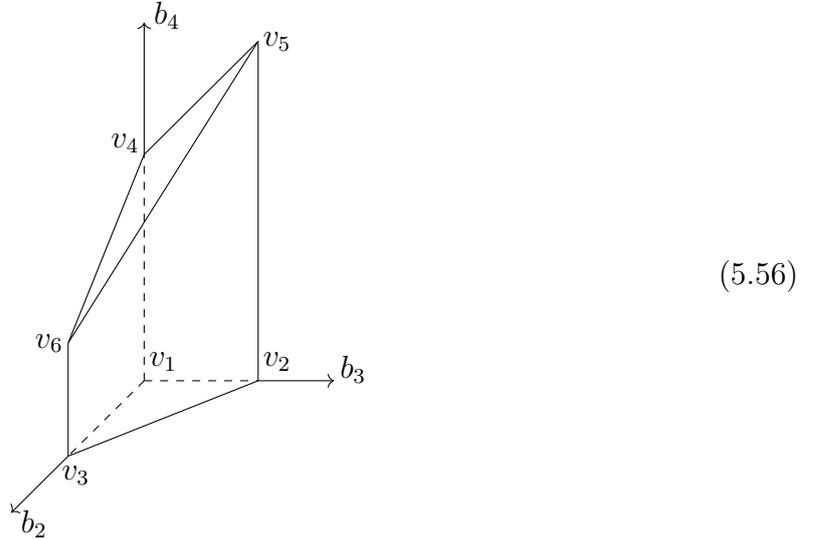
\bea
 \label{toric:N=8:SYM}
 \begin{tikzpicture}
  [scale=0.5 ]
  
  \draw[-,dashed] (0,0) -- (3,0) node[below] {};
  \draw[->,solid] (3,0) -- (5,0) node[below] {};
  
  \draw[-,dashed] (0,0) -- (0,6) node[below] {};
  \draw[->,solid] (0,6) -- (0,9.5) node[below] {};
  
  \draw[-,dashed] (0,0) -- (-2,-2) node[below] {};
  \draw[->,solid] (-2,-2) -- (-3.5,-3.5) node[below] {};
  
  \draw (-2.9,-3.8) node {$b_2$};
  \draw (5.5,0.3) node {$b_3$};
  \draw (0.6,9.7) node {$b_4$};
  
  \draw[-,solid] (3,0) -- (-2,-2) node[below] {};
  \draw[-,solid] (-2,-2) -- (-2,1) node[below] {};
  \draw[-,solid] (3,0) -- (3,9) node[below] {};
  \draw[-,solid] (3,9) -- (-2,1) node[below] {};
  \draw[-,solid] (3,9) -- (0,6) node[below] {};
  \draw[-,solid] (0,6) -- (-2,1) node[below] {};
  
  \draw (0.5,0.5) node {$v_1$};
  \draw (3.5,0.5) node {$v_2$};
  \draw (-1.8,-2.5) node {$v_3$};
  \draw (-0.5,6.3) node {$v_4$};
  \draw (3.5,9.) node {$v_5$};
  \draw (-2.5,1) node {$v_6$};
  
 \end{tikzpicture}
\eea
The baryonic symmetries are characterized by \eqref{baryonic}. They read
\bea
 B_1^{(1)} = \frac{r_3}{r_1} \, , ~~\quad & B_2^{(1)} = 0 \, , ~~&& B_3^{(1)} = -1 \, , &&& B_4^{(1)} = -\frac{r_3}{r_1} \, , &&&& B_5^{(1)} = 0 \, , &&&&& B_6^{(1)} = 1 \, , \\
 B_1^{(2)} = \frac{r_2}{r_1} \, , ~~\quad & B_2^{(2)} = -1 \, , ~~&& B_3^{(2)} = 0 \, , &&& B_4^{(2)} = -\frac{r_2}{r_1} \, , &&&& B_5^{(2)} = 1 \, , &&&&& B_6^{(2)} = 0 \, .
\eea
The dual polytope associated with the K\"ahler parameters $\lambda_a$ is given by
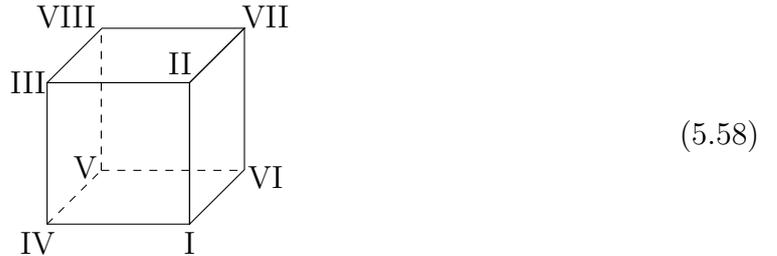
\bea
 \label{dual:polytope:N=8:SYM}
 \begin{tikzpicture}
 [scale=1.25 ]
  \pgfmathsetmacro{\cubex}{1.5}
  \pgfmathsetmacro{\cubey}{1.5}
  \pgfmathsetmacro{\cubez}{1.5}
  \draw[black] (0,0,0) -- ++(-\cubex,0,0) -- ++(0,-\cubey,0) -- ++(\cubex,0,0) -- cycle;
  \draw[black] (0,0,0) -- ++(0,0,-\cubez) -- ++(0,-\cubey,0) -- ++(0,0,\cubez) -- cycle;
  \draw[black] (0,0,0) -- ++(-\cubex,0,0) -- ++(0,0,-\cubez) -- ++(\cubex,0,0) -- cycle;
    
  \draw[-,dashed] (-1.5,-1.5) -- (-0.93,-0.93) node[below] {};
  \draw[-,dashed] (-0.93,-0.93) -- (0.55,-0.93) node[below] {};
  \draw[-,dashed] (-0.93,-0.93) -- (-0.93,0.55) node[below] {};
  
  \draw (0,-1.7) node {\rom{1}};
  \draw (-0.1,0.2) node {\rom{2}};
  \draw (-1.7,0.) node {\rom{3}};
  \draw (-1.6,-1.7) node {\rom{4}};
  \draw (-1.1,-.9) node {\rom{5}};
  \draw (0.8,-1) node {\rom{6}};
  \draw (0.8,0.7) node {\rom{7}};
  \draw (-1.3,0.7) node {\rom{8}};

 \end{tikzpicture}
\eea
whose vertices correspond to the facets of the toric diagram \eqref{toric:N=8:SYM} as follows
\bea
 \rom{1} & = (645) \, , ~~~~ && \rom{2} = (634) \, , &&& \rom{3} = (431) \, , &&&& \rom{4} = (415) \, , \\
 \rom{5} & = (251) \, , && \rom{6} = (265) \, , &&& \rom{7} = (623) \, , &&&& \rom{8} = (132) \, .
\eea
It is now easy to compute the master volume, whose explicit expression is given in \eqref{N=8:SYM:master:volume}.
We work in the gauge \eqref{gauge:lambda:main}. Then, the remaining $\lambda_a$ and $A$ can be found, using the ansatz \eqref{lambda:solution:toric}, and solving \eqref{constraints} explicitly.
In particular,
\be
 \begin{aligned}
  & \lambda_{i+3} = - \frac{\lambda r_{i}}{2 b_4} \, , \quad \text{ for } i = 1,2,3 \, , \\
  & \lambda = \frac{\sqrt{N}}{\sqrt{2} \pi^2} \sqrt{\frac{b_2 b_3 b_4( b_1 - b_2 - b_3) \left( b_1 r_1 - b_3 ( r_1 - r_2 ) - b_2 ( r_1 - r_3 ) - b_4 \right)}{b_1 \left( b_1 r_1 - b_3 ( r_1 - r_2 ) - b_2 ( r_1 - r_3 ) \right)}} \, .
 \end{aligned}
\ee
Notice that the above solution satisfies \eqref{Decoupling:lambda}.
Plugging these expressions into \eqref{N=8:SYM:master:volume}, we find that
\be
 \cV_{\text{on-shell}} = \frac{N^{3/2}}{6 \sqrt{2} \pi^2 b_1} \sqrt{\frac{b_2 b_3 ( b_1 - b_2 - b_3 ) b_4 \left( b_1 r_1 - b_3 (r_1-r_2) - b_2 (r_1-r_3) - b_4 \right)}{b_1 \left( b_1 r_1 - b_3 (r_1-r_2) - b_2 (r_1-r_3) \right)}} \, .
\ee
For the entropy functional \eqref{entropy:functional} we obtain
\bea
 \label{entropy:functional:N=8:SYM}
 S (b_i & , \fn_a) = - \frac{2 \sqrt{2} \pi N^{3/2}}{3} \sqrt{\frac{b_2 b_3 b_4( b_1 - b_2 - b_3) \left( b_1 r_1 - b_3 ( r_1 - r_2 ) - b_2 ( r_1 - r_3 ) - b_4 \right)}{b_1 \left( b_1 r_1 - b_3 ( r_1 - r_2 ) - b_2 ( r_1 - r_3 ) \right)}} \\
 & \times \bigg(
 \frac{( (b_1 - b_2 - b_3) r_1 - 2 b_4) \fn_1}{( b_1 - b_2 - b_3) ( b_1 r_1 - b_3 (r_1-r_2) - b_2 (r_1-r_3) - b_4)} \\
 & + \frac{(b_1 r_1-b_3 (r_1-2 r_2) - b_2 (r_1-r_3)) \fn_2}{b_3 ( b_1 r_1 - b_3 (r_1-r_2) - b_2 (r_1-r_3) - b_4)}
 + \frac{(b_1 r_1 - b_3 (r_1-r_2) - b_2 (r_1-2 r_3)) \fn_3}{b_2 ( b_1 r_1 - b_3 (r_1-r_2) - b_2 (r_1-r_3) - b_4)} \\
 & + \frac{(b_1 r_1 - b_3 (r_1-r_2) - b_2 (r_1-r_3) - 2 b_4) \fn_4}{( b_1 - b_2 - b_3) b_4}
 \bigg) \, .
\eea
We assign the R-charge $\Delta_a$ and the flux $\fn_a$, $a=1,\ldots,6$, to each vertex of the toric diagram \eqref{toric:N=8:SYM}.
Recall that supersymmetry requires
\be
 \sum_{a=1}^6 \Delta_a = 2 \, , \qquad \sum_{a=1}^6 \fn_a = 2 - 2 \fg \, .
\ee
The R-charges associated to the vertices of the toric diagram are mapped to those of the chiral fields and the monopoles by \cite[(7.10)]{Jafferis:2011zi}
\be
 \begin{aligned}
 \Delta_{\phi_1} = \Delta_1 + \Delta_4 \, , \qquad & \Delta_{\phi_2} = \Delta_2 + \Delta_5 \, , \qquad && \Delta_{\phi_3} = \Delta_3 + \Delta_6 \, , \\
 \Delta_{T_1} = \sum_{i=1}^3 r_i \Delta_i \, , \qquad & \Delta_{T_2} = \sum_{i=1}^3 r_i \Delta_{i+3} \, .
 \end{aligned}
\ee
Using the parameterization \eqref{normRcharges} for the R-charges, we also find
\bea
 \label{Delta:b:N=8:SYM}
 \Delta_1 & = \frac{2}{b_1} \left( b_1-b_2-b_3 - \frac{(b_1 - b_2 - b_3) b_4}{b_1 r_1 - b_3 (r_1-r_2) - b_2 (r_1-r_3)}\right) \, ,\\
 \Delta_2 & = \frac{2 b_3}{b_1} \left(1-\frac{b_4}{b_1 r_1 - b_3 (r_1-r_2) - b_2 (r_1-r_3)}\right) \, , \\
 \Delta_3 & = \frac{2 b_2}{b_1} \left(1-\frac{b_4}{b_1 r_1 - b_3 (r_1-r_2) - b_2 (r_1-r_3)}\right) \, , \\
 \Delta_4 & = \frac{2 (b_1-b_2-b_3) b_4}{b_1 \left( b_1 r_1 - b_3 (r_1-r_2) - b_2 (r_1-r_3) \right)} \, , \\
 \Delta_5 & = \frac{2 b_3 b_4}{b_1 \left( b_1 r_1 - b_3 (r_1-r_2) - b_2 (r_1-r_3) \right)} \, , \\
 \Delta_6 & = \frac{2 b_2 b_4}{b_1 \left( b_1 r_1 - b_3 (r_1-r_2) - b_2 (r_1-r_3) \right)} \, .
\eea
The R-charges satisfy
\be
 \Delta_1 \Delta_5 - \Delta_2 \Delta_4 = 0 \, , \qquad \Delta_1 \Delta_6 - \Delta_3 \Delta_4 = 0 \, .
\ee
These constraints are equivalent to
\be
 \sum_{a=1}^{6} B_a^{(i)} \frac{\partial a_{3\rd} (\Delta_a)}{\partial \Delta_a} = 0 \, , \quad \text{ for } i = 1, 2 \, ,
\ee
where $a_{3\rd} (\Delta_a)$ is given by
\bea
 a_{3\rd} (\Delta_a) & \equiv \frac{1}{24} \sum_{a , b , c , e = 1}^{6} \left| \det ( v_a , v_b , v_c , v_e ) \right| \Delta_a \Delta_b \Delta_c \Delta_e \\
 & = r_1 \Delta_1 \Delta_4 (\Delta_2 + \Delta_5) (\Delta_3 + \Delta_6) \\
 & + r_2 \Delta_2 \Delta_5 (\Delta_1 + \Delta_4) (\Delta_3 + \Delta_6) \\
 & + r_3 \Delta_3 \Delta_6 (\Delta_1 + \Delta_4) (\Delta_2 + \Delta_5) \, .
\eea
Finally, the second equation in \eqref{constraints} imposes the following constraints on the fluxes
\bea
 \label{Q111:flux:sim:N=8:SYM}
  0 & = \sum_{a=1}^{6} \fn_a \frac{\partial}{\partial \Delta_a} ( \Delta_1 \Delta_5 - \Delta_2 \Delta_4 ) = \Delta_1 \fn_5 - \Delta_2 \fn_4 - \Delta_4 \fn_2 + \Delta_5 \fn_1 \, , \\
  0 & = \sum_{a=1}^{6} \fn_a \frac{\partial}{\partial \Delta_a} ( \Delta_1 \Delta_6 - \Delta_3 \Delta_4 ) = \Delta_1 \fn_6 - \Delta_3 \fn_4 - \Delta_4 \fn_3 + \Delta_6 \fn_1 \, .
\eea
These constraints can also be obtained from \eqref{flux constraints}.
The $S^3$ free energy of this theory was computed in \cite[(4.10)]{Jafferis:2011zi} and it can be written as
\be
 F_{S^3} (\Delta_a) = \frac{4 \pi \sqrt{2} N^{3/2}}{3} \sqrt{\frac{\left( \sum_{i=1}^{3} r_i \Delta_i \right) \left( \sum_{i=1}^{3} r_i \Delta_{i+3} \right) \prod_{i=1}^{3}(\Delta_i+\Delta_{i+3})}
 {\sum_{i=1}^{3} r_i (\Delta_i+\Delta_{i+3})}} \, .
\ee
The topologically twisted index was also computed in \cite[(5.56)]{Hosseini:2016ume} and it is given by
\be
 \label{index:N=8:SYM}
 \cI(\Delta_a,\fn_a) = - \frac{1}{2} \sum_{a = 1}^{6} \fn_a \frac{\partial F_{S^3}(\Delta_a)}{\partial \Delta_a} \, .
\ee
It is easy to see that, using \eqref{Delta:b:N=8:SYM} and \eqref{Q111:flux:sim:N=8:SYM}, and setting $b_1=1$,
\be
 \begin{aligned}
  & S (b_i , \fn_a) = \cI (\Delta_a , \fn_a)\bigg|_{\Delta_a(b_i)}
  = - \frac{2 \sqrt{2} \pi N^{3/2}}{3} \sum_{a=1}^{6} \fn_a \frac{\partial \sqrt{a_{3\rd}(\Delta_a)}}{\partial \Delta_a} \bigg|_{\Delta_a(b_i)} \, , \\
  & \cV_{\text{on-shell}} (b_i) = \frac{1}{64 \pi^3} F_{S^3} (\Delta_a) \bigg|_{\Delta_a(b_i)} = \frac{N^{3/2}}{24 \sqrt{2} \pi^2} \sqrt{a_{3\text{d}} (\Delta_a)} \bigg|_{\Delta_a(b_i)} \, .
 \end{aligned}
\ee

\subsection[The SPP theory]{The SPP theory}

We now consider the quiver gauge theory \cite{Hanany:2008fj} 
\be
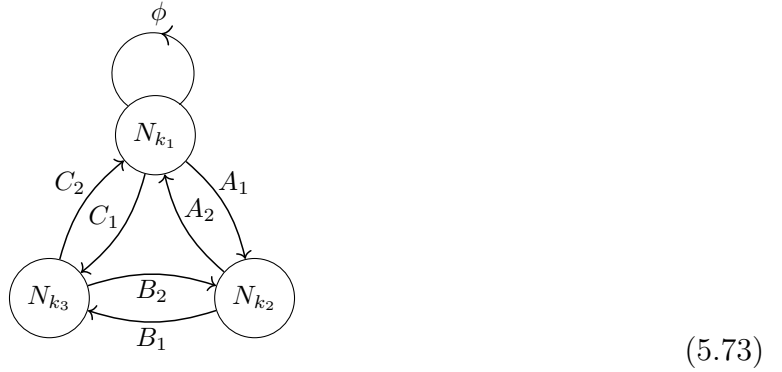
\label{SPP:quiver}
\begin{tikzpicture}[font=\footnotesize, scale=0.9]
\begin{scope}[auto,
  every node/.style={draw, minimum size=0.5cm}, node distance=2cm];
\node[circle]  (UN)  at (0.35,1.7) {$N_{k_1}$};
\node[circle]  (UN2)  at (-1.2,-.7) {$N_{k_3}$};
\node[circle]  (UN3)  at (1.8,-.7) {$N_{k_2}$};
\end{scope}\draw[decoration={markings, mark=at position 0.45 with {\arrow[scale=1.5]{>}}}, postaction={decorate}, shorten >=0.7pt] (0.7,2.15) arc (-50:235:0.6cm);
\draw[draw=black,solid,line width=0.2mm,<-]  (UN) to[bend right=18] (UN2) ;
\draw[draw=black,solid,line width=0.2mm,->]  (UN) to[bend left=18]  (UN2) ;
\draw[draw=black,solid,line width=0.2mm,<-]  (UN) to[bend right=18] (UN3) ; 
\draw[draw=black,solid,line width=0.2mm,->]  (UN) to[bend left=18]  (UN3) ;  
\draw[draw=black,solid,line width=0.2mm,<-]  (UN2) to[bend right=18](UN3) ;
\draw[draw=black,solid,line width=0.2mm,->]  (UN2) to[bend left=18] (UN3) ;
\node at (1.5,1.0) {$A_1$};
\node at (1.0,0.6) {$A_2$};
\node at (0.3,-1.3) {$B_1$};
\node at (0.3,-0.6) {$B_2$};
\node at (-0.4,0.5) {$C_1$};
\node at (-0.9,1.0) {$C_2$};
\node at (0.4,3.5) {$\phi$};
\end{tikzpicture}
\ee
describing the dynamics of $N$ M2-branes at a fibration over the suspended pinch point (SPP) singularity \cite{Morrison:1998cs}.
In the following, we choose  Chern-Simons levels  $(k_1 , k_2, k_3) = (2, -1 , -1)$. The superpotential reads
\begin{equation}
 \label{W:SPP}
 W = \Tr \left[\phi \left( A_1 A_2 - C_1 C_2 \right) - A_2 A_1 B_1 B_2 + C_2 C_1 B_2 B_1 \right] \, .
\end{equation}
Therefore, the R-charges of the chiral fields must satisfy
\be
 \Delta_{\phi} + \sum_{i=1}^{2} \Delta_{A_i} = 2 \, , \qquad \Delta_{\phi} + \sum_{i=1}^{2} \Delta_{C_i} = 2 \, , \qquad \sum_{i=1}^{2} ( \Delta_{A_i} + \Delta_{B_i} ) = 2 \, .
\ee
We also introduce the magnetic fluxes $(\fn_{\phi} , \fn_{A_i} , \fn_{B_i} , \fn_{C_i})$ for the chiral fields $(\phi , A_i, B_i, C_i)$, respectively. They satisfy
\be
 \fn_{\phi} + \sum_{i=1}^{2} \fn_{A_i} = 2 - 2 \fg \, , \qquad \fn_{\phi} + \sum_{i=1}^{2} \fn_{C_i} = 2 - 2 \fg \, , \qquad \sum_{i=1}^{2} ( \fn_{A_i} + \fn_{B_i} ) = 2 - 2 \fg \, .
\ee
The toric cone is determined by the vectors
\bea
 \vec{v}_1 & = ( 0, 0, 0 ) \, , ~~~ && \vec{v}_2 = ( 1, -1, 0 ) \, , &&& \vec{v}_3 = ( 2, 0, 0 ) \, , \\
 \vec{v}_4 & = ( 1, 1, 0) \, , && \vec{v}_5 = ( 0, 0, 1) \, , &&& \vec{v}_6 = ( 1, 0, 1 ) \, .
\eea
The toric diagram is depicted below.
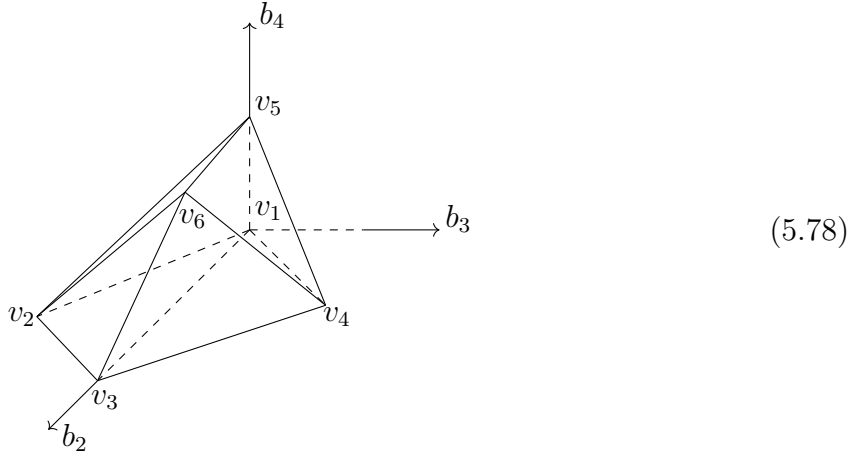
\bea
 \label{toric:SPP}
 \begin{tikzpicture}
  [scale=0.5 ]
  
  \draw[-,dashed] (0,0) -- (3,0) node[below] {};
  \draw[->,solid] (3,0) -- (5,0) node[below] {};
  
  \draw[-,dashed] (0,0) -- (0,3) node[below] {};
  \draw[->,solid] (0,3) -- (0,5.5) node[below] {};
  
  \draw[-,dashed] (0,0) -- (-4,-4) node[below] {};
  \draw[->,solid] (-4,-4) -- (-5.3,-5.3) node[below] {};
  
  \draw (-4.6,-5.5) node {$b_2$};
  \draw (5.5,0.3) node {$b_3$};
  \draw (0.6,5.7) node {$b_4$};
  
  \draw[-,solid] (-5.6,-2.3) -- (-1.7,1) node[below] {};
  \draw[-,solid] (-4,-4) -- (-1.7,1) node[below] {};
  \draw[-,solid] (2,-2) -- (-1.7,1) node[below] {};
  \draw[-,solid] (0,3) -- (-1.7,1) node[below] {};
  \draw[-,solid] (0,3) -- (2,-2) node[below] {};
  \draw[-,solid] (-4,-4) -- (2,-2) node[below] {};
  \draw[-,solid] (-4,-4) -- (-5.6,-2.3) node[below] {};
  \draw[-,dashed] (0,0) -- (-5.6,-2.3) node[below] {};
  \draw[-,dashed] (0,0) -- (2,-2) node[below] {};
  \draw[-,solid] (0,3) -- (-5.6,-2.3) node[below] {};
  
  \draw (0.5,0.5) node {$v_1$};
  \draw (-6,-2.3) node {$v_2$};
  \draw (-3.8,-4.5) node {$v_3$};
  \draw (2.3,-2.3) node {$v_4$};
  \draw (0.5,3.3) node {$v_5$};
  \draw (-1.5,0.3) node {$v_6$};
  
 \end{tikzpicture}
\eea
The baryonic symmetries are characterized by \eqref{baryonic} and they are given by
\bea
 B_1^{(1)} = 1 \, , ~~\quad & B_2^{(1)} = 0 \, , ~~&& B_3^{(1)} = -1 \, , &&& B_4^{(1)} = 0 \, , &&&& B_5^{(1)} = - 2 \, , &&&&& B_6^{(1)} = 2 \, , \\
 B_1^{(2)} = - 1 \, , ~~\quad & B_2^{(2)} = 1 \, , ~~&& B_3^{(2)} = - 1 \, , &&& B_4^{(2)} = 1 \, , &&&& B_5^{(2)} = 0 \, , &&&&& B_6^{(2)} = 0 \, .
\eea
The dual polytope associated with the K\"ahler parameters $\lambda_a$, $a=1,\ldots,6$ is given by
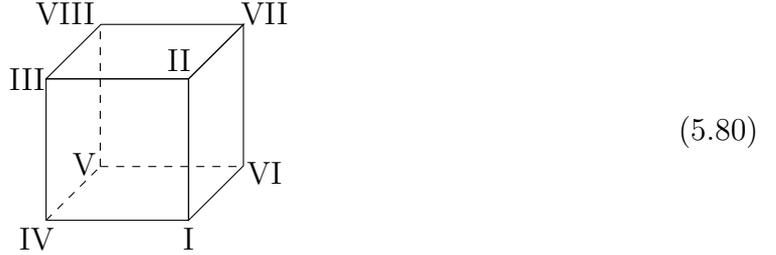
\bea
 \label{dual:polytope:SPP}
 \begin{tikzpicture}
 [scale=1.25 ]
  \pgfmathsetmacro{\cubex}{1.5}
  \pgfmathsetmacro{\cubey}{1.5}
  \pgfmathsetmacro{\cubez}{1.5}
  \draw[black] (0,0,0) -- ++(-\cubex,0,0) -- ++(0,-\cubey,0) -- ++(\cubex,0,0) -- cycle;
  \draw[black] (0,0,0) -- ++(0,0,-\cubez) -- ++(0,-\cubey,0) -- ++(0,0,\cubez) -- cycle;
  \draw[black] (0,0,0) -- ++(-\cubex,0,0) -- ++(0,0,-\cubez) -- ++(\cubex,0,0) -- cycle;
    
  \draw[-,dashed] (-1.5,-1.5) -- (-0.93,-0.93) node[below] {};
  \draw[-,dashed] (-0.93,-0.93) -- (0.55,-0.93) node[below] {};
  \draw[-,dashed] (-0.93,-0.93) -- (-0.93,0.55) node[below] {};
  
  \draw (0,-1.7) node {\rom{1}};
  \draw (-0.1,0.2) node {\rom{2}};
  \draw (-1.7,0.) node {\rom{3}};
  \draw (-1.6,-1.7) node {\rom{4}};
  \draw (-1.1,-.9) node {\rom{5}};
  \draw (0.8,-1) node {\rom{6}};
  \draw (0.8,0.7) node {\rom{7}};
  \draw (-1.3,0.7) node {\rom{8}};

 \end{tikzpicture}
\eea
whose vertices correspond to the facets of the toric diagram \eqref{toric:SPP} as follows
\bea
 \rom{1} & = (645) \, , ~~~~ && \rom{2} = (634) \, , &&& \rom{3} = (431) \, , &&&& \rom{4} = (415) \, , \\
 \rom{5} & = (251) \, , && \rom{6} = (265) \, , &&& \rom{7} = (623) \, , &&&& \rom{8} = (132) \, .
\eea
The master volume is then easily computed and its explicit form can be found in \eqref{SPP:master:volume}.
As before we work in the gauge \eqref{gauge:lambda:main}. Using \eqref{constraints} and \eqref{lambda:solution:toric} we may fix $\lambda_a$ and $A$.
We will not report the long resulting expressions here.
We assign the R-charge $\Delta_a$ and the flux $\fn_a$, to each vertex of the toric diagram \eqref{toric:SPP}.
Supersymmetry then requires
\be
 \sum_{a=1}^6 \Delta_a = 2 \, , \qquad \sum_{a=1}^6 \fn_a = 2 - 2 \fg \, .
\ee
The R-charges associated to the vertices of the toric diagram are mapped to those of the chiral fields by \cite{Hanany:2008fj}
\bea
 \Delta_{A_1} = \Delta_1 + \Delta_2 \, , \qquad & \Delta_{A_2} = \Delta_3 + \Delta_4 \, , \qquad && \Delta_{C_1} = \Delta_1 + \Delta_4 \, , \\
 \Delta_{C_2} = \Delta_3 + \Delta_2 \, ,\qquad & \Delta_{B_1} = \Delta_5 \, , && \Delta_{B_2} = \Delta_6 \, .
\eea
Employing the parameterization \eqref{normRcharges} for the R-charges, we obtain
\bea
 \label{Delta:b:SPP}
 \Delta_1 & = \frac{2 \left(b_1-b_3-b_4\right) \left(2 b_1-b_2-b_3-b_4\right) \left(b_1+b_3-b_4\right) \left(2 b_1-b_2+b_3-b_4\right)}
 {b_1 \left(4 b_1^3-8 b_4 b_1^2+\left(b_4 \left(2 b_2+5 b_4\right)-4 b_3^2\right) b_1-b_4 \left(b_2^2+b_4 b_2-3 b_3^2+b_4^2\right)\right)} \, , \\
 \Delta_2 & = \frac{\left(b_2-b_3\right) \left(2 \left(b_1+b_3\right)-b_4\right) \left( b_1 - b_3 - b_4 \right) \left(2 b_1 - b_2 - b_3 - b_4\right)}
 {b_1 \left(4 b_1^3 - 4 b_3^2 b_1 - b_4^3 - \left(b_2-5 b_1\right) b_4^2 - \left(8 b_1^2-2 b_2 b_1+b_2^2-3 b_3^2\right) b_4\right)} \, , \\
 \Delta_3 & = \frac{2 \left(b_2^2-b_3^2\right) \left(\left(b_1-b_4\right)^2-b_3^2\right)}{b_1 \left(4 b_1^3-8 b_4 b_1^2+\left(b_4 \left(2 b_2+5 b_4\right)-4 b_3^2\right) b_1-b_4 \left(b_2^2+b_4 b_2-3 b_3^2+b_4^2\right)\right)} \, , \\
 \Delta_4 & = \frac{\left(b_2+b_3\right) \left(b_1+b_3-b_4\right) \left( 2 b_1 - b_2 + b_3 - b_4\right) \left( 2 b_1 - 2 b_3 - b_4 \right)}
 {b_1 \left( 4 b_1^3 - 4 b_3^2 b_1 - b_4^3 - \left(b_2-5 b_1\right) b_4^2 - \left(8 b_1^2-2 b_2 b_1+b_2^2-3 b_3^2\right) b_4\right)} \, , \\
 \Delta_5 & = \frac{2 \left(b_1+b_2-b_4\right) b_4 \left( 2 b_1 - b_2 + b_3 - b_4 \right) \left( 2 b_1 - b_2 - b_3 - b_4 \right)}
 {b_1 \left( 4 b_1^3 - 4 b_3^2 b_1 - b_4^3 - \left(b_2-5 b_1\right) b_4^2 - \left(8 b_1^2-2 b_2 b_1+b_2^2-3 b_3^2\right) b_4\right)} \, , \\
 \Delta_6 & = \frac{2 \left(b_2^2-b_3^2\right) b_4 \left( 3 b_1 - b_2 - 2 b_4 \right)}{b_1 \left(4 b_1^3-8 b_4 b_1^2+\left(b_4 \left(2 b_2+5 b_4\right)-4 b_3^2\right) b_1-b_4 \left(b_2^2+b_4 b_2-3 b_3^2+b_4^2\right)\right)} \, ,
\eea
that are independent of the magnetic fluxes (as we are dealing with the mesonic twist).
To compare with the results in \cite{Martelli:2011qj,Hosseini:2016tor}, using the symmetries of the quiver \eqref{SPP:quiver}, we restrict
\bea
 \label{SPP:limit}
 & \Delta_1 = \Delta_3 \equiv \frac{2 ( 1 - \Delta )^2}{4 - 3 \Delta} \, , \quad && \Delta_2 = \Delta_4 \equiv \frac{( 2 - \Delta ) ( 1- \Delta )}{4 - 3 \Delta}\, , &&& \Delta_5 = \Delta_6 \equiv \Delta \, , \\
 & \fn_1 = \fn_3 \equiv ( 1 - \fg ) ( 1 - \fn ) - \fn_4 \, , && \qquad \fn_2 = \fn_4  \, , &&& \fn_5 = \fn_6 \equiv ( 1 - \fg ) \fn \, .
\eea
This is the consequence of choosing the Reeb vector $b = ( 1, 1 - \Delta / 2, 0, \Delta )$.
The second equation in \eqref{constraints} also imposes the constraint
\be
 \label{flux:constraint:SPP:simp}
 ( 1 - \fg ) ( \Delta  ( 12 - 5 \Delta ) + \left( \Delta  ( 3 \Delta - 8 ) + 6 ) \fn - 8 \right) + ( 4 - 3 \Delta )^2 \fn_4 = 0 \, .
\ee
The $S^3$ free energy of this theory was computed in \cite[(5.20)]{Martelli:2011qj} and it is given by
\be
 F_{S^3} (\Delta) = \frac{8 \pi N^{3/2}}{3} ( 2- \Delta) (1 - \Delta) \sqrt{\frac{\Delta}{4 - 3 \Delta}} \, .
\ee
The topologically twisted index was also computed in \cite[(B.19)]{Hosseini:2016tor} and it reads
\be
 \cI (\Delta , \fn) = - \frac{4 \pi ( 1 - \fg ) N^{3/2}}{3 \sqrt{\Delta} (4  - 3 \Delta )^{3/2}} \left( 7 \Delta^3 - 18 \Delta ^2 + 12 \Delta - \left( 6 \Delta^3 - 19 \Delta^2 + 18 \Delta - 4 \right) \fn \right) \, .
\ee
It is easy to evaluate the entropy functional \eqref{entropy:functional}; using \eqref{Delta:b:SPP}, \eqref{SPP:limit}, \eqref{flux:constraint:SPP:simp}, and setting $b_1=1$, we find that
\be
 S (b_i , \fn_a) = \cI (\Delta , \fn)\Big|_{\Delta (b_i)} \, .
\ee
Notice also that
\be
 \cV_{\text{on-shell}} (b_i) = \frac{1}{64 \pi^3} F_{S^3}(\Delta) \bigg|_{\Delta(b_i)} = \frac{N^{3/2}}{24 \sqrt{2} \pi^2} \sqrt{a_{3\text{d}} (\Delta)} \bigg|_{\Delta(b_i)} \, ,
\ee
where \cite{Amariti:2011uw}
\be
 a_{3\rd} (\Delta_a) \equiv \frac{1}{24} \sum_{a , b , c , e = 1}^{6} \left| \det ( v_a , v_b , v_c , v_e ) \right| \Delta_a \Delta_b \Delta_c \Delta_e - \frac{1}{2} ( \Delta_3 \Delta_5 - \Delta_1 \Delta_6 )^2 \, .
\ee
Once again the constraints on the R-charges and the fluxes can be written as in \eqref{Rcharge constraints} and \eqref{flux constraints}, respectively.

\subsection[The cone over \texorpdfstring{$M^{1,1,1}$}{M[111]}]{The cone over $M^{1,1,1}$}

Consider the cone over the $Y_7 = M^{1,1,1}$, \ie\;$C(M^{1,1,1})$.
The gauge theory dual to AdS$_4 \times M^{1,1,1}$ has a \emph{chiral} quiver (in a four-dimensional sense) \cite{Martelli:2008si},
and thus the large $N$ limit of its partition functions on $S^3$ and $\Sigma_\fg \times S^1$ are not known \cite{Jafferis:2011zi,Hosseini:2016tor}.
However, it is interesting to evaluate the entropy functional \eqref{entropy:functional} for $M^{1,1,1}$ and provide a prediction for the large $N$ topologically twisted index of the dual gauge theory.

The $C(M^{1,1,1})$ determines a polytope with six vertices
\bea
 \vec{v}_1 & = ( -1, 0, 0 ) \, , ~~~ && \vec{v}_2 = ( 0, -1, 0 ) \, , &&& \vec{v}_3 = ( 1, 1, 0 ) \, , \\
 \vec{v}_4 & = ( 0, 0, -3) \, , && \vec{v}_5 = ( 0, 0, 3) \, , &&& \vec{v}_6 = ( 0, 0, 0 ) \, ,
\eea
where $\vec{v}_6$ is an internal point.  The toric diagram is shown below.
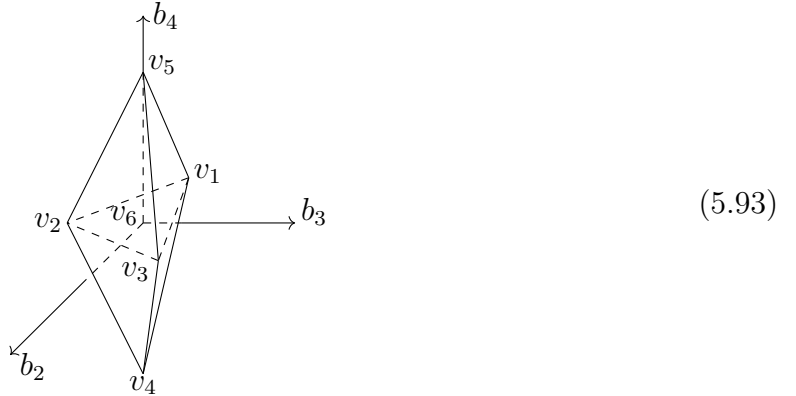
\bea
 \label{toric:M111}
 \begin{tikzpicture}
  [scale=0.5 ]
  
  \draw[-,dashed] (0,0) -- (1,0) node[below] {};
  \draw[->,solid] (1,0) -- (4,0) node[below] {};
  
  \draw[-,dashed] (0,0) -- (0,4) node[below] {};
  \draw[->,solid] (0,4) -- (0,5.5) node[below] {};
  
  \draw[-,dashed] (0,0) -- (-1.5,-1.5) node[below] {};
  \draw[->,solid] (-1.5,-1.5) -- (-3.5,-3.5) node[below] {};
  
  \draw (-2.9,-3.8) node {$b_2$};
  \draw (4.5,0.3) node {$b_3$};
  \draw (0.6,5.5) node {$b_4$};
  
  \draw[-,dashed] (1.2,1.2) -- (-2,0) node[below] {};
  \draw[-,dashed] (1.2,1.2) -- (0.4,-1) node[below] {};
  \draw[-,solid] (1.2,1.2) -- (0,4) node[below] {};
  \draw[-,dashed] (0.4,-1) -- (-2,0) node[below] {};
  \draw[-,solid] (1.2,1.2) -- (0,-4) node[below] {};
  \draw[-,solid] (0.4,-1) -- (0,-4) node[below] {};
  \draw[-,solid] (0.4,-1) -- (0,4) node[below] {};
  \draw[-,solid] (-2,0) -- (0,4) node[below] {};
  \draw[-,solid] (-2,0) -- (0,-4) node[below] {};
  
  \draw (1.7,1.3) node {$v_1$};
  \draw (-2.5,0) node {$v_2$};
  \draw (-0.2,-1.3) node {$v_3$};
  \draw (0,-4.3) node {$v_4$};
  \draw (0.5,4.2) node {$v_5$};
  \draw (-0.5,0.2) node {$v_6$};
  
 \end{tikzpicture}
\eea
The baryonic symmetry is characterized by \eqref{baryonic} and it is given by
\be
 B_1 = -2 \, , \qquad B_2 = -2 \, , \qquad B_3 = -2 \, , \qquad B_4 = 3 \, , \qquad B_5 = 3 \, .
\ee
The dual polytope associated with the K\"ahler parameters $\lambda_a$ is given by
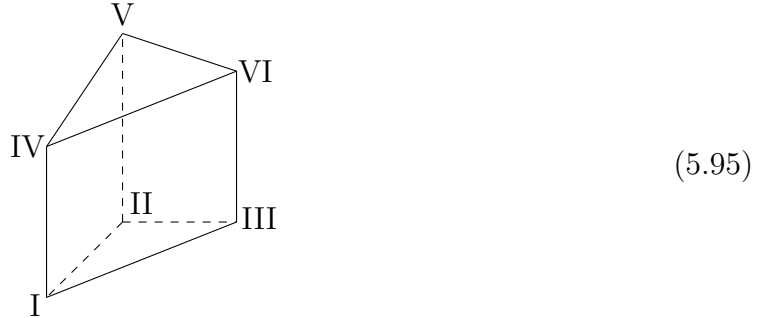
\bea
 \label{dual:polytope:M111}
 \begin{tikzpicture}
  [scale=0.5 ]
  
  \draw[-,dashed] (0,0) -- (3,0) node[below] {};
  \draw[-,dashed] (0,0) -- (0,5) node[below] {};
  \draw[-,dashed] (0,0) -- (-2,-2) node[below] {};
  
  \draw[-,solid] (3,0) -- (-2,-2) node[below] {};
  \draw[-,solid] (-2,-2) -- (-2,2) node[below] {};
  \draw[-,solid] (3,0) -- (3,4) node[below] {};
  \draw[-,solid] (3,4) -- (-2,2) node[below] {};
  \draw[-,solid] (3,4) -- (0,5) node[below] {};
  \draw[-,solid] (0,5) -- (-2,2) node[below] {};
  
  \draw (0.5,0.5) node {\rom{2}};
  \draw (3.6,0.1) node {\rom{3}};
  \draw (-2.3,-2.2) node {\rom{1}};
  \draw (0.,5.5) node {\rom{5}};
  \draw (3.5,4) node {\rom{6}};
  \draw (-2.5,2) node {\rom{4}};
  
 \end{tikzpicture}
\eea
whose vertices correspond to the facets of the toric diagram \eqref{toric:M111} as follows
\bea
 \rom{1} & = (124) \, , ~~~~ && \rom{2} = (234) \, , &&& \rom{3} = (134) \, , \\
 \rom{4} & = (135) \, , && \rom{5} = (235) \, , &&& \rom{6} = (125) \, .
\eea
Having determined the dual polytope \eqref{dual:polytope:M111} it is now straightforward to compute the master volume, whose explicit expression can be found in \eqref{M111:master:volume}.
As before we work in the gauge \eqref{gauge:lambda:main}. Using the first equation in \eqref{constraints} and \eqref{lambda:solution:toric} we can fix the remaining $\lambda_a$ as
\be
 \lambda_4 = - \lambda_5 = \frac{3 \lambda}{2 b_4} \, ,
\ee
where
\be
 \lambda = \frac{\sqrt{N}}{18 \sqrt{3} \pi^2}
 \sqrt{\frac{( 9 (b_1+b_2-2 b_3)^2 - b_4^2 ) (9 (b_1-2 b_2+b_3)^2 - b_4^2 ) ( 9 (b_1+b_2+b_3)^2 - b_4^2 )}{b_1 \left( 27 \left( b_1^2 - b_2^2 + b_3 b_2 - b_3^2 \right) + b_4^2 \right)}} \, .
\ee
The last equation of \eqref{constraints} can also be solved for $A$.
Substituting these expressions into \eqref{M111:master:volume} we obtain
\be
 \label{V:on-shell:M111}
 \cV_{\text{on-shell}} = \frac{N^{3/2}}{108 \sqrt{3} \pi^2 b_1}
 \sqrt{\frac{( 9 (b_1+b_2-2 b_3)^2 - b_4^2 ) (9 (b_1-2 b_2+b_3)^2 - b_4^2 ) ( 9 (b_1+b_2+b_3)^2 - b_4^2 )}{b_1 \left( 27 \left( b_1^2 - b_2^2 + b_3 b_2 - b_3^2 \right) + b_4^2 \right)}}\, .
\ee
For the entropy functional \eqref{entropy:functional} we find that
\bea
 \label{entropy:M111}
 S (b_i , \fn_a) & = - 2 \pi \sqrt{3} N^{3/2} \sqrt{\frac{b_1 \left(9 (b_1+b_2-2 b_3)^2-b_4^2\right) \left(9 (b_1-2 b_2+b_3)^2-b_4^2\right) \left(9 (b_1+b_2+b_3)^2-b_4^2\right)}{27 \left(b_1^2-b_2^2-b_3^2+b_2 b_3\right) + b_4^2}} \\
 & \times \bigg( \frac{(b_2-b_3) \fn_2}{b_1 \left(9 (b_1+b_2-2 b_3)^2-b_4^2\right)}
 + \frac{b_2 \fn_3}{b_1 \left(9 (b_1+b_2+b_3)^2-b_4^2\right)} \\
 & + \frac{2 \left(2 b_4^2-3 (4 b_1-2 b_2+b_3) b_4+27 \left(2 b_1^2-2 b_2 b_1-b_2^2-b_3^2+(b_1+b_2) b_3\right)\right) \fn_4}{27 b_1 (3 (b_1+b_2-2 b_3)-b_4) (3 (b_1-2 b_2+b_3)-b_4) (3 (b_1+b_2+b_3)-b_4)} \\
 & + \frac{2  \left(2 b_4^2+3 (4 b_1-2 b_2+b_3) b_4+27 \left(2 b_1^2-2 b_2 b_1-b_2^2-b_3^2+(b_1+b_2) b_3\right)\right) \fn_5}{27 b_1 (3 (b_1+b_2-2 b_3)+b_4) (3 (b_1-2 b_2+b_3)+b_4) (3 (b_1+b_2+b_3)+b_4)} \bigg) \, .
\eea
As usual, we associate the R-charge $\Delta_a$ and the flux $\fn_a$ to the vertex $v_a$, $a=1,\ldots,5$, of the toric diagram \eqref{toric:M111} \cite{Hanany:2008fj}. They satisfy
\be
 \sum_{a=1}^5 \Delta_a = 2 \, , \qquad \sum_{a=1}^5 \fn_a = 2 - 2 \fg \, .
\ee
Note that we did \emph{not} include the internal point $v_6$.
Using the parameterization \eqref{normRcharges} for the R-charges, we also obtain
\bea
 \label{Delta:b:M111}
 \Delta_1 & = \frac{2 (2 (b_1+b_2)-b_3) (3 (b_1-2 b_2+b_3)-b_4) (3 (b_1-2 b_2+b_3)+b_4)}{3 b_1 \left(27 b_1^2+b_4^2-27 \left(b_2^2-b_3 b_2+b_3^2\right)\right)} \, , \\
 \Delta_2 & = \frac{2 (2 b_1-b_2+2 b_3) (3 (b_1+b_2-2 b_3)-b_4) (3 (b_1+b_2-2 b_3)+b_4)}{3 b_1 \left(27 b_1^2+b_4^2-27 \left(b_2^2-b_3 b_2+b_3^2\right)\right)} \, , \\
 \Delta_3 & = \frac{2 (2 b_1-b_2-b_3) (3 (b_1+b_2+b_3)-b_4) (3 (b_1+b_2+b_3)+b_4)}{3 b_1 \left(27 b_1^2+b_4^2-27 \left(b_2^2-b_3 b_2+b_3^2\right)\right)} \, , \\
 \Delta_4 & = \frac{(3 (b_1+b_2-2 b_3)-b_4) (3 (b_1-2 b_2+b_3)-b_4) (3 (b_1+b_2+b_3)-b_4)}{3 b_1 \left(27 b_1^2+b_4^2-27 \left(b_2^2-b_3 b_2+b_3^2\right)\right)} \ ,\\
 \Delta_5 & = \frac{(3 (b_1+b_2-2 b_3)+b_4) (3 (b_1-2 b_2+b_3)+b_4) (3 (b_1+b_2+b_3)+b_4)}{3 b_1 \left(27 b_1^2+b_4^2-27 \left(b_2^2-b_3 b_2+b_3^2\right)\right)} \, .
\eea
Notice that the R-charges \eqref{Delta:b:M111} are independent of the fluxes and they fulfill the following relation
\be
 \label{Delta:constraint:M111}
 3 \Delta_1 (9 \Delta_2 \Delta_3-4 \Delta_4 \Delta_5) - 4 \Delta_4 \Delta_5 \left( 3 ( \Delta_2 + \Delta_3 )+ 2 (\Delta_4+\Delta_5) \right) = 0 \, .
\ee
This constraint can be obtained by looking at
\be
 \sum_{a=1}^{5} B_a \frac{\partial a_{3\rd} (\Delta_a)}{\partial \Delta_a} = 0 \, ,
\ee
where $a_{3\rd} (\Delta_a)$ is given by \cite{Amariti:2011uw}
\bea
 a_{3\rd} (\Delta_a) & \equiv \frac{1}{24} \sum_{a , b , c , e = 1}^{5} \left| \det ( v_a , v_b , v_c , v_e ) \right| \Delta_a \Delta_b \Delta_c \Delta_e - \frac{8}{3} ( \Delta_4 \Delta_5 )^2 \\
 & = \Delta_1 \left( 9 \Delta_2 \Delta_3 \Delta_4 + 6 ( \Delta_2 + \Delta_3 ) \Delta_5 \Delta_4 + 9 \Delta_2 \Delta_3 \Delta_5 \right) + \frac{2}{3} \Delta_4 \Delta_5 ( 9 \Delta_2 \Delta_3 - 4 \Delta_4 \Delta_5 ) \, .
\eea 
The second equation in \eqref{constraints} also imposes the following constraint on the magnetic fluxes
\bea
 \label{flux:constraint:M111}
 0 & = \sum_{a=1}^{5} \fn_a \frac{\partial}{\partial \Delta_a} \left[ 3 \Delta_1 (9 \Delta_2 \Delta_3-4 \Delta_4 \Delta_5) - 4 \Delta_4 \Delta_5 \left( 3 ( \Delta_2 + \Delta_3 )+ 2 (\Delta_4+\Delta_5) \right) \right] \\
 & = 3 \left(4 \Delta_4 \Delta_5-9 \Delta_2 \Delta_3\right) \fn_1
 + 3 \left(4 \Delta_4 \Delta_5-9 \Delta_1 \Delta_3\right) \fn_2
 + 3 \left(4 \Delta_4 \Delta_5-9 \Delta_1 \Delta_2\right) \fn_3 \\
 & + 4 \Delta_5 \left(3 \Delta_1+3 \Delta_2+3 \Delta_3+4 \Delta_4+2 \Delta_5\right) \fn_4
 + 4 \Delta_4 \left(3 \Delta_1+3 \Delta_2+3 \Delta_3+2 \Delta_4+4 \Delta_5\right) \fn_5 \, .
\eea
This is equivalent to \eqref{flux constraints}.
Finally, defining the quantity, 
\be
 \begin{aligned}
  \cF (\Delta_a) & \equiv  N^{3/2} \sqrt{\frac{2 \pi^6}{27 \text{Vol}_{\text{S}}(M^{1,1,1})}} \\
  & = \frac{4 \pi N^{3/2}}{3 \sqrt{3}} \sqrt{\frac{\prod_{i=1}^{3} \prod_{j=4}^{5} (3 \Delta_i + 2 \Delta_j)}
  {9 \sum_{i<j}^{3} \Delta_i \Delta_j  + 6 (\Delta_4 + \Delta_5 ) \sum_{i=1}^3 \Delta_i + 4 \left( \Delta_4 ^2 +\Delta_4 \Delta_5 + \Delta_5^2 \right)}} \, ,
 \end{aligned}
\ee
we may rewrite \eqref{entropy:M111} as 
\bea
 S (b_i , \fn_a) & = - \frac12 \sum_{a=1}^{5} \fn_a \frac{\partial \cF (\Delta_a)}{\partial \Delta_a} \bigg|_{\Delta_a (b_i)}
 & = - \frac{2 \sqrt{2} \pi N^{3/2}}{3} \sum_{a=1}^{5} \fn_a \frac{\partial \sqrt{a_{3\rd}(\Delta_a)}}{\partial \Delta_a} \bigg|_{\Delta_a(b_i)} \, .
\eea
Also, observe that
\be
 \cV_{\text{on-shell}} (b_i) = \frac{1}{64 \pi^3} \cF (\Delta_a) \bigg|_{\Delta_a (b_i)} = \frac{N^{3/2}}{24 \sqrt{2} \pi^2} \sqrt{a_{3\text{d}} (\Delta_a)} \bigg|_{\Delta_a(b_i)} \, .
\ee

\section{Discussion and conclusions}
\label{discussion}

In this paper we investigated the relation between $\cI$-extremization and its gravitational dual, recently proposed  in \cite{Couzens:2018wnk,Gauntlett:2018dpc}, and, we provided many examples and a large class  
of twisted compactifications where the two extremizations agree off-shell. Our analysis also raises many questions and open problems. 

In particular, as noticed in \cite{Hosseini:2016tor,Hosseini:2016ume,Azzurli:2017kxo}, baryonic symmetries disappear  in the large $N$ limit of the topologically twisted index for known three-dimensional quiver gauge theories with holographic duals and $N^{3/2}$ scaling. In order to match the existing field theory computation with the formalism of \cite{Couzens:2018wnk,Gauntlett:2018dpc}, we restricted to a particular class of twisted compactifications, characterized by the absence of a twist in the baryonic directions. However, there certainly exist AdS$_2$ solutions with baryonic fluxes \cite{Donos:2012sy,Halmagyi:2013sla,Azzurli:2017kxo}. It would be very interesting to understand if there is a way to introduce baryonic charges in the large $N$ limit considered  in \cite{Hosseini:2016tor,Hosseini:2016ume,Azzurli:2017kxo} or to find more general saddle points and compare the result with 
the construction of \cite{Couzens:2018wnk,Gauntlett:2018dpc}.

Other obvious questions concern the  interpretation of the cubic constraints \eqref{Rcharge constraints}. Notice that the constraints are a consequence of the Sasakian parameterization \eqref{Rcharges}.
Therefore they already show up in  the three-dimensional aspects of the physics and in the discussion about the equivalence between $F$-maximization and  volume minimization. From the physical point of view, 
 $F_{S^3}$ is only function of the mesonic directions, and its extremization  leads to a prediction for the R-charges of the mesonic operators of the theory, corresponding to the KK modes of the compactification. The $d-4$ constraints \eqref{Rcharge constraints} allow to determine the on-shell value of all the R-charges $\Delta_a$ and lead to a prediction for the R-charges and dimensions of the baryonic operators also, since these are usually obtained by wrapping M5-branes on certain linear combinations of the cycles $S_\alpha$. As we saw in various examples, some of the constraints arise in the large $N$ limit of QFT partition functions when we impose that the theory has gauge group $\SU(N)$ and the distribution of eigenvalues is traceless. However, not all baryonic symmetries for the known quivers are related to a $\U(1)$ subgroup of a $\U(N)$ gauge symmetry. Some of them appear as symmetries rotating the flavors. In all these cases, a field theory interpretation is still missing.

Similarly, the role of the quartic polynomial $a_{3\rd}(\Delta)$ is still unclear both from the geometrical and physical point of view. It is a quartic polynomial with the property that, when restricted to the Sasakian parameterization \eqref{Rcharges}, it coincides with $F_{S^3}^2$ as a function of $b_i$. It has been originally found   by analyzing examples in \cite{Amariti:2011uw,Amariti:2012tj} but a general formula is still lacking. It is known to be of the form \cite{Amariti:2011uw,Amariti:2012tj} 
\be
 \label{a3d:discussion}
 a_{3\text{d}} (\Delta_a) = \frac{1}{24} \sum_{a , b , c , e = 1}^{d} \left| \det ( v_a , v_b , v_c , v_e ) \right| \Delta_a \Delta_b \Delta_c \Delta_e + \text{quartic corrections} \, ,
\ee
where the correction terms are related to internal lines in the toric diagram. Without the corrections terms, this expression would be the analogue of 
the well-known expression for the trial $a$ central charge of the quiver associated to D3-branes at Calabi-Yau toric three-folds \cite{Benvenuti:2006xg}.
In this paper, we further noticed that, quite remarkably,  the constraints among R-charges can be written in terms of $a_{3\rd}(\Delta)$ using \eqref{Rcharge constraints}.
It would be interesting to find a direct geometric interpretation for $a_{3\rd}(\Delta)$.
Even from the physical point of view, the analogy of $a_{3\rd}(\Delta)$ with the four-dimensional trial central charge  $a_{4\rd}(\Delta)$ is quite intriguing.
As its four-dimensional analogue \cite{Butti:2005vn}, $a_{3\rd}(\Delta)$ is automatically extremized with respect to the baryonic symmetries
and it coincides off-shell with the inverse of the volume functional --- which is also $F_{S^3}^2(\Delta_a)$ --- when imposing the Sasakian parameterization \eqref{Rcharges}.
Therefore, $F$-maximization is also equivalent to the extremization of $a_{3\rd}(\Delta)$ with respect to all the directions, including the baryonic ones. It would be interesting to see if there is a field theory interpretation of this observation. 

Similar questions arise for the twisted compactifications of the three-dimensional theories on $\Sigma_\fg$. In particular, it would be nice to find a purely field theory interpretation of the constraints \eqref{S20}, or equivalently \eqref{flux constraints}. 
It would be also interesting to see if $a_{3\rd}(\Delta)$ plays some role in the solution to the equations in \cite{Couzens:2018wnk,Gauntlett:2018dpc} for a generic choice of fluxes. In particular, in all our examples for the mesonic twist it is true that
\be \cI (\fn_a , \Delta_a) = - \frac{2 \sqrt{2} \pi N^{3/2}}{3} \sum_{a=1}^{d} \fn_a \frac{\partial \sqrt{a_{3\rd}(\Delta_a)}}{\partial \Delta_a} \, .\ee
It would be interesting to see if there is a similar expression for an arbitrary twist. 

\section*{Acknowledgements}

We would like to thank Noppadol Mekareeya, Francesco Sala, and Yuji Tachikawa for very useful discussions and collaboration on related topics.
The work of SMH was supported by World Premier International Research Center Initiative (WPI Initiative), MEXT, Japan.
AZ is partially supported by the INFN and ERC-STG grant 637844-HBQFTNCER.

\appendix

\section{Simplifying the supersymmetry conditions for toric manifolds}
\label{sec:app}

In this appendix we discuss some geometrical aspects of the toric manifolds $Y_7$ considered in the main text and we prove the results presented in sections \ref{subsec:UT} and \ref{subsec:mesonic:twist}.

\subsection{Master and Sasaki volumes} \label{sec:appgeo}

The cone over a Sasaki manifold $Y_7(b_i)$, $C(Y_7(b_i))$,  is  a K\"ahler manifold but it is not in general   Calabi-Yau. 
Considering the Reeb vector 
\bea \xi = b_1 \partial_z = \sum_{i=1}^4 b_i \partial_{\varphi_i}\, ,\eea
and the dual one-form $\eta$, $i_{\xi} \eta =1$, for a Sasaki manifold  we have $\rd\eta =2 \omega_{\text{S}}$, where $\omega_{\text{S}}$ is the K\"ahler form transverse to the Reeb foliation. For a Sasaki manifold, we also have $\rho=2 b_1 \omega_{\text{S}}$, where $\rho$ is the Ricci two-form on $Y_7(b_i)$.
The volumes of $Y_7(b_i)$ and of its cycles $S_a(b_i)$ are given by the explicit formulae \cite{Martelli:2005tp,Hanany:2008fj}
\bea
 \label{Sasaki:volumesII}
 &\text{Vol}_{\text{S}}(Y_7) =  \int_{Y_7} \eta \wedge \frac{\omega_{\text{S}}^3}{6} = \frac{\pi}{3 b_1} \sum_{a=1}^d \text{Vol}_{\text{S}}(S_a) \, , \\
 & \text{Vol}_{\text{S}}(S_a) =   \int_{S_a} \eta \wedge \frac{\omega_{\text{S}}^2}{2} = \pi^3 \sum_{k = 2}^{\ell_a-1} \frac{ (v_{a}, w_{k-1},w_{k},w_{k+1}) (v_{a}, w_{k},w_{1},w_{\ell_a})}{(v_{a}, b,w_{k},w_{k+1})(v_{a}, b,w_{k-1},w_{k})(v_{a}, b,w_{1},w_{\ell_a})}  \, ,
\eea
where $w_a$,  $k = 1 , \ldots , \ell_a$, is a  counterclockwise ordered sequence of vectors  adjacent to $v_a$.

For the backgrounds in \cite{Couzens:2018wnk,Gauntlett:2018dpc},\footnote{We use the same symbols, $\eta$, $\omega$ and $\rho$ for the forms  on the fibration $Y_9$, and their restriction to the manifold $Y_7$.}  it is still true that $\rd\eta=\rho/b_1$ but now the restriction of $\rho$ and $\omega$ to $Y_7$ are no more proportional. However, it is still true that
\bea
 \label{Sasaki:volumes2}
 &  \frac{1}{(2b_1)^3}\int_{Y_7} \eta \wedge \frac{\rho^3}{6} =  \text{Vol}_{\text{S}}(Y_7)  \, , \\
 &  \frac{1}{(2b_1)^2} \int_{S_a} \eta \wedge \frac{\rho^2}{2} = \text{Vol}_{\text{S}}(S_a)  \, ,
\eea
where  the subscript S indicates us that the volume are computed for the Sasaki metric on $Y_7(b_i)$. Indeed, the integrals in \eqref{Sasaki:volumes2} can be computed using the Duistermaat-Heckman localization formula  using a resolution of the complex cone $C(Y_7)$  and $\rd\eta=\rho/b_1$ \cite{Martelli:2006yb,Couzens:2018wnk}. Since we can use a fixed point formula for evaluating the integrals, the result coincides with the formulae given in \cite{Martelli:2005tp,Martelli:2006yb}.

Consider now the master volume \eqref{master:volume:1D}. Using the expression \eqref{omega:rho:toric} for $\omega$, we see that $\cV$ is a cubic form in $\lambda_a$
\be\label{mastercubic}
 \cV =  \frac{1}{6} \int_{Y_7} \eta \wedge \omega^3=\frac16 \sum_{a , b , c = 1}^{d} J_{abc} \lambda_a \lambda_b \lambda_c \, ,
\ee
where
\bea\label{intersection} J_{abc}= -(2\pi)^3 \int_{Y_7} \eta \wedge c_a \wedge c_b \wedge c_c \, .\eea 
For $\lambda_a=-1/(2b_1)$, the metric becomes Sasaki and the master volume coincides with $ \text{Vol}_{\text{S}}(Y_7)$. More generally,
using the expression \eqref{omega:rho:toric} for $\rho$, we also  see that the Sasakian volumes \eqref{Sasaki:volumes2} are related to the intersection numbers by\footnote{We used $\int_{Y_7} \eta \wedge c_a \wedge \rho^2= \int_{S_a} \eta \wedge \rho^2$.} 
\be
 \label{Sasakian:volumes:J}
 \text{Vol}_{\text{S}}(Y_7) = - \frac{1}{48 b_1^3} \sum_{a,b,c=1}^{d} J_{abc} \, , \qquad \text{Vol}_{\text{S}} (S_a) = - \frac{1}{16 \pi b_1^2} \sum_{b,c=1}^{d} J_{abc} \, .
\ee

\subsection{Supersymmetry conditions: universal twist}\label{sec:appUT}

The universal twist is defined by 
\be \label{UTfl}
 n_i = \frac{b_i}{b_1} n_1 \, , \quad \forall i = 1,\ldots , 4 \, .
\ee
where $n_1 = 2 - 2 \fg$. We can solve \eqref{constraints} by taking all the $\lambda_a$ equal
\be
 \label{lambda:universal:twist}
 \lambda_a = - \frac{1}{2b_1} \lambda \, , \quad \forall a = 1,\ldots , d \, .
\ee
Using \eqref{mastercubic} and \eqref{Sasakian:volumes:J} we find 
\be
 \begin{aligned}
  & \hskip 4truecm  \cV = -\frac{\lambda^3}{6 (2 b_1)^3} \sum_{a,b,c=1}^d J_{abc} =  \lambda^3 \text{Vol}_{\text{S}}(Y_7) \, , \\
  & \qquad \frac{\partial \cV}{\partial \lambda_a}= \frac{\lambda^2}{2 (2 b_1)^2} \sum_{b,c=1}^d J_{abc} = - 2 \pi \lambda^2 \text{Vol}_{\text{S}}(S_a) \, , \qquad
  \sum_{a = 1}^{d} \frac{\partial \cV}{\partial \lambda_a} = - 6 b_1 \lambda^2 \text{Vol}_{\text{S}}(Y_7) \, , \\
  & \sum_{b = 1}^d \frac{\partial^2 \cV}{\partial \lambda_a \partial \lambda_b} =-\frac{\lambda}{2 b_1} \sum_{b,c=1}^d J_{abc}  = 8 \pi b_1 \lambda \text{Vol}_{\text{S}}(S_a) \, , \qquad
  \sum_{a , b = 1}^d \frac{\partial^2 \cV}{\partial \lambda_a \partial \lambda_b}  = 24 b_1^2 \lambda \text{Vol}_{\text{S}}(Y_7) \, .
 \end{aligned}
\ee
Note that $\cV$, $\text{Vol}_{\text{S}}(Y_7)$, and $\text{Vol}_{\text{S}}(S_a)$ are homogeneous functions of $b_i$ of degree $-1$, $-4$, and $-3$, respectively.
The set of equations \eqref{constraints} can then be rewritten as
\bea
 \label{constraints:UTwist:I}
 & N = 6 b_1 \lambda^2 \text{Vol}_{\text{S}}(Y_7) \, , \\
 & \fn_a N = - 2 ( 2 A b_1 + \pi n_1 \lambda ) \lambda \text{Vol}_{\text{S}}(S_a) \, ,\\
 & A = - \frac{\pi n_1}{b_1} \lambda \, ,
\eea
where we used \eqref{UTfl} and  $\sum_{i=1}^4 b_i \partial_{b_i} \cV= -\cV$.
Let us emphasize that the above set of equations depends on the choice of the Reeb vector through $\text{Vol}_{\text{S}}(Y_7)$ and $\text{Vol}_{\text{S}}(S_a)$.
We thus obtain
\be\label{solUT}
 \begin{aligned}
  & \lambda = \pm \sqrt{\frac{N}{6 b_1 \text{Vol}_{\text{S}}(Y_7)}} \, , \qquad
  A = - \frac{\pi n_1}{b_1} \lambda \, , \\
  & \fn_a =\frac{n_1}{2} \left( \frac{2\pi}{3 b_1} \frac{\text{Vol}_{\text{S}}(S_a)}{\text{Vol}_{\text{S}}(Y_7)} \right) \equiv \frac{n_1}{2} \Delta_a \, ,
 \end{aligned}
\ee
where we introduced the set of basic R-charges \eqref{Rcharges}. Notice that $\sum_{a=1}^d \Delta_a(b_i)=2$.
Note also that $\sum_{a=1}^d \fn_a = n_1$ hence \eqref{twisting} is correctly satisfied. Evaluating the entropy functional \eqref{entropy:functional} and using the plus sign in \eqref{solUT}, we find 
\bea
 S (b_i,\fn_a) 
  = - \frac{8 n_1}{b_1^{3/2}} N^{3/2} \sqrt{\frac{2 \pi^6}{27 \text{Vol}_{\text{S}}(Y_7)}} \, ,
\eea
thus reproducing \eqref{SBH1}. 

\subsection{Supersymmetry conditions: mesonic twist}
\label{sec:appMT}

The mesonic twist is characterized by the condition
\be
 \label{Decoupling:lambda2}
 \sum_{a = 1}^{d} B_a^{(i)} \lambda_a = 0 \, , \qquad \forall i = 1, \ldots, d - 4 \, ,
\ee
where $B_a^{(i)}$ are baryonic symmetries satisfying \eqref{baryonic}. 
We can use the invariance \eqref{gauge} to choose  the gauge
\be
 \label{gauge:lambda}
 \lambda_1 = \lambda_2 = \lambda_3 = 0 \, .
\ee
We now prove that  there exists a solution to the set of equations \eqref{constraints}, compatible with \eqref{Decoupling:lambda2}, such that 
\be
 \label{lambda:solution:toric2}
 \lambda_a = -\frac12 \frac{(v_1,v_2,v_3,v_a)}{(v_1,v_2,v_3,b)} \lambda \, , \quad \forall a = 1, \ldots , d \, .
\ee

We will use repeatedly the identity \cite{Gauntlett:2018dpc}
\be
 \label{id:J:lambda}
 \sum_{b=1}^{d} J_{abc} v^b_i =  \frac{b_i}{b_1} \sum_{b=1}^{d} J_{abc} \, .
\ee
Hence, for the R-charges \eqref{normRcharges} we find
\bea
 \Delta_a (b_i , \fn_a) & = - \frac{1}{N} \sum_{b,c=1}^{d} J_{abc} \lambda_b \lambda_c
 = - \frac{\lambda^2}{4 N b_1^2} \sum_{b,c=1}^{d} J_{abc} \, .
\eea
Imposing $\sum_{a=1}^d \Delta_a =2$ and using \eqref{Sasakian:volumes:J}, 
we fix the value of the K\"ahler parameter $\lambda$ 
\be
 \label{lambda:Jabc}
 \lambda^2 = - \frac{8 N b_1^2}{\sum_{a,b,c} J_{abc}} = \frac{N}{6 b_1 \text{Vol}_{\text{S}} (Y_7 )} \, ,
\ee
which depends on the choice of the Reeb vector through $\text{Vol}_{\text{S}}(Y_7)$.
Furthermore, we  discover that the $\Delta_a$ are actually independent of the fluxes $\fn_a$ and are given by
\be
 \label{Delta:J:ratio2}
 \Delta_a (b_i) = \frac{2 \sum_{b,c} J_{abc}}{\sum_{a,b,c} J_{abc}} = \frac{2 \pi}{3 b_1} \frac{\text{Vol}_{\text{S}} (S_a)}{\text{Vol}_{\text{S}}(Y_7)} \, .
\ee
We can also evaluate the on-shell value of the master volume (see \eqref{master:volume:on-shell}). It reads
\be
 \cV_{\text{on-shell}} (b_i) = -\frac{\lambda^3}{48 b_1^3} \sum_{a , b , c = 1}^{d} J_{abc} 
 =  \frac{N^{3/2}}{4 \pi^3 b_1^{3/2}}  \sqrt{\frac{2 \pi^6}{27 \text{Vol} (Y_7(b_i))}} \, ,
\ee
thus reproducing \eqref{masterMT}.
From \eqref{id:J:lambda}, we can also derive the useful identity 
\be
 \label{Reeb}
 2 \frac{b_k}{b_1} = \sum_{a=1}^d v_a^k \Delta_a(b_i) \, , \qquad \forall k = 1, \ldots , 4 \, ,
\ee
which is actually the simplest way of comparing the two sides of \eqref{free}.

Let us now move and solve the equations \eqref{constraints}. We already used the first equation in order to find $\lambda$.
The other equations can be written as
\bea
 \label{constraints:mesonic:twist}
 & \fn_a N = -\frac{A}{2\pi} \sum_{b,c=1}^d J_{abc} \lambda_c - \frac{b_1}{2} \sum_{b,c=1}^{d} \nabla J_{abc} \lambda_b \lambda_c \, ,\\
 & \frac{A}{2 \pi} \sum_{a,b,c=1}^d J_{abc} \lambda_c = - N n_1 - \frac{b_1}{2} \sum_{a,b,c=1}^d  \nabla J_{abc} \lambda_b \lambda_c
 \, ,
\eea
where  we introduced the operator $\nabla \equiv \sum_{i=1}^{4} n_i \partial_{b_i}$. We can write
\bea
 \label{id:nabla:J:lambda}
 \sum_{c=1}^d \nabla J_{abc} \lambda_c
  = -  \frac{\lambda}{2 b_1} \left[ \sum_{c=1}^d \nabla J_{abc} + \sum_{c=1}^d J_{abc} \left( \frac{(v_1,v_2,v_3,n)}{(v_1,v_2,v_3,b)} - \frac{n_1}{b_1} \right) \right] \, ,
\eea
where we used the identity \cite{Gauntlett:2018dpc}
\be
 \sum_{a=1}^d v_a^k \nabla J_{abc} = \frac{b_k}{b_1} \sum_{a=1}^d \nabla J_{abc} + \left( \frac{n^k}{b_1} - \frac{n_1 b_k}{(b_1)^2} \right) \sum_{a=1}^d J_{abc} \, , \qquad \forall k = 1, \ldots, 4 \, ,
\ee
that follows from \eqref{id:J:lambda}. Thus, we solve the second equation in \eqref{constraints:mesonic:twist} for $A$
\bea
 \label{A:mesonic:twist}
 A = -  \frac{\pi \lambda}{2} \left( \frac{3 n_1}{b_1} - \frac{\sum_{a,b,c} \nabla J_{abc}}{\sum_{a,b,c} J_{abc}} - 2 \frac{(v_1,v_2,v_3,n)}{(v_1,v_2,v_3,b)} \right) \, ,
\eea
where we used \eqref{lambda:Jabc}. Using  \eqref{lambda:Jabc} and \eqref{A:mesonic:twist} we can now evaluate the entropy functional \eqref{entropy:functional}. It reads 
\bea
 S (b_i,\fn_a) & = 8 \pi^2 \bigg ( N A - \frac{\pi b_1}{3} \sum_{a,b,c=1}^{d} \nabla J_{abc} \lambda_a \lambda_b \lambda_c \bigg ) 
 = - \frac{16 \sqrt{2} \pi^3}{3 \sqrt{b_1}} N^{3/2} \nabla \frac{b_1^{3/2}}{\sqrt{-\sum_{a,b,c} J_{abc}}} \, .
\eea
Using \eqref{Sasakian:volumes:J}, we can finally write
\be\label{S102}
 S (b_i,\fn_a) = - \frac{4}{\sqrt{b_1}} \nabla \sqrt{\frac{2 \pi^6}{27 \text{Vol}_{{\text S}}(Y_7)}} N^{3/2} \, ,
\ee
thus reproducing \eqref{S10}.

Now let us go back and see what constraints the mesonic twist imposes on the fluxes (as the equations \eqref{constraints:mesonic:twist} depend on both mesonic and baryonic fluxes).
Consider the first equation in \eqref{constraints:mesonic:twist}.
Using \eqref{id:J:lambda}, \eqref{lambda:Jabc}, \eqref{id:nabla:J:lambda}, and \eqref{A:mesonic:twist}, we find that
\bea
 \label{S201}
 \fn_a N = n_1 N \frac{\sum_{b,c} J_{abc}}{\sum_{a,b,c} J_{abc}} + b_1 N \nabla \frac{\sum_{b,c} J_{abc}}{\sum_{a,b,c} J_{abc}} \, .
\eea
Using \eqref{Delta:J:ratio2} we then obtain
\bea
 \label{S202}
 \fn_a = \frac{n_1}{2} \Delta_a + \frac{b_1}{2} \nabla \Delta_a = \frac{1}{2} \nabla ( b_1 \Delta_a) \, , \qquad \forall a = 1, \ldots, d \, ,
\eea
thus reproducing \eqref{S20}. Notice that, as required by consistency, 
\be
 \label{S203}
 \sum_{a=1}^d v_a \fn_a
 = \frac{1}{2} \sum_{a=1}^d \nabla ( b_1 v_a \Delta_a)
 = \nabla b =n \, ,
\ee
where we used \eqref{Reeb}. 

\section{Explicit expressions of master volumes}\label{app:mastervol} 

In this appendix we collect the expressions for the master volume \eqref{master:volume:1D} of the toric examples discussed in section \ref{sec:Examples}.
\paragraph*{The cone over $Q^{1,1,1}$.}
{\footnotesize
\bea
 \label{Q111:master:volume}
 \cV_{Q^{1,1,1}} & = \frac{8 \pi^4 (b_3+b_4) (b_1 - b_3-b_4)^2 \lambda_1^3}{3 (b_1-b_2) b_3 b_4 (b_1-b_2-b_3-b_4)}
 + \frac{8 \pi^4 (b_2+b_4) (b_1-b_2-b_4)^2 \lambda_2^3}{3 b_2 (b_1-b_3) b_4 (b_1-b_2-b_3-b_4)} \\
 & + \frac{8 \pi^4 (b_2+b_3) (b_1-b_2-b_3)^2 \lambda_3^3}{3 b_2 b_3 (b_1-b_4) (b_1-b_2-b_3-b_4)}
 - \frac{8 \pi^4 (2 b_1-b_2-b_4) (b_1-b_2-b_4)^2 \lambda_4^3}{3 (b_1-b_2) b_3 (b_1-b_4) (2 b_1-b_2-b_3-b_4)} \\
 & - \frac{8 \pi^4 (2 b_1-b_2-b_3) (b_1-b_2-b_3)^2 \lambda_5^3}{3 (b_1-b_2) (b_1-b_3) b_4 (2 b_1-b_2-b_3-b_4)}
 - \frac{8 \pi^4 (2 b_1-b_3-b_4) (b_1-b_3-b_4)^2 \lambda_6^3}{3 b_2 (b_1-b_3) (b_1-b_4) (2 b_1-b_2-b_3-b_4)} \\
 & + 8 \pi^4 (b_1-b_3-b_4) \bigg( \frac{\lambda_2}{b_4 (b_1-b_2-b_3-b_4)} + \frac{\lambda_3}{b_3 (b_1-b_2-b_3-b_4)}
 - \frac{\lambda_4}{(b_1-b_2) b_3} - \frac{\lambda_5}{(b_1-b_2) b_4} \bigg) \lambda_1^2 \\
 & - \left(\frac{8 \pi^4 (b_1-b_2-b_3) \lambda_4}{b_3 (b_1-b_4)} + \frac{8 \pi^4 (b_1-b_2-b_3) \lambda_6}{b_2 (b_1-b_4)}\right) \lambda_3^2 \\
 & + \left(\frac{8 \pi^4 (b_1-b_2-b_4) \lambda_3}{b_2 (b_1-b_2-b_3-b_4)} - \frac{8 \pi^4 (b_1-b_2-b_4) \lambda_5}{(b_1-b_3) b_4} - \frac{8 \pi^4 (b_1-b_2-b_4) \lambda_6}{b_2 (b_1-b_3)}\right) \lambda_2^2 \\
 & + \left( \frac{8 \pi^4 (b_1-b_2-b_4) \lambda_5}{(b_1-b_2) (2 b_1-b_2-b_3-b_4)} + \frac{8 \pi^4 (b_1-b_2-b_4) \lambda_6}{(b_1-b_4) (2 b_1-b_2-b_3-b_4)}\right) \lambda_4^2 \\
 & + \bigg( \frac{8 \pi^4 (b_1-b_2-b_4) \lambda_2^2}{b_4 (b_1-b_2-b_3-b_4)} + \left( \frac{16 \pi^4 \lambda_3}{b_1 - b_2-b_3-b_4} - \frac{16 \pi^4 \lambda_5}{b_4}\right) \lambda_2
 + \frac{8 \pi^4 (b_1-b_2-b_3) \lambda_3^2}{b_3 (b_1-b_2-b_3-b_4)} \\
 & + \frac{8 \pi^4 (b_1-b_2-b_4) \lambda_4^2}{(b_1-b_2) b_3}
 + \frac{8 \pi^4 (b_1-b_2-b_3) \lambda_5^2}{(b_1-b_2) b_4} - \frac{16 \pi^4 \lambda_4 \lambda_5}{b_1-b_2} - \frac{16 \pi^4 \lambda_3 \lambda_4}{b_3} \bigg) \lambda_1 \\
 & + 8 \pi^4 \left(\frac{(b_1-b_2-b_3) \lambda_3^2}{b_2 (b_1-b_2-b_3-b_4)} - \frac{2 \lambda_6 \lambda_3}{b_2} + \frac{(b_1-b_2-b_3) \lambda_5^2}{(b_1-b_3) b_4}
 + \frac{(b_1-b_3-b_4) \lambda_6^2}{b_2 (b_1-b_3)} - \frac{2 \lambda_5 \lambda_6}{b_1-b_3}\right) \lambda_2 \\
 & + 8 \pi^4 \left( \frac{(b_1-b_2-b_4) \lambda_4^2}{b_3 (b_1-b_4)} - \frac{2 \lambda_6 \lambda_4}{b_1-b_4} + \frac{(b_1-b_3-b_4) \lambda_6^2}{b_2 (b_1-b_4)}\right) \lambda_3 \\
 & + \frac{8 \pi^4}{2 b_1-b_2-b_3-b_4} \left( \frac{(b_1-b_2-b_3) \lambda_5^2}{b_1-b_2} - 2 \lambda_6 \lambda_5 + \frac{(b_1-b_3-b_4) \lambda_6^2}{b_1-b_4}\right) \lambda_4 \\
 & + \frac{8 \pi^4 (b_1-b_3-b_4) \lambda_5 \lambda_6^2}{(b_1-b_3) (2 b_1-b_2-b_3-b_4)}
 + \frac{8 \pi^4 (b_1-b_2-b_3) \lambda_5^2 \lambda_6}{(b_1-b_3) (2 b_1-b_2-b_3-b_4)} \, .
\eea
}
\paragraph*{Flavoring $\cN = 8$ SYM.}
{\footnotesize
\bea
 \label{N=8:SYM:master:volume}
 \cV_{\cN=8\text{ SYM}} & = \frac{8 \pi^4 \left(b_3 b_4 r_2 - \left( ( b_1-b_2-b_3 ) r_1 - b_4 \right)^2 \right) \lambda_1^3}{3 b_2 b_3 b_4 r_1^2} 
 - \frac{8 \pi^4 \left( (b_3 r_2 - b_4 )^2-b_2 b_4 r_3\right) \lambda_2^3}{3 b_2 (b_1-b_2-b_3 ) b_4 r_2^2} \\
 & + \frac{8 \pi^4 \left(b_4 \left( (b_1-b_2-b_3) r_1 + 2 b_2 r_3 \right) - b_2^2 r_3^2 - b_4^2\right) \lambda_3^3}{3 (b_1-b_2-b_3) b_3 b_4 r_3^2} \\
 & + \frac{8 \pi^4 \left( (b_3 r_2 - b_4 )^2 - b_2 r_3 \left( (b_1-b_2-b_3) r_1 - b_3 r_2 + b_4 \right) \right) \lambda_4^3}{3 b_2 b_3 r_1^2 \left(-b_1 r_1+b_3 ( r_1-r_2 )+b_2 (r_1-r_3)+b_4\right)} \\
 & - \frac{8 \pi^4 \lambda_5^3}{3 (b_1-b_2-b_3)} \left( \frac{r_3}{r_2^2} - \frac{b_4}{b_2 r_2^2} + \left(\frac{b_3}{b_1 r_1 - b_3 (r_1-r_2) - b_2 (r_1-r_3) - b_4} + \frac{1}{r_2} \right) \frac{b_3}{b_2} \right) \\
 & +\frac{8 \pi^4 \left(\left(b_4-(b_1-b_2-b_3) r_1\right) \left(b_4-b_3 r_2-(b_1-b_2-b_3) r_1 \right)-b_2 b_3 r_2 r_3\right) \lambda_6^3}{3 (b_1-b_2-b_3) b_3 r_3^2 \left(b_4 -b_1 r_1+b_3 (r_1-r_2)+b_2 (r_1-r_3)\right)} \\
 & - 8 \pi^4 \left(\frac{(b_1 - b_2 - b_3 ) \lambda_2}{b_2 b_4}- \frac{\left(b_4-(b_1 - b_2 - b_3 ) r_1\right) \lambda_3}{b_3 b_4 r_1}
 - \frac{\left(b_4-(b_1 - b_2 - b_3 ) r_1-b_3 r_2\right) \lambda_4}{b_2 b_3 r_1^2}+\frac{\lambda_5}{b_2 r_1}\right) \lambda_1^2 \\
 & - 8 \pi^4 \left( \frac{(b_3 r_2 - b_4 ) \lambda_2^2}{b_2 b_4 r_2} + 2 \left(\frac{\lambda_3}{b_4} + \frac{\lambda_5}{b_2 r_2}\right) \lambda_2
 - \frac{(b_3 r_2-b_4) \lambda_4^2}{b_2 b_3 r_1^2} - \frac{\lambda_5^2}{b_2 r_2} + \frac{b_2 \lambda_3^2}{b_3 b_4}
 + \frac{2 \lambda_4 \lambda_5}{b_2 r_1} + \frac{2 \lambda_3 \lambda_4}{b_3 r_1}\right) \lambda_1 \\
 & - \frac{8 \pi^4}{b_1 - b_2 - b_3} \left(\frac{b_3 \lambda_3}{b_4} - \frac{(b_4-b_3 r_2-b_2 r_3) \lambda_5}{b_2 r_2^2} - \frac{\lambda_6}{r_2}\right) \lambda_2^2 \\
 & + 8 \pi^4 \left(\frac{\lambda_4}{b_3 r_3} + \frac{\left(b_4-(b_1 - b_2 - b_3 ) r_1-b_2 r_3\right) \lambda_6}{\left(b_1-b_2-b_3\right) b_3 r_3^2}\right) \lambda_3^2 \\
 & + \frac{8 \pi^4}{ \left(b_4-b_1 r_1+b_3 (r_1-r_2)+b_2 (r_1-r_3)\right)}
 \left(\frac{(b_4-b_3 r_2-b_2 r_3) \lambda_5}{b_2 r_1} + \frac{(b_1 - b_2 - b_3 ) \lambda_6}{b_3}\right) \lambda_4^2 \\
 & - \frac{8 \pi^4}{b_1 - b_2 - b_3} \left( \lambda_3^2 \left(\frac{b_2}{b_4}-\frac{1}{r_3}\right)
 +\frac{\lambda_5^2 \left(b_4-b_2 r_3\right)}{b_2 r_2^2}+\frac{2 \lambda_6 \lambda_3}{r_3}+\frac{2 \lambda_5 \lambda_6}{r_2}-\frac{\lambda_6^2}{r_3} \right) \lambda_2 \\
 & + 8 \pi^4 \left(\frac{\lambda_4^2}{b_3 r_1}-\frac{2 \lambda_6 \lambda_4}{b_3 r_3}-\frac{\left(b_4-(b_1 - b_2 - b_3 ) r_1\right) \lambda_6^2}{\left(b_1-b_2-b_3\right) b_3 r_3^2}\right) \lambda_3 \\
 & + \frac{8 \pi^4 \lambda_4}{b_4-b_1 r_1+b_3 (r_1-r_2)+b_2 (r_1-r_3)} \left(\frac{b_3 \lambda_5^2}{b_2} + 2 \lambda_6 \lambda_5 + \frac{b_2 \lambda_6^2}{b_3} \right) + \frac{8 \pi^4 \lambda_6^2 \lambda_4}{b_3 r_3} \\
 & + \frac{8 \pi^4 b_2 \lambda_5 \lambda_6^2}{(b_1 - b_2 - b_3 ) \left(b_4-b_1 r_1+b_3 (r_1-r_2)+b_2 (r_1-r_3)\right)} \\
 & + \frac{8 \pi^4 \left(b_4-(b_1 - b_2 - b_3 ) r_1-b_2 r_3\right) \lambda_5^2 \lambda_6}{(b_1-b_2-b_3) r_2 \left(b_4 - b_1 r_1 + b_3 (r_1-r_2) - b_2 (r_1-r_3)\right)} \, .
\eea
}
\paragraph*{The SPP theory.}
{\footnotesize
\bea
 \label{SPP:master:volume}
 \cV_{\text{SPP}} & = -\frac{4 \pi^4 \left(b_3^2+(2 b_1-b_2-2 b_4) (2 b_1-3 b_2-2 b_4)\right) \lambda_1^3}{3 (b_2^2-b_3^2) b_4}
 - \frac{16 \pi^4 (b_1-b_2) \lambda_5^2 \lambda_6}{(b_1-b_4 )^2 - b_3^2}
 - \frac{16 \pi^4 ( b_1 - b_2 - b_4 ) \lambda_5 \lambda_6^2} {b_3^2 - (b_1-b_4)^2} \\
 & - \frac{8 \pi^4 b_3^2 \left(2 (b_1+b_3)-b_4\right) \lambda_2^3}{3 (b_2+b_3) (b_1+b_3-b_4) b_4 (2 b_1 - b_2 + b_3 - b_4)}
 + \frac{4 \pi^4 \left( (4 b_1-3 b_2-b_4) (b_2-b_4)-b_3^2\right) \lambda_3^3}{3 (2 b_1-b_2+b_3-b_4) b_4 (2 b_1 - b_2 - b_3 - b_4)} \\
 & - \frac{8 \pi^4 b_3^2 \left( 2 b_1 - 2 b_3 - b_4\right) \lambda_4^3}{3 (b_2-b_3) b_4 ( b_1 - b_3 - b_4) ( 2 b_1 - b_2 - b_3 - sb_4)}
 - \frac{16 \pi^4 \left( (b_2-b_4 ) (b_2^2-b_3^2+b_4^2 ) - b_1 ( b_2^2 - b_3^2 - b_4^2 ) \right) \lambda_5^3}{3 (b_2-b_3 ) (b_2+b_3 ) (b_1+b_3-b_4 ) ( b_1 - b_3 - b_4 )} \\
 & + \frac{16 \pi^4 \left(4 b_4^3+6 (b_2-2 b_1) b_4^2+2 (6 b_1^2-7 b_2 b_1+2 b_2^2-b_3^2) b_4 - (b_1-b_2) \left( (b_2-2 b_1 )^2 - b_3^2 \right) \right) \lambda_6^3}
 {3 (b_1-b_3-b_4) (b_1+b_3-b_4) ( 2 b_1-b_2+b_3-b_4 ) ( 2 b_1 - b_2 - b_3 - b_4 )} \\
 & - 4 \pi^4 \lambda_1^2 \left(\frac{(2 b_1 - b_2 + b_3 - 2 b_4) \lambda_2}{(b_2+b_3) b_4} - \frac{\lambda_3}{b_4}
 + \frac{\left( 2 b_1 - b_2 - b_3 - 2 b_4 \right) \lambda_4}{(b_2-b_3) b_4}+\frac{4 ( b_1 - b_2 - b_4 ) \lambda_5}{b_2^2-b_3^2}\right) \\
 & + 8 \pi^4 \lambda_2^2 \left(\frac{b_3 \lambda_3}{(2 b_1-b_2+b_3-b_4) b_4} + \frac{b_3 \lambda_5}{(b_2+b_3) (b_1+b_3-b_4)}
 + \frac{b_3 \lambda_6}{(b_1+b_3-b_4) (2 b_1-b_2+b_3-b_4)}\right) \\
 & - 4 \pi^4 \lambda_3^2 \left( \frac{ ( b_2 - b_3 - b_4) \lambda_4}{b_4 ( 2 b_1 - b_2 - b_3 - b_4)}
 - \frac{4 (b_1-b_2) \lambda_6}{(2 b_1-b_2+b_3-b_4) ( 2 b_1 - b_2 - b_3 - b_4)}\right) \\
 & - 8 \pi^4 \lambda_4^2 \left(\frac{ b_3 \lambda_5}{(b_2-b_3) ( b_1 - b_3 - b_4)}
 + \frac{b_3 \lambda_6}{( b_1 - b_3 - b_4) ( 2 b_1 - b_2 - b_3 - b_4)} \right) \\
 & + 4 \pi^4 \lambda_1 \left(\frac{2 b_3 \lambda_2^2}{(b_2+b_3) b_4} - 2 \lambda_2 \left( \frac{\lambda_3}{b_4} + \frac{2 \lambda_5}{b_2+b_3}\right)
 + \frac{\lambda_3^2}{b_4} - \frac{2 b_3 \lambda_4^2}{(b_2-b_3) b_4} - \frac{4 \lambda_4 \lambda_5}{b_2-b_3} - \frac{4 b_4 \lambda_5^2}{b_2^2-b_3^2} - \frac{2 \lambda_3 \lambda_4}{b_4}\right) \\
 & + 4 \pi^4 \lambda_2 \bigg( - \frac{ (b_2+b_3-b_4 ) \lambda_3^2}{ (2 b_1-b_2+b_3-b_4 ) b_4} - \frac{4 \lambda_6 \lambda_3}{ 2 b_1 - b_2 + b_3 - b_4}
 + \frac{2 (b_2+b_3-b_4) \lambda_5^2}{(b_2+b_3) (b_1+b_3-b_4)} \\
 & + \frac{2 ( 2 b_1-b_2+b_3-2 b_4 ) \lambda_6^2}{ (b_1+b_3-b_4 ) ( 2 b_1-b_2+b_3-b_4 )} - \frac{4 \lambda_5 \lambda_6}{b_1+b_3-b_4} \bigg) \\
 & - \frac{8 \pi^4 \lambda_3}{2 b_1 - b_2 - b_3 - b_4} \left( \frac{b_3 \lambda_4^2}{b_4}
 + 2 \lambda_6 \lambda_4 + \frac{2 b_4 \lambda_6^2}{(2 b_1-b_2+b_3-b_4)} \right) \\
 & + \frac{8 \pi^4 \lambda_4}{b_1 - b_3 - b_4} \left(\frac{( b_2 - b_3 - b_4 ) \lambda_5^2}{b_2-b_3}
 - 2 \lambda_6 \lambda_5 + \frac{( 2 b_1-b_2-b_3-2 b_4) \lambda_6^2}{2 b_1 - b_2 - b_3 - b_4}\right)
 \, .
\eea
}
\paragraph*{The cone over $M^{1,1,1}$.}
{\footnotesize
\bea
 \label{M111:master:volume}
 \cV_{M^{1,1,1}} & = \frac{16 \pi^4 \left((b_3-2 b_2) b_4^2+9 (b_1-2 b_2+b_3) \left(-b_3^2-(b_1+2 b_2) b_3+2 b_2 (b_1+b_2)\right)\right) \lambda_1^3}
 {\left(3 (b_1+b_2-2 b_3)-b_4\right) \left(3 (b_1+b_2+b_3)-b_4\right) \left(3 (b_1+b_2-2 b_3)+b_4\right) \left(3 (b_1+b_2+b_3)+b_4\right)} \\
 & + \frac{16 \pi^4 \left(\left(b_2-2 b_3\right) b_4^2-9 (b_1+b_2-2 b_3) \left(-2 b_3^2+2 \left(b_2-b_1\right) b_3+b_2 (b_1+b_2)\right)\right) \lambda_2^3}
 {\left(9 (b_1-2 b_2+b_3)^2-b_4^2\right) \left(9 (b_1+b_2+b_3)^2-b_4^2\right)} \\
 & + \frac{16 \pi^4 \left((b_2+b_3) b_4^2-9 (b_1+b_2+b_3) \left(b_2^2-4 b_3 b_2+b_3^2+b_1 (b_2+b_3)\right)\right) \lambda_3^3}
 {\left(9 (b_1+b_2-2 b_3)^2-b_4^2\right) \left(9 (b_1-2 b_2+b_3)^2-b_4^2\right)} \\
 & - \frac{24 \pi^4 b_4^2 \lambda_4^3}{\left(3 (b_1+b_2-2 b_3)+b_4\right) \left(3 (b_1-2 b_2+b_3)+b_4\right) \left(3 (b_1+b_2+b_3)+b_4\right)} \\
 & - \frac{24 \pi^4 b_4^2 \lambda_5^3}{\left(3 (b_1+b_2-2 b_3)-b_4\right) \left(3 (b_1-2 b_2+b_3)-b_4\right) \left(3 (b_1+b_2+b_3)-b_4\right)} \\
 & + \bigg(\frac{48 \pi^4 b_2 \lambda_2}{9 (b_1+b_2+b_3)^2-b_4^2} + \frac{48 \pi^4 (b_2-b_3) \lambda_3}{9 (b_1+b_2-2 b_3)^2-b_4^2}
 + \frac{8 \pi^4 \left(b_4-3 (b_1-2 b_2+b_3)\right) \lambda_5}{\left(3 (b_1+b_2-2 b_3)-b_4\right) \left(3 (b_1+b_2+b_3)-b_4\right)} \\
 & - \frac{8 \pi^4 \left(3 (b_1-2 b_2+b_3)+b_4\right) \lambda_4}{\left(3 (b_1+b_2-2 b_3)+b_4\right) \left(3 (b_1+b_2+b_3)+b_4\right)} \bigg) \lambda_1^2 \\
 & + 8 \pi^4 \bigg(\frac{6 (b_3-b_2) \lambda_3}{9 (b_1-2 b_2+b_3)^2-b_4^2} + \frac{\left(b_4-3 (b_1+b_2-2 b_3)\right) \lambda_5}{\left(3 (b_1-2 b_2+b_3)-b_4\right) \left(3 (b_1+b_2+b_3)-b_4\right)} \\
 & - \frac{\left(3 (b_1+b_2-2 b_3)+b_4\right) \lambda_4}{\left(3 (b_1-2 b_2+b_3)+b_4\right) \left(3 (b_1+b_2+b_3)+b_4\right)} \bigg) \lambda_2^2 \\
 & + 8 \pi^4 \left(\frac{\left(b_4-3 (b_1+b_2+b_3)\right) \lambda_5}{\left(3 (b_1+b_2-2 b_3)-b_4\right) \left(3 (b_1-2 b_2+b_3)-b_4\right)}
 - \frac{\left(3 (b_1+b_2+b_3)+b_4\right) \lambda_4}{\left(3 (b_1+b_2-2 b_3)+b_4\right) \left(3 (b_1-2 b_2+b_3)+b_4\right)}\right) \lambda_3^2 \\
 & + 8 \pi^4 \bigg( \frac{6 b_3 \lambda_2^2}{9 (b_1+b_2+b_3)^2-b_4^2} - 2 \left(\frac{\lambda_4}{3 (b_1+b_2+b_3)+b_4} + \frac{\lambda_5}{3 (b_1+b_2+b_3)-b_4}\right) \lambda_2 \\
 & + \frac{3 b_4 \lambda_4^2}{\left(3 (b_1+b_2-2 b_3)+b_4\right) \left(3 (b_1+b_2+b_3)+b_4\right)}
 + 2 \left(\frac{\lambda_5}{b_4-3 (b_1+b_2-2 b_3)}-\frac{\lambda_4}{3 (b_1+b_2-2 b_3)+b_4}\right) \lambda_3 \\
 & - \frac{3 b_4 \lambda_5^2}{\left(3 (b_1+b_2-2 b_3)-b_4\right) \left(3 (b_1+b_2+b_3)-b_4\right)}-\frac{6 b_3 \lambda_3^2}{9 (b_1+b_2-2 b_3)^2-b_4^2} \bigg) \lambda_1 \\
 & - 8 \pi^4 \bigg( \frac{6 b_2 \lambda_3^2}{9 (b_1-2 b_2+b_3)^2-b_4^2} - 2 \left(\frac{\lambda_5}{b_4-3 (b_1-2 b_2+b_3)} - \frac{\lambda_4}{3 (b_1-2 b_2+b_3)+b_4}\right) \lambda_3 \\
 & - \frac{3 b_4 \lambda_4^2}{\left(3 (b_1-2 b_2+b_3)+b_4\right) \left(3 (b_1+b_2+b_3)+b_4\right)} + \frac{3 b_4 \lambda_5^2}{\left(3 (b_1-2 b_2+b_3)-b_4\right) \left(3 (b_1+b_2+b_3)-b_4\right)} \bigg) \lambda_2 \\
 & + 24 \pi^4 b_4 \left(\frac{\lambda_4^2}{\left(3 (b_1+b_2-2 b_3)+b_4\right) \left(3 (b_1-2 b_2+b_3)+b_4\right)}
 - \frac{\lambda_5^2}{\left(3 (b_1+b_2-2 b_3)-b_4\right) \left(3 (b_1-2 b_2+b_3) - b_4\right)}\right) \lambda_3 \, .
\eea
}

\bibliographystyle{ytphys}

\bibliography{I-extremization}

\end{document}